\documentclass[12pt]{article}
\usepackage{amsmath}
\usepackage{amsfonts}
\usepackage{enumerate}

\pdfoutput=1

\title{Spinors in extended Minkowski space}
\author{jason hanson}

\begin{document}
\maketitle

\begin{abstract}
The exterior algebra of Minkowski space naturally has the structure of a sixteen--dimensional Clifford algebra representation, and so can be used as the space of spinors.  We examine plane, circular, and spherical solutions to the free Dirac equation in this extended notion of Minkowski space, and indicate how they can be used to obtain solutions to the standard four--dimensional Dirac equation.
\end{abstract}

%%%%%%%%%%%%%%%%%%%%%%%%%%%%%%%%%%%%%%%%%%%%%%%%%%%%%%%%%%%%%%%%%
\section{Introduction}

Let ${\mathbb M}$ denote Minkowski space.  We choose the sign convention such that the metric in rectangular coordinates is $\eta={\it diag}(1,-1,-1,-1)$.  And for gamma matrices $\gamma_\alpha$, we choose the sign convention with $\gamma_\alpha\gamma_\beta+\gamma_\beta\gamma_\alpha=-2\eta_{\alpha\beta}$.  With these choices, the free Dirac equation is $\gamma^\alpha\partial_\alpha\psi=m\psi$.

In the standard formulation of Dirac theory, spinors are functions $\psi:{\mathbb M}\rightarrow V$ with values in a four--dimensional Clifford algebra representation.  However, the Dirac equation is under--determined in the sense that the gamma matrices are endomorphisms of $V$, but only defined up to similarity transform: we can replace $\gamma_\alpha$ with $\gamma_\alpha'=S\gamma_\alpha S^{-1}$ for any non--singular matrix $S$.  As as a consequence, solutions of the Dirac equation are similarly ambiguous.
  
Nonetheless, once we have chosen a set of gamma matrices, the Lorentz group acts on $V$ via the spin representation constructed from them.  Namely if $L=({L^\alpha}_\beta)$ is a Lorentz algebra element, so that $L^T\eta+\eta L=0$, the spin representation of $L$ is defined to be
\begin{equation}\label{eq:spinrep}
  \sigma(L)\doteq-\tfrac{1}{4}L^{\alpha\beta}\gamma_\alpha\gamma_\beta.
\end{equation}
And if $\Lambda$ is a Lorentz transformation with $\Lambda=\exp(L)$, then a spinor $\psi$ is transformed according to $\psi'=\exp(\sigma(L))\psi$.  The Dirac equation is invariant under the spin action.

The exterior bundle $\bigwedge_\ast M$ over space--time $M$ was introduced in \cite{me0} as a framework for coupling the Dirac and Einstein equations.  In this article, we restrict our attention to flat space--time.  That is, to Minkowski space ${\mathbb M}$.  The exterior bundle in this case is just the complexified exterior algebra $\bigwedge{\mathbb M}$, which we will call {\em extended Minkowski space.}  Spinors here are functions $\psi:{\mathbb M}\rightarrow\bigwedge{\mathbb M}$.  One benefit of working in extended Minkowski space is there is a natural choice for gamma matrices.  Namely $\gamma_\alpha\psi={\bf e}_\alpha\wedge\psi-\iota_\alpha\psi$.  Here ${\bf e}_\alpha$ is a basis vector for ${\mathbb M}$, and the two operations $\wedge$ and $\iota$ are the exterior and interior products, which are standard constructions defined on an exterior algebra.  In particular, the Dirac equation on extended Minkowski space is now unambiguous.

On the other hand, the spin action now becomes an ad hoc construction.  The standard Lorentz group action on ${\mathbb M}$ extends to an action on $\bigwedge{\mathbb M}$, so gives the natural action on extended Minkowski space.  We can still use equation \eqref{eq:spinrep} to define a secondary action of the Lorentz group.  As discussed in \cite{me0}, the Dirac equation on extended Minkowski space is in fact invariant under both the extended and spin actions.

Although the extended Minkowski space formulation of the Dirac equation involves spinors with sixteen components, as opposed to only four in the standard formulation, the underlying Lorentz geometry is preserved by virtue of the extended Lorentz action.  This can make searches for symmetric solutions of the Dirac equation more natural.

{\it Outline of article.}  We first review the constructions from \cite{me0} tailored to Minkowski space.  In subsequent sections, we find solutions to the free Dirac equation in extended Minkowski space that have plane, circular, and spherical symmetries.  In the case of plane solutions, we classify the solutions in terms of their spin and extended helicities.  And in the case of circular and spherical solutions, we find their plane wave decompositions.  In all cases, we show how the solutions can be projected to solutions of the standard free Dirac equation.

All computations in the article are straight--forward and can be verified either using a symbolic algebra package or by manual computation.  In the interest of completeness, I have uploaded to arXiv.org an extended version of this article which includes detailed computations of the statements made here.

%%%%%%%%%%%%%%%%%%%%%%%%%%%%%%%%%%%%%%%%%%%%%%%%%%%%%%%%%%%%%%%%%
\section{Extended Minkowski space}

%::::::::::::::::::::::::::::::::::::::::::::::::::::::::::::::::
\subsection{Basic constructions}

Here we give an overview of the basic exterior bundle constructions, but restricted to a flat space--time.  More details are given in \cite{me0}, although the constructions in this subsection are standard, and can be found for instance in \cite{Sternberg}, \cite{Lawson}, and \cite{Fulton}.

We extend Minkowski space ${\mathbb M}={\mathbb R}\{{\bf e}_0,{\bf e}_1,{\bf e}_2,{\bf e}_3\}$ by forming the complexified exterior algebra $\bigwedge{\mathbb M}$.  For a multi--index $I=\alpha_1\alpha_2\cdots\alpha_k$, define
$${\bf e}_I
  \doteq{\bf e}_{\alpha_1}\wedge{\bf e}_{\alpha_2}\wedge\cdots\wedge{\bf e}_{\alpha_k}.
$$
And we write $|I|=k$ for the length of $I$.  As a vector space, extended Minkowski space is sixteen--dimensional and has basis elements ${\bf e}_I$, where
\begin{equation}\label{eq:indices}
  I\in\{\emptyset,0,1,2,3,01,02,03,12,13,23,012,013,023,123,0123\}.
\end{equation}
Elements of $\bigwedge{\mathbb M}$ are of the form $\psi=\psi^I{\bf e}_I$, where summation over the basis element multi--indices in \eqref{eq:indices} is implied for repeated upper and lower multi--index pairs (Einstein convention).

A metric $\eta$ on ${\mathbb M}$ extends to a Hermitian metric $\hat\eta$ on $\bigwedge{\mathbb M}$.  In the case when $\eta$ is diagonal, $\hat\eta$ is also diagonal.  Specifically, we have $\hat\eta_{\emptyset\emptyset}=1$ and
$$\hat\eta_{II}=\eta_{\alpha_1\alpha_1}\eta_{\alpha_2\alpha_2}\cdots\eta_{\alpha_k\alpha_k}$$
for $I=\alpha_1\cdots\alpha_k\neq\emptyset$ and containing no repeated indices.

Gamma matrices can be defined on any exterior algebra by the general construction $\gamma_\alpha\psi\doteq{\bf e}_\alpha\wedge\psi-\iota_\alpha\psi$.  Here $\iota_\alpha$ denotes the interior product, which can be computed by the linear extension of (i) $\iota_\alpha{\bf e}_\beta=\eta_{\alpha\beta}{\bf e}_\emptyset$ and the graded derivation rule (ii) $\iota_\alpha({\bf e}_I\wedge{\bf e}_J)=(\iota_\alpha{\bf e}_I)\wedge{\bf e}_J+(-1)^{|I|}{\bf e}_I\wedge(\iota_\alpha{\bf e}_J)$.  In particular $\iota_\alpha{\bf e}_\emptyset=0$.  Thus defined, gamma matrices satisfy the Clifford algebra relation $\gamma_\alpha\gamma_\beta+\gamma_\beta\gamma_\alpha=-2\eta_{\alpha\beta}{\mathcal I}$, with ${\mathcal I}$ the $16\times 16$ identity matrix.  Moreover, we have $\gamma_\alpha^\dagger\hat\eta+\hat\eta\gamma_\alpha=0$, where $()^\dagger$ denotes the Hermitian conjugate of a matrix.

In rectangular coordinates, the covariant derivative on Minkowski space is equal to the usual coordinate derivative: $\nabla_\alpha=\partial_\alpha$.  However, in general we have $\nabla_\alpha (f^\beta{\bf e}_\beta)=(\partial_\alpha f^\beta+f^\rho{\Gamma^\beta}_{\rho\alpha}){\bf e}_\beta$, where the ${\Gamma^\beta}_{\rho\alpha}$ are the Christoffel symbols of the metric $\eta$.  We extend the covariant derivative to $\bigwedge{\mathbb M}$ via the Leibniz rule: $\hat\nabla_\alpha(\psi\wedge\varphi)=(\hat\nabla_\alpha\psi)\wedge\varphi+\psi\wedge(\hat\nabla_\alpha\varphi)$.  The Dirac equation on extended Minkowski space is then
\begin{equation}\label{eq:dirac-rect}
  \gamma^\alpha\hat\nabla_\alpha\psi=m\psi
\end{equation}
for (possibly) non--rectangular coordinates.

Any transformation, Lorentz or otherwise, on Minkowski space extends naturally to a transformation on $\bigwedge{\mathbb M}$.  In particular, if $L=({L^\alpha}_\beta)$ is a Lorentz algebra element, then the definition $L\cdot{\bf e}_\alpha\doteq{L^\beta}_\alpha{\bf e}_\beta$ and extended using the Leibniz rule, defines the action of $L$ on extended Minkowski space, which we denote by $\hat{L}$.  This in fact defines a Lie algebra representation of the Lorentz group, and we call this the {\em extended representation.}  On the other hand, we can use the gamma matrices to construct the {\em spin representation} exactly as in equation \eqref{eq:spinrep}.

%::::::::::::::::::::::::::::::::::::::::::::::::::::::::::::::::
\subsection{Theta matrices}

For a multi--index $I=\alpha_1\cdots\alpha_k$, the $16\times 16$ matrix $\Theta_I$ is defined by the two properties
\begin{enumerate}[(i)]
  \item $\Theta_I$ commutes with gamma matrices, and
  \item $\Theta_I{\bf e}_\emptyset={\bf e}_I$.
\end{enumerate}
These two properties completely determine $\Theta_I$.  Indeed, set $\gamma_I\doteq\gamma_{\alpha_1}\cdots\gamma_{\alpha_k}$.  If $\eta$ is diagonal and $I$ contains no duplicated indices, then $\gamma_I{\bf e}_\emptyset={\bf e}_I$.  It follows that $\Theta_I{\bf e}_J=\gamma_J{\bf e}_I$, provided that $J$ also contains no duplicated indices.

It should be noted that even though $\gamma_I{\bf e}_\emptyset={\bf e}_I=\Theta_I{\bf e}_\emptyset$, $\Theta_I$ is necessarily distinct from $\gamma_I$.  In particular, $\Theta_I\neq\Theta_{\alpha_1}\cdots\Theta_{\alpha_k}$ in general.  And if $I$ does contain duplicated indices, then $\Theta_I=0$, while $\gamma_I$ is not necessarily trivial.  On the other hand just like single--index gamma matrices, single--index theta matrices also satisfy the Clifford relation: $\Theta_\alpha\Theta_\beta+\Theta_\beta\Theta_\alpha=-2\eta_{\alpha\beta}{\mathcal I}$.

In rectangular coordinates $(x^\alpha)=(t,x,y,z)$, the Minkowski metric is
$$\eta={\it diag}(1,-1,-1,-1).$$
In this case, theta matrices enjoy some other properties that we will find useful.  The matrix ${\mathcal J}\doteq\Theta_{txyz}$ acts as an auxiliary complex structure on extended Minkowski space in the sense that ${\mathcal J}^2=-{\mathcal I}$.  It is possible to show that ${\mathcal J}^\dagger\hat\eta=\hat\eta{\mathcal J}$.  In general, we have $\Theta_I^\dagger\hat\eta=s_I\hat\eta\Theta_I$, where $s_I=-1$ if $|I|=1,2$, and $s_I=1$ if $|I|=0,3,4$.  For the (possibly complex) vector $\vec{v}=(v^x,v^y,v^z)$, define
$$\Theta(\vec{v})\doteq v^x\Theta_{yz}-v^y\Theta_{xz}+v^z\Theta_{xy}.
$$
Then $\Theta(\vec{v})^\dagger\hat\eta=-\hat\eta\Theta(\vec{v}^{\,\ast})$, where $\vec{v}^{\,\ast}$ is the complex conjugate of $\vec{v}$.  Moreover, $\Theta(\vec{v})^2=-(\vec{v}\cdot\vec{v}){\mathcal I}$ and ${\mathcal J}\Theta(\vec{v})=-(v^x\Theta_{tx}+v^y\Theta_{ty}+v^z\Theta_{tz})=\Theta(\vec{v}){\mathcal J}$.

%::::::::::::::::::::::::::::::::::::::::::::::::::::::::::::::::
\subsection{Projection onto an irreducible summand}
\label{sec:proj}

As a Clifford algebra representation, $\bigwedge{\mathbb M}$ is reducible: it can be decomposed into four isomorphic irreducible summands, each of dimension 4.  A projection $\pi$ onto an irreducible summand should commute with the gamma matrices and have trace ${\it tr}\,\pi=4$.  If we further require that $\pi$ commutes with ${\mathcal J}$, then $\pi$ must have the form
$$\pi_{\vec{w}}^\epsilon
  \doteq\tfrac{1}{4}({\mathcal I}-i\epsilon{\mathcal J})
                    \bigl({\mathcal I}-{\mathcal J}\Theta(\vec{w})\bigr)
$$
with $\epsilon=\pm 1$ and $\vec{w}$ a complex vector with $\vec{w}\cdot\vec{w}=1$.  One shows that $\pi_{\vec{w}}^\epsilon{\mathcal J}=i\epsilon\pi_{\vec{w}}^\epsilon$.  I.e., up to sign ${\mathcal J}$ is the same as the usual complex structure when restricted to the image of $\pi_{\vec{w}}^\epsilon$.  One also shows that $\pi_{\vec{w}}^+$ and $\pi_{\vec{w}\,'}^-$ commute, and $(\pi_{\vec{w}}^\epsilon)^\dagger\,\hat\eta\,\pi_{\vec{w}\,'}^\epsilon=0$, although $(\pi_{\vec{w}}^+)^\dagger\,\hat\eta\,\pi_{\vec{w}\,'}^-$ is not necessarily zero.  In particular, the extended metric is trivial when restricted to an irreducible summand, and more generally, between summands lying within the same eigenspace of ${\mathcal J}$. 

In order to recover the standard formulation of the Dirac equation, we must choose a particular projection.  And we must choose a basis for the image of that projection.  For definiteness we take
$$\pi_z\doteq\pi_{(0,0,1)}^+
  =\tfrac{1}{4}({\mathcal I}-i{\mathcal J}+\Theta_{tz}-i\Theta_{xy}),
$$
and the image of $\pi_z$ to be ${\mathbb C}\{{\bf f}_0,{\bf f}_1,{\bf f}_2,{\bf f}_3\}$, where ${\bf f}_0\doteq\sqrt{2}\,\pi_z({\bf e}_\emptyset-i{\bf e}_t)$, ${\bf f}_1\doteq\sqrt{2}\,\pi_z({\bf e}_{tx}+i{\bf e}_x)$, ${\bf f}_2\doteq\sqrt{2}\,\pi_z({\bf e}_\emptyset+i{\bf e}_t)$, ${\bf f}_3\doteq\sqrt{2}\,\pi_z({\bf e}_{tx}-i{\bf e}_x)$.  More explicitly,
\begin{equation}\label{eq:summandbasis}
\begin{aligned}
  {\bf f}_0&=\tfrac{1}{2\sqrt{2}}({\bf e}_\emptyset+{\bf e}_{tz}-i{\bf e}_{xy}-i{\bf e}_{txyz}
               -i{\bf e}_t+i{\bf e}_z-{\bf e}_{txy}+{\bf e}_{xyz})\\
  {\bf f}_1&=\tfrac{1}{2\sqrt{2}}({\bf e}_{tx}-i{\bf e}_{ty}+{\bf e}_{xz}-i{\bf e}_{yz}
               +i{\bf e}_x+{\bf e}_y-i{\bf e}_{txz}-{\bf e}_{tyz})\\
  {\bf f}_2&=\tfrac{1}{2\sqrt{2}}({\bf e}_\emptyset+{\bf e}_{tz}-i{\bf e}_{xy}-i{\bf e}_{txyz}
               +i{\bf e}_t-i{\bf e}_z+{\bf e}_{txy}-{\bf e}_{xyz})\\
  {\bf f}_3&=\tfrac{1}{2\sqrt{2}}({\bf e}_{tx}-i{\bf e}_{ty}+{\bf e}_{xz}-i{\bf e}_{yz}
               -i{\bf e}_x-{\bf e}_y+i{\bf e}_{txz}+{\bf e}_{tyz}).
\end{aligned}
\end{equation}
Although $\hat\eta$ is trivial on this summand, this basis is mutually orthonormal with respect to the standard Hermitian interior product: ${{\bf f}_\alpha}^\dagger{\bf f}_\beta=\delta_{\alpha\beta}$.  In terms of this basis, we obtain the standard gamma matrices in the Dirac representation:
$$\gamma_t=i\left(\begin{smallmatrix} {\it id} & 0\\ 0 & -{\it id}\end{smallmatrix}\right)
  \quad
  \gamma_x=-i\left(\begin{smallmatrix} 0 & \sigma_x\\ -\sigma_x & 0\end{smallmatrix}\right)
  \quad
  \gamma_y=-i\left(\begin{smallmatrix} 0 & \sigma_y\\ -\sigma_y & 0\end{smallmatrix}\right)
  \quad
  \gamma_z=-i\left(\begin{smallmatrix} 0 & \sigma_z\\ -\sigma_z & 0\end{smallmatrix}\right),
$$
where $\sigma_x$, $\sigma_y$, $\sigma_z$ are the Pauli matrices, and ${\it id}$ is the $2\times 2$ identity (the overall factor of $i$ is due to our choice of sign in the Clifford relation).  Choosing any other basis for the image of $\pi_z$ changes the gamma matrices by a similarity transformation.

Two remarks are in order.  First as pointed out in \cite{me0}, the natural choice for the adjoint of a spinor $\psi$ in extended Minkowski space is $\psi^\dagger\hat\eta$.  But because the extended metric $\hat\eta$ is trivial on irreducible summands, one is forced into making a somewhat arbitrary choice for adjoint spinor on summands, such as $\psi^\dagger\gamma^0$.  Second, the extended Lorentz action does not commute with projection, so that the extended structure is lost when restricting to an irreducible summand.

%%%%%%%%%%%%%%%%%%%%%%%%%%%%%%%%%%%%%%%%%%%%%%%%%%%%%%%%%%%%%%%%%
\section{Plane waves}

Using rectangular coordinates, the plane wave solutions of the free Dirac equation \eqref{eq:dirac-rect} are of the form
\begin{equation}\label{eq:plane-gen}
  \psi=e^{-ip_\alpha x^\alpha}\chi
  \quad\text{with}\quad
  p_\alpha p^\alpha=m^2,
\end{equation}
and $\chi$ any constant spinor that satisfies $-ip^\alpha\gamma_\alpha\chi=m\chi$.  This constraint on $\chi$ can be rephrased as saying that $\chi$ is in the kernel of $K_{\bf p}\doteq m{\mathcal I}+ip^\alpha\gamma_\alpha$.  The matrix
$$\Pi_{\bf p}\doteq\tfrac{1}{2m}(m{\mathcal I}-ip^\alpha\gamma_\alpha)
$$
gives a projection onto the the kernel of $K_{\bf p}$.  And in fact we have ${\it tr}\,\Pi_{\bf p}=8$, so that $\chi$ belongs to a complex vector space of dimension 8.

The image space of $\Pi_{\bf p}$ can be further decomposed into two summands of dimension 4.  These summands are characterized by the sign of the scalar invariant $\psi^\dagger\hat\eta\psi=\chi^\dagger\hat\eta\chi$.  The spinors
\begin{equation}\label{eq:chi-plus}
\begin{aligned}
    \chi_0^+({\bf p})
      &\doteq\tfrac{1}{\sqrt{2}m}(m{\bf e}_\emptyset-ip^\alpha{\bf e}_\alpha)
  & \chi_1^+({\bf p})
      &\doteq\Theta_{yz}\chi_0^+({\bf p})\\
  \chi_2^+({\bf p})
      &\doteq\Theta_{xz}\chi_0^+({\bf p})
  & \chi_3^+({\bf p})
      &\doteq\Theta_{xy}\chi_0^+({\bf p})
\end{aligned}
\end{equation}
are independent, in the kernel of $K_{\bf p}$, and have scalar invariants that are equal to $1$.  Moreover, the spinors
\begin{equation}\label{eq:chi-minus}
  \chi_\alpha^-({\bf p})\doteq{\mathcal J}\chi_\alpha^+({\bf p})
\end{equation}
are also in the kernel of $K_{\bf p}$ and have $-1$ as scalar invariant.  The $\chi_\alpha^+({\bf p})$ and $\chi_\beta^-({\bf p})$ taken together are mutually orthogonal with respect to the extended metric.

We will say that a nontrivial spinor in the subspace of the image of $\Pi_{\bf p}$ spanned by the $\chi_\alpha^+({\bf p})$ is {\em positive.}  One in the subspace spanned by the $\chi_\alpha^-({\bf p})$ is {\em negative.}  And any spinor $\chi$ with $\chi^\dagger\hat\eta\chi=0$ is {\em null.}

If we assume that the plane wave of equation \eqref{eq:plane-gen} carries a charge, then its electromagnetic field $A^\alpha$ satisfies the equation $\partial_\beta\partial^\beta A^\alpha-\partial^\alpha\partial_\beta A^\beta=j^\alpha$.  And the charge current is given by
$$j^\alpha=-ie\psi^\dagger\hat\eta\gamma^\alpha\psi.$$
For the positive and negative plane waves in equations \eqref{eq:chi-plus} and \eqref{eq:chi-minus}, we have $j^\alpha=-ie(\chi_\beta^\pm)^\dagger\hat\eta\gamma^\alpha\chi_\beta^\pm=\pm ep^\alpha/m$.  In this sense, we can view ${\mathcal J}$ as a charge conjugation operator for plane waves.  A null spinor has zero charge current.

%::::::::::::::::::::::::::::::::::::::::::::::::::::::::::::::::
\subsection{Helicity operators}

Define $\vec{p}\doteq(p^x,p^y,p^z)$, and $\vec{u}\doteq\vec{p}/|\vec{p}|$ its normalization.  Rotation about $\vec{p}$ is generated by the Lorentz algebra element
$$L_{\vec{u}}=\left(\begin{smallmatrix} 0 & 0 & 0 & 0\\ 0 & 0 & -u^z & u^y\\
                        0 & u^z & 0 & -u^x\\ 0 & -u^y & u^x & 0\end{smallmatrix}\right).
$$
The standard helicity operator is taken to be the spin representation of $L_{\vec{u}}$.  As per equation \eqref{eq:spinrep},
$$\sigma(L_{\vec u})
   =-\tfrac{1}{2}(u^x\gamma_y\gamma_z-u^y\gamma_x\gamma_z+u^z\gamma_x\gamma_y).
$$
On the other hand, we also have the extended representation $\hat{L}_{\vec{u}}$.  We call $\sigma(L_{\vec{u}})$ the {\em spin helicity operator} and $\hat{L}_{\vec{u}}$  the {\em extended helicity operator.}

These two representations commute, so that they possess simultaneous eigenvectors.  For ${\bf p}=(p^\alpha)=(p^t,\vec{p})$, the positive spinors
\begin{equation}\label{eq:heleigen}
\begin{aligned}
  \varphi_0^+({\bf p})
    &\doteq\tfrac{1}{\sqrt{2}}\bigl({\mathcal I}+i\Theta(\vec{u})\bigr)\chi_0^+({\bf p})
  & \varphi_1^+({\bf p})
    &\doteq\tfrac{1}{\sqrt{2}}\bigl({\mathcal I}-i\Theta(\vec{u})\bigr)\chi_0^+({\bf p})\\
  \varphi_2^+({\bf p})
    &\doteq\Theta(\vec{v})\chi_0^+({\bf p})
  & \varphi_3^+({\bf p})
    &\doteq\Theta(\vec{v}^{\,\ast})\chi_0^+({\bf p})
\end{aligned}
\end{equation}
are eigenvectors of $\sigma(L_{\vec{u}})$ with respective eigenvalues $i/2$, $-i/2$, $i/2$, $-i/2$, and eigenvectors of $\hat{L}_{\vec{u}}$ with eigenvalues $0$, $0$, $i$, $-i$.  Here $\vec{v}=(v^x,v^y,v^z)$ is any {\em complex} vector such that $\vec{u}\times\vec{v}=i\vec{v}$ and $\vec{v}\cdot\vec{v}^{\,\ast}=1$.  For instance if $\vec{u}\neq(\pm 1,0,0)$, we may take
$$\vec{v}=\tfrac{1}{\sqrt{2}}\bigl(i\sqrt{1-u_x^2},
             \tfrac{u^z-iu^xu^y}{\sqrt{1-u_x^2}},
             -\tfrac{u^y+iu^xu^z}{\sqrt{1-u_x^2}}\bigr)
$$
or any phase shift of this.  Moreover, both $\sigma(L_{\vec{u}})$ and $\hat{L}_{\vec{u}}$ commute with ${\mathcal J}$, so that the negative spinors
$$\varphi_\alpha^-({\bf p})
  \doteq{\mathcal J}\varphi_\alpha^+({\bf p})
$$
are also eigenvectors with the same eigenvalues as their positive counterparts.

%::::::::::::::::::::::::::::::::::::::::::::::::::::::::::::::::
\subsection{Projections}

To obtain solutions to the standard Dirac equation, we apply the projection $\pi_z\doteq\pi_{(0,0,1)}^+$ from subsection \ref{sec:proj} to the above constructions.  In particular, we have that $\pi_z\chi_2^+({\bf p})=-i\pi_z\chi_1^+$, $\pi_z\chi_3^+({\bf p})=i\pi_z\chi_0^+$, and $\pi_z\chi_\alpha^-({\bf p})=i\pi_z\chi_\alpha^+({\bf p})$.  So that the projection of any plane wave solution of the extended Dirac equation is of the form
$$e^{-ip_\alpha x^{\alpha}}\tilde\varphi({\bf p})$$
where $\tilde\varphi({\bf p})$ is a complex linear combination of $\pi_z\chi_0^+({\bf p})$ and $\pi_z\chi_1^+({\bf p})$.  Or equivalently, a combination of $\pi_z\varphi_0^+({\bf p})$ and $\pi_z\varphi_1^+({\bf p})$.  Because $\pi_z$ commutes with gamma matrices, $\pi_z\varphi_0^+({\bf p})$ and $\pi_z\varphi_1^+({\bf p})$ are necessarily in the $i/2$ and $-i/2$--eigenspaces of the spin helicity operator.  Explicitly, up to scalar multiple
\begin{align*}
  \pi_z\varphi_\epsilon^+({\bf p})
  &=(E+\epsilon|\vec{p}|+m)(|\vec{p}|-\epsilon p^z){\bf f}_0
    -\epsilon(E+\epsilon|\vec{p}|+m)(p^x+ip^y){\bf f}_1\\
  &\quad\quad
   -(E+\epsilon|\vec{p}|-m)(|\vec{p}|-\epsilon p^z){\bf f}_2
   +\epsilon(E+\epsilon|\vec{p}|-m)(p^x+ip^y){\bf f}_3,
\end{align*}
where $\varphi_+^+({\bf p})\doteq\varphi_0^+({\bf p})$ and $\varphi_-^+({\bf p})\doteq\varphi_1^+({\bf p})$, and $E\doteq p^t$.  However, these are not normalized.  Indeed, $[\pi_z\varphi_\epsilon^+({\bf p})]^\dagger\pi_z\varphi_\epsilon^+({\bf p})=8E|\vec{p}|(E+\epsilon|\vec{p}|)(|\vec{p}|-\epsilon p^z)$.
Normalizing to $\sqrt{2E}$, we can write the spin helicity eigenfunctions as
\begin{align*}
  \tilde\varphi_\epsilon({\bf p})
  &=\tfrac{(E+\epsilon|\vec{p}|+m)\sqrt{|\vec{p}|-\epsilon p^z}}
          {2\sqrt{|\vec{p}|(E+\epsilon|\vec{p}|)}}\,{\bf f}_0
    -\tfrac{\epsilon e^{i\theta}(E+\epsilon|\vec{p}|+m)\sqrt{|\vec{p}|+\epsilon p^z}}
           {2\sqrt{|\vec{p}|(E+\epsilon|\vec{p}|)}}\,{\bf f}_1\\
  &\quad\quad
    -\tfrac{(E+\epsilon|\vec{p}|-m)\sqrt{|\vec{p}|-\epsilon p^z}}
          {2\sqrt{|\vec{p}|(E+\epsilon|\vec{p}|)}}\,{\bf f}_0
    +\tfrac{\epsilon e^{i\theta}(E+\epsilon|\vec{p}|-m)\sqrt{|\vec{p}|+\epsilon p^z}}
           {2\sqrt{|\vec{p}|(E+\epsilon|\vec{p}|)}}\,{\bf f}_1
\end{align*}
with $\tilde\varphi_\epsilon({\bf p})$ the normalization of $\pi_z\varphi_\epsilon^+({\bf p})$, and $\theta$ is such that $(p^x+ip^y)=e^{i\theta}\sqrt{p_x^2+p_y^2}$.  Note that $(E+\epsilon|\vec{p}|\pm m)/\sqrt{E+\epsilon|\vec{p}|}=\sqrt{E+\epsilon|\vec{p}|}\pm\sqrt{E-\epsilon|\vec{p}|}$, as $m^2=E^2-|\vec{p}|^2$.

%%%%%%%%%%%%%%%%%%%%%%%%%%%%%%%%%%%%%%%%%%%%%%%%%%%%%%%%%%%%%%%%%
\section{Circular beams}

We use cylindrical coordinates $(x^\alpha)=(t,s,\xi,z)$ on Minkowski space, where $x=s\cos\xi$ and $y=s\sin\xi$.  The Minkowski metric is then
$$\eta={\rm diag}(1,-1,-s^2,-1).$$
We can convert from cylindrical to rectangular coordinates via the equations
$${\bf e}_s=\tfrac{x}{s}{\bf e}_x+\tfrac{y}{s}{\bf e}_y
  \quad\text{and}\quad
  {\bf e}_\xi=-y{\bf e}_x+x{\bf e}_y
$$
where $s=\sqrt{x^2+y^2}$.  Other basis elements of extended Minkowski space can be computed from these.  For instance, ${\bf e}_{ts}={\bf e}_t\wedge{\bf e}_s=\tfrac{x}{s}{\bf e}_{tx}+\tfrac{y}{s}{\bf e}_{ty}$.

%::::::::::::::::::::::::::::::::::::::::::::::::::::::::::::::::
\subsection{Dirac equation}

The nontrivial Christoffel symbols are ${\Gamma^s}_{\xi\xi}=-s$ and ${\Gamma^\xi}_{s\xi}=\tfrac{1}{s}$, so that the nontrivial effects of the extended connection on rank one basis elements of $\bigwedge{\mathbb M}$ are
$$\hat\Gamma_s{\bf e}_\xi=\tfrac{1}{s}{\bf e}_\xi
  \quad\quad
  \hat\Gamma_\xi{\bf e}_s=\tfrac{1}{s}{\bf e}_\xi
  \quad\quad
  \hat\Gamma_\xi{\bf e}_\xi=-s{\bf e}_s.
$$
The effect of the extended connection on other basis elements are computed using the Leibniz rule.  In particular, $\hat\Gamma_t=\hat\Gamma_z=0$.

We assume that the spinor field $\psi=\psi(t,s,\xi,z)$ is separable of the form
$$\psi=e^{-i(Et-kz)}\varphi,
  \quad\text{with}\quad
  \varphi=\varphi(s)=\varphi^I(s){\bf e}_I
$$
and $E,k$ constants.  The Dirac operator $D\psi=\gamma^\alpha(\partial_\alpha\psi+\hat\Gamma_\alpha\psi)$ is then
$$D\psi=\gamma^s\partial_s\psi
        +(-iE\gamma^t+ik\gamma^z+\gamma^s\hat\Gamma_s+\gamma^\xi\hat\Gamma_\xi)\psi.
$$
The Dirac equation $D\psi=m\psi$ yields sixteen coupled equations.

%::::::::::::::::::::::::::::::::::::::::::::::::::::::::::::::::
\subsection{Solutions}

There are four independent non--singular solutions, and four independent singular solutions, to the Dirac equation.  The non--singular solutions are:
\begin{equation}\label{eq:circ}
\begin{aligned}
  \psi_0(E,k)&\doteq e^{-i(Et-kz)}\varphi_0(E,k)
  & \psi_1(E,k)&\doteq e^{-i(Et-kz)}\varphi_1(E,k)\\
  \psi_2(E,k)&\doteq\Theta_{tz}\psi_0(E,k)
  & \psi_3(E,k)&\doteq\Theta_{tz}\psi_1(E,k)
\end{aligned}
\end{equation}
where
\begin{align*}
  \varphi_0(E,k)
    &\doteq\tfrac{1}{\sqrt{2}m}[mJ_0(ps){\bf e}_\emptyset - iEJ_0(ps){\bf e}_t
                                + pJ_1(ps){\bf e}_s - ikJ_0(ps){\bf e}_z]\\
  \varphi_1(E,k)
    &\doteq\tfrac{1}{\sqrt{2}m}[mJ_1(ps){\bf e}_{sz} + ikJ_1(ps){\bf e}_s
                                - pJ_0(ps){\bf e}_z - iEJ_1(ps){\bf e}_{tsz}]
\end{align*}
and $p\doteq\sqrt{E^2-k^2-m^2}$.  Here $J_0$ and $J_1$ are the zeroth and first order Bessel functions of the first kind.  Normalization is chosen such that $\psi_\alpha^\dagger\hat\eta\psi_\alpha=\pm 1$ when $p=0$.  Note that the matrix $\Theta_{tz}$ commutes with the extended Dirac operator (in fact, both matrices $\Theta_t$ and $\Theta_z$ do as well).

The charge currents of these solutions have trivial $s,\xi$ components, and nontrivial $t,z$ components.  The nontrivial components are
\begin{align*}
  -ie\psi_\alpha^\dagger\hat\eta\gamma^t\psi_\alpha &=\tfrac{eE}{m}J_\alpha^2(ps)
    & -ie\psi_\alpha^\dagger\hat\eta\gamma^z\psi_\alpha &=\tfrac{ek}{m}J_\alpha^2(ps)
\end{align*}
for $\alpha=0,1$.  The currents for $\psi_2$ and $\psi_3$ have opposite signs to those of $\psi_0$ and $\psi_1$.  So in the context of circular beams, $\Theta_{tz}$ functions as a charge conjugation operator.

We remark that ${\mathcal J}\psi_\alpha$ is necessarily also a solution of the Dirac equation.  However, the resulting solution will be singular.  Indeed, ${\mathcal J}{\bf e}_z=-{\bf e}_{txy}=-\tfrac{1}{s}{\bf e}_{ts\xi}$.  So that the result of applying ${\mathcal J}$ to $\psi_\alpha(E,k)$ will have a singular term involving $\tfrac{1}{s}J_0(ps)$.

%::::::::::::::::::::::::::::::::::::::::::::::::::::::::::::::::
\subsection{Projections}

Again we use $\pi_z\doteq\pi_{(0,0,1)}^+$ to obtain solutions to the standard Dirac equation.  We find that $\pi_z\psi_2(E,k)=\pi_z\psi_0(E,k)$ and $\pi_z\psi_3(E,k)=\pi_z\psi_1(E,k)$.  So that there are two independent non--singular circular beam solutions of the standard Dirac equation.  Using the the basis $\{{\bf f}_0,{\bf f}_1,{\bf f}_2,{\bf f}_3\}$ in equation \eqref{eq:summandbasis}, we have up to scalar multiple:
\begin{align*}
  \pi_z\psi_0(E,k)
    &=e^{-i(Et-kz)}
      [(E-k+m)J_0(ps){\bf f}_0 - ip\tfrac{x+iy}{s}J_1(ps){\bf f}_1\\
    &\quad\quad\quad\quad\quad\quad\quad\quad
        -(E-k-m)J_0(ps){\bf f}_2 + ip\tfrac{x+iy}{s}J_1(ps){\bf f}_3]\\
  \pi_z\psi_1(E,k)
    &=e^{-i(Et-kz)}
      [ipJ_0(ps){\bf f}_0 + (E+k+m)\tfrac{x+iy}{s}J_1(ps){\bf f}_1\\
    &\quad\quad\quad\quad\quad\quad\quad\quad
        -ipJ_0(ps){\bf f}_2 - (E+k-m)\tfrac{x+iy}{s}J_1(ps){\bf f}_3].
\end{align*}
In the absence of the preceding discussion, one can verify by hand that these are indeed solutions of the standard Dirac equation, provided that the Dirac representation of gamma matrices is used.

%::::::::::::::::::::::::::::::::::::::::::::::::::::::::::::::::
\subsection{Plane wave decomposition}

If we decompose the solutions in equation \eqref{eq:circ} into plane waves, we find that
\begin{align*}
  \psi_0(E,k)
    &=\int_{\mathbb M}\Psi_c(E,k,{\bf q})e^{-iq_\alpha x^\alpha}\chi_0^+({\bf q})\,d^4{\bf q}\\
  \psi_1(E,k)
    &=\int_{\mathbb M}\Psi_c(E,k,{\bf q})e^{-iq_\alpha x^\alpha}\chi_C^+({\bf q})\,d^4{\bf q}.
\end{align*}
Here the integrals are over all four--vectors ${\bf q}=(q^\alpha)=(q^t,q^x,q^y,q^z)$ in Minkowski space, and the distribution function is given by
\begin{align*}
  \Psi_c(E,k,{\bf q})
    &\doteq\tfrac{1}{2\pi p}\,\delta(q^t-E)\,\delta(q^z-k)\,\delta(q-p)
\end{align*}
where $q\doteq\sqrt{q_x^2+q_y^2}$, and $p\doteq\sqrt{E^2-m^2-k^2}$.  In the second decomposition equation, we have set
$$\chi_C^+({\bf p})\doteq\tfrac{1}{i\sqrt{p_x^2+p_y^2}}
                            [p^y\chi_1^+({\bf p})+p^x\chi_2^+({\bf p})].
$$
Note that the delta functions guarantee $q_\alpha q^\alpha=m^2$, so that $\psi_0(E,k)$ is a composition exclusively of plane waves involving spinors $\chi_0^+({\bf q})$, with $\chi_0^+$ as in equation \eqref{eq:chi-plus}.  Similarly, $\psi_2(E,k)$ is composed of plane waves with spinors of type $\chi_3^-$, and $\psi_3(E,k)$ with type $\chi_C^-\doteq(p^x\chi_1^--p^y\chi_2^-)/\sqrt{p_x^2+p_y^2}$.

The circular beam solutions in equation \eqref{eq:circ} are essentially radially symmetric solutions of the three--dimensional (two spacial dimensions) free Dirac equation.  In this case, semi--numerical solutions were obtained in Appendix~A of \cite{FillionGourdeau} for a Gaussian distribution of plane waves.  In contrast, the closed form solutions presented here have a pink noise distribution.

%%%%%%%%%%%%%%%%%%%%%%%%%%%%%%%%%%%%%%%%%%%%%%%%%%%%%%%%%%%%%%%%%
\section{Spherical waves}

Using spherical polar coordinates $(x^\alpha)=(t,r,\theta,\phi)$ on Minkowski space, where $x=r\sin\theta\cos\phi$, $y=r\sin\theta\cos\phi$, and $z=r\cos\theta$, the space--time metric becomes
$$\eta={\it diag}(1,-1,-r^2,-r^2\sin^2\theta).$$
For basis elements of extended Minkowski space, we can convert from spherical to rectangular coordinates using the identities
$${\bf e}_r=\tfrac{x}{r}{\bf e}_x+\tfrac{y}{r}{\bf e}_y+\tfrac{z}{r}{\bf e}_z,\quad
  {\bf e}_\theta=\tfrac{xz}{s}{\bf e}_x+\tfrac{yz}{s}{\bf e}_y-s{\bf e}_z,\quad
  {\bf e}_\phi=-y{\bf e}_x+x{\bf e}_y
$$
where $r=\sqrt{x^2+y^2+z^2}$ and $s=\sqrt{x^2+y^2}$.

%::::::::::::::::::::::::::::::::::::::::::::::::::::::::::::::::
\subsection{Dirac equation}

The metric compatible connection $\Gamma_\alpha$ on Minkowski space has non--trivial components given by
\begin{align*}
  \Gamma_r{\bf e}_\theta&=\tfrac{1}{r}{\bf e}_\theta
  & \Gamma_r{\bf e}_\phi&=\tfrac{1}{r}{\bf e}_\phi
  & \Gamma_\theta{\bf e}_r&=\tfrac{1}{r}{\bf e}_\theta
  & \Gamma_\theta{\bf e}_\theta &= -r{\bf e}_r
  & \Gamma_\theta{\bf e}_\phi &= \cot\theta\,{\bf e}_\phi\\
  \Gamma_\phi{\bf e}_r &=\tfrac{1}{r}{\bf e}_\phi
  & \Gamma_\phi{\bf e}_\theta &= \cot\theta\,{\bf e}_\phi
  & \Gamma_\phi{\bf e}_\phi &= \makebox[0pt][l]{$-r\sin^2\theta\,{\bf e}_r-\sin\theta\cos\theta\,{\bf e}_\theta$.}
\end{align*}
In particular $\Gamma_t=0$.  Again, the extended connection $\hat\Gamma_\alpha$ is computed using the Leibniz rule.

We seek a separable solution of the Dirac equation of the form
$$\psi(t,r)=e^{-iEt}\varphi(r).$$
Because we do not allow $\varphi$ to depend on $\theta$ or $\phi$, the spinor $\varphi$ must have the form
$$\varphi(r)=\varphi^\emptyset{\bf e}_\emptyset
            +\varphi^{tr}{\bf e}_{tr}
            +\varphi^t{\bf e}_t
            +\varphi^r{\bf e}_r.
$$
The Dirac equation $\gamma^\alpha\partial_\alpha\psi+\gamma^\alpha\hat\Gamma_\alpha\psi=m\psi$ thus yields the four coupled equations
\begin{align*}
    -\partial_r\varphi^r-\tfrac{2}{r}\varphi^r+iE\varphi^t&=m\varphi^\emptyset
  &
    \partial_r\varphi^t-iE\varphi^r&=m\varphi^{tr}\\
    \partial_r\varphi^{tr}+\tfrac{2}{r}\varphi^{tr}-iE\varphi^\emptyset&=m\varphi^t
  &
  -\partial_r\varphi^\emptyset+iE\varphi^{tr}&=m\varphi^r.
\end{align*}

%::::::::::::::::::::::::::::::::::::::::::::::::::::::::::::::::
\subsection{Solutions}

There are two independent non--singular solutions of the Dirac equation, namely
\begin{equation}\label{eq:spherespin}
\begin{aligned}
  \psi_0(E)&=\tfrac{1}{\sqrt{2}m}e^{-iEt}
               [m\,j_0(pr)\,{\bf e}_\emptyset
                -iE\,j_0(pr)\,{\bf e}_t
                +p\,j_1(pr)\,{\bf e}_r]\\
  \psi_1(E)&=\tfrac{1}{\sqrt{2}m}e^{-iEt}
               [m\,j_1(pr)\,{\bf e}_{tr}
                +p\,j_0(pr)\,{\bf e}_t
                +iE\,j_1(pr)\,{\bf e}_r].
\end{aligned}
\end{equation}
Here $j_0(x)\doteq\tfrac{\sin{x}}{x}$ and $j_1(x)\doteq\tfrac{\sin{x}-x\cos{x}}{x^2}$ are the zeroth and first order spherical Bessel functions of the first kind, and $p\doteq\sqrt{E^2-m^2}$.  The normalization factor is chosen such that $\psi_0(E)=e^{-ip_\alpha x^\alpha}\chi_0^+({\bf p})$ when $p=0$ ($E=m$).

The charge currents are concentrated in the $t$--component only, and we have
$$-ie\psi_0^\dagger\hat\eta\gamma^t\psi_0
    =\tfrac{eE}{m}j_0^2(pr)
  \quad\text{and}\quad
  -ie\psi_1^\dagger\hat\eta\gamma^t\psi_1
    =-\tfrac{eE}{m}j_1^2(pr).
$$

We remark that ${\mathcal J}\psi_0(E)$ and ${\mathcal J}\psi_1(E)$ will necessarily also solve the Dirac equation.  However, these solutions are not spherically symmetric.  Indeed, ${\mathcal J}{\bf e}_t=-{\bf e}_{xyz}=-\tfrac{1}{r^2\sin\theta}{\bf e}_{r\theta\phi}$.  So ${\mathcal J}\psi_0(E)$ and ${\mathcal J}\psi_1(E)$ both have terms that depend on $\theta$ as well as $r$

%::::::::::::::::::::::::::::::::::::::::::::::::::::::::::::::::
\subsection{Projections}

The results of using $\pi\doteq\pi_{(0,0,1)}^+$ and the basis $\{{\bf f}_0,{\bf f}_1,{\bf f}_2,{\bf f}_3\}$ from equation \eqref{eq:summandbasis} to project the spinors in eqation \eqref{eq:spherespin} onto an irreducible summand of extended Minkowski space are
\begin{align*}
  \pi\psi_0(E)
    &=e^{-iEt}\{[(E+m)j_0(pr)-ip\tfrac{z}{r}j_1(pr)]\,{\bf f}_0
                -ip\tfrac{x+iy}{r}j_1(pr)\,{\bf f}_1\\
    &\quad\quad\quad\quad
                -[(E-m)j_0(pr)-ip\tfrac{z}{r}j_1(pr)]\,{\bf f}_2
                +ip\tfrac{x+iy}{r}j_1(pr)\,{\bf f}_3\}\\
  \pi\psi_1(E)
    &=e^{-iEt}\{[(E+m)\tfrac{z}{r}j_1(pr)+ipj_0(pr)]\,{\bf f}_0
                +(E+m)\tfrac{x+iy}{r}j_1(pr)\,{\bf f}_1\\
    &\quad\quad\quad\quad
                -[(E-m)\tfrac{z}{r}j_1(pr)+ipj_0(pr)]\,{\bf f}_2
                -(E-m)\tfrac{x+iy}{r}j_1(pr)\,{\bf f}_3\}
\end{align*}
(where $p=\sqrt{E^2-m^2}$) up to scalar multiple.  Again, it can be verified by hand that these are indeed solutions of the standard Dirac equation, provided that the Dirac representation of gamma matrices is used.

Although I do not know if explicit radially symmetric solutions of the standard Dirac equation have appeared previously, a general method for generating solutions of the free Dirac equation in spherical polar coordinates appears in \cite{Carlson}, and predicts the overall form of such solutions.

%::::::::::::::::::::::::::::::::::::::::::::::::::::::::::::::::
\subsection{Plane wave decomposition}

The spherical solutions in equation \eqref{eq:spherespin} can be decomposed into plane waves.  Namely,
\begin{align*}
  \psi_0(E)
    &=\int_{\mathbb M}\Psi(E,{\bf q})\,e^{-iq_\alpha x^\alpha}\chi_0^+({\bf q})\,d^4{\bf q}\\
  \psi_1(E)
    &=\int_{\mathbb M}\Psi(E,{\bf q})\,e^{-iq_\alpha x^\alpha}\chi_S^-({\bf q})\,d^4{\bf q}.
\end{align*}
Here the distribution function is now given by
$$\Psi(E,{\bf q})
  \doteq\tfrac{1}{4\pi p^2}\,\delta(E-q^t)\,\delta(p-|\vec{q}|)
$$
where ${\bf q}=(q^\alpha)=(q^t,\vec{q})$ and $p=\sqrt{E^2-m^2}$.  The auxiliary plane wave spinor in the plane wave expansion for $\psi_1(E)$ is defined as
$$\chi_S^-({\bf p})
  \doteq iu^x\chi_1^-({\bf p}) - iu^y\chi_2^-({\bf p}) + iu^z\chi_3^-({\bf p})
  =\tfrac{1}{\sqrt{2}m}(-im{\bf e}_t\wedge\vec{u} + |\vec{p}|{\bf e}_t + p^t\vec{u}),
$$
where $\vec{u}\doteq\vec{p}/|\vec{p}|$.

%::::::::::::::::::::::::::::::::::::::::::::::::::::::::::::::::
\subsection{Moving spherical solutions}

The solutions in equation \eqref{eq:spherespin} are for a wave packet at rest.  We apply an extended Lorentz boost to get solutions of the Dirac equation for a wave packet in motion.  The results are
\begin{multline*}
  \psi_0(E,\vec{v})
   =\tfrac{1}{\sqrt{2}m}e^{-iE\gamma(t-\vec{r}\cdot\vec{v})}
    \{mj_0(p\sigma){\bf e}_\emptyset
      -iE\gamma j_0(p\sigma)({\bf e}_t+\vec{v})\hfill\\
      +\tfrac{p}{\sigma}j_1(p\sigma)
         [\vec{r}+\gamma^2(\vec{r}\cdot\vec{v}-t)\vec{v}
          +\gamma^2(\vec{r}\cdot\vec{v}-|\vec{v}|^2t){\bf e}_t]\}
\end{multline*}
\begin{multline*}
  \psi_1(E,\vec{v})
  =\tfrac{1}{\sqrt{2}m}e^{-iE\gamma(t-\vec{r}\cdot\vec{v})}
    \{\tfrac{m\gamma}{\sigma}j_1(p\sigma)
        [{\bf e}_t\wedge(\vec{r}-t\vec{v})+\vec{v}\wedge\vec{r}]
      +p\gamma j_0(p\sigma)({\bf e}_t+\vec{v})\\
      +\tfrac{iE}{\sigma}j_1(p\sigma)
       [\vec{r}+\gamma^2(\vec{r}\cdot\vec{v}-t)\vec{v}
          +\gamma^2(\vec{r}\cdot\vec{v}-|\vec{v}|^2t){\bf e}_t]\}.
\end{multline*}
Here $\vec{v}=(v^x,v^y,v^z)=v^k{\bf e}_k$ is the velocity of the packet in the observer's frame, $\vec{r}=(x,y,z)=x^k{\bf e}_k$ is its position, $\gamma\doteq 1/\sqrt{1-|\vec{v}|^2}$, and
$$\sigma\doteq\sqrt{|\vec{r}|^2+\gamma^2(\vec{r}\cdot\vec{v})^2
                       -2\gamma^2(\vec{r}\cdot\vec{v})t
                       +\gamma^2|\vec{v}|^2t^2}.
$$
If we apply the projection $\pi_z$ to $\psi_0(E,\vec{v})$ and $\psi_1(E,\vec{v})$ we obtain two independent spherically symmetric moving solutions to the standard free Dirac equation.  However, the resulting expressions are unpleasant and are omitted here.

%%%%%%%%%%%%%%%%%%%%%%%%%%%%%%%%%%%%%%%%%%%%%%%%%%%%%%%%%%%%%%%%%

\end{document}

\maketitle

\begin{abstract}
This is meant as a companion to the article {\it Spinors in extended Minkowski space.}  Its purpose is to provide details on the computations underlying the statements presented there.
\end{abstract}

%%%%%%%%%%%%%%%%%%%%%%%%%%%%%%%%%%%%%%%%%%%%%%%%%%%%%%%%%%%%%%%%%
\section{Introduction}

Throughout we let $\eta$ denote the Minkowski metric with sign convention $(+,-,-,-)$.  And we assume that gamma matrices obey the Clifford relation
\begin{equation}\label{eq:clifford}
  \gamma_\alpha\gamma_\beta+\gamma_\beta\gamma_\alpha=-2\eta_{\alpha\beta}{\mathcal I},
\end{equation}
where ${\mathcal I}$ is the identity matrix.

The Lorentz group is the set of all real $4\times 4$ matrices $\Lambda$ that preserve the Minkowski metric: $\Lambda^T\eta\Lambda=\eta$.  The Lorentz algebra ${\rm so}(\eta)$ is the Lie algebra of the Lorentz group, and consists of all real matrices $L$ for which $L^T\eta+\eta L=0$.  Define the {\em spin representation} of ${\rm so}(\eta)$ by
$$\sigma(L)\doteq-\tfrac{1}{4}L^{\alpha\beta}\gamma_\alpha\gamma_\beta$$
for any $L=({L^\alpha}_\beta)$ in ${\rm so}(\eta)$.

\begin{fact}
$\sigma$ defines a Lie algebra representation of ${\rm so}(\eta)$.
\end{fact}

\begin{proof}
We need to verify that commutators are preserved.  That is for any $L,M\in{\rm so}(\eta)$, we have $[\sigma(L),\sigma(M)]=\sigma([L,M])$.  Repeated applications of equation \eqref{eq:clifford} yields
\begin{align*}
  \gamma_\alpha\gamma_\beta\gamma_\mu\gamma_\nu
         -\gamma_\mu\gamma_\nu\gamma_\alpha\gamma_\beta
  &=-\gamma_\alpha\gamma_\mu\gamma_\beta\gamma_\nu
       -2\eta_{\beta\mu}\gamma_\alpha\gamma_\nu
    +\gamma_\mu\gamma_\alpha\gamma_\nu\gamma_\beta
       +2\eta_{\nu\alpha}\gamma_\mu\gamma_\beta\\
  &=(\gamma_\mu\gamma_\alpha\gamma_\beta\gamma_\nu
       +2\eta_{\alpha\mu}\gamma_\beta\gamma_\nu)
       -2\eta_{\beta\mu}\gamma_\alpha\gamma_\nu\\
  &\mspace{72mu}
    +(-\gamma_\mu\gamma_\alpha\gamma_\beta\gamma_\nu
       -2\eta_{\nu\beta}\gamma_\mu\gamma_\alpha)
       +2\eta_{\nu\alpha}\gamma_\mu\gamma_\beta\\
  &=2(\eta_{\alpha\mu}\gamma_\beta\gamma_\nu
      -\eta_{\beta\mu}\gamma_\alpha\gamma_\nu
      -\eta_{\nu\beta}\gamma_\mu\gamma_\alpha
      +\eta_{\nu\alpha}\gamma_\mu\gamma_\beta).
\end{align*}
Observe that $(\eta L)_{\alpha\beta}=\eta_{\alpha\tau}{L^\tau}_\beta=L_{\alpha\beta}$ and $L^T\eta=(\eta L)^T$.  It follows that $L^T\eta+\eta L=0$ implies that $L^{\alpha\beta}=-L^{\beta\alpha}$.  Consequently,
\begin{align*}
  [\sigma(L),\sigma(M)]
    &=\sigma(L)\sigma(M)-\sigma(M)\sigma(L)\\
    &=\tfrac{1}{16}L^{\alpha\beta}M^{\mu\nu}
        (\gamma_\alpha\gamma_\beta\gamma_\mu\gamma_\nu
         -\gamma_\mu\gamma_\nu\gamma_\alpha\gamma_\beta)\\
    &=\tfrac{1}{8}L^{\alpha\beta}M^{\mu\nu}
        (\eta_{\alpha\mu}\gamma_\beta\gamma_\nu
         -\eta_{\beta\mu}\gamma_\alpha\gamma_\nu
         -\eta_{\nu\beta}\gamma_\mu\gamma_\alpha
         +\eta_{\nu\alpha}\gamma_\mu\gamma_\beta)\\
    &=\tfrac{1}{4}L^{\alpha\beta}M^{\mu\nu}
        (-\eta_{\beta\mu}\gamma_\alpha\gamma_\nu
         +\eta_{\nu\alpha}\gamma_\mu\gamma_\beta)\\
    &=\tfrac{1}{4}(-L^{\alpha\beta}\eta_{\beta\mu}M^{\mu\nu}\gamma_\alpha\gamma_\nu
                   +M^{\mu\nu}\eta_{\nu\alpha}L^{\alpha\beta}\gamma_\mu\gamma_\beta)\\
    &=-\tfrac{1}{4}[M,L]^{\alpha\nu}\gamma_\alpha\gamma_\nu
     =\sigma([M,L]).\qedhere
\end{align*}
\end{proof}

%%%%%%%%%%%%%%%%%%%%%%%%%%%%%%%%%%%%%%%%%%%%%%%%%%%%%%%%%%%%%%%%%
\section{Extended Minkowski space}

{\em Standing assumption.}
Unless otherwise stated, we will assume that all multi--indices are {\em unique.} That is, they contain no duplicated indices.  Recall that $|I|$ denotes the length of the multi--index $I$.

%::::::::::::::::::::::::::::::::::::::::::::::::::::::::::::::::
\subsection{Basic constructions}

Extended Minkowski space, denoted $\bigwedge{\mathbb M}$, is the complexified exterior algebra of Minkowski space ${\mathbb M}$.  Spinors are taken to be functions with values in $\bigwedge{\mathbb M}$.  We obtain a Hermitian metric $\hat\eta$ by extending the metric $\eta$.  That is, for spinors $\psi,\varphi$ we have $\psi^\dagger\hat\eta\varphi=\bar\psi^I\hat\eta_{IJ}\varphi^J$.  The components of the extended metric are defined in general by $\hat\eta_{IJ}=0$ if $|I|\neq|J|$, and
$$\hat\eta_{IJ}=\det(\eta_{\alpha_i\beta_j})
  \quad\text{for}\quad
  I=\alpha_1\cdots\alpha_k, J=\beta_1\cdots\beta_k
$$
if $|I|=|J|=k\neq 0$.  In the case of a diagonal metric, this formula takes a simple form.

\begin{fact}
If $\eta$ is diagonal, then $\hat\eta$ is also diagonal with $\hat\eta_{\emptyset\emptyset}=1$ and $\hat\eta_{II}=\eta_{\alpha_1\alpha_1}\cdots\eta_{\alpha_k\alpha_k}$ for $I=\alpha_1\cdots\alpha_k\neq\emptyset$.
\end{fact}

\begin{proof}
Suppose the sets $\{\alpha_1,\dots,\alpha_k\}$ and $\{\beta_1,\dots,\beta_k\}$ are not equal.  Say $\alpha_l\not\in\{\beta_1,\dots,\beta_k\}$.  Then the $l$--th row of the $k\times k$ matrix $(\eta_{\alpha_i\beta_j})$ consists entirely of 0 values.  Hence $\hat\eta_{IJ}=0$.  On the other hand, the matix $(\eta_{\alpha_i\alpha_j})$ is diagonal, and its determinant is as advertised.
\end{proof}

Gamma matrices on extended Minkowski space are defined as $\gamma_\alpha\psi\doteq{\bf e}_\alpha\wedge\psi-\iota_\alpha\psi$.  Here $\iota_\alpha$ is the usual interior product on an exterior algebra.  Recall that $\iota_\alpha{\bf e}_\beta=\eta_{\alpha\beta}$, and $\iota_\alpha({\bf e}_I\wedge{\bf e}_J)=(\iota_\alpha{\bf e}_I)\wedge{\bf e}_J+(-1)^{|I|}{\bf e}_I\wedge(\iota_\beta{\bf e}_J)$.

\begin{fact}\label{fact:gammaclifford}
$\gamma_\alpha\gamma_\beta+\gamma_\beta\gamma_\alpha=-2\eta_{\alpha\beta}{\mathcal I}$.
\end{fact}

\begin{proof}
Compute for any spinor $\psi$:
\begin{align*}
  \gamma_\alpha\gamma_\beta\psi
  &=\gamma_\alpha({\bf e}_\beta\wedge\psi-\iota_\beta\psi)
   ={\bf e}_{\alpha\beta}\wedge\psi-\iota_\alpha({\bf e}_\beta\wedge\psi)
    -{\bf e}_\alpha\wedge\iota_\beta\psi+\iota_\alpha\iota_\beta\psi\\
  &={\bf e}_{\alpha\beta}\wedge\psi-\eta_{\alpha\beta}\psi+{\bf e}_\beta\wedge\iota_\alpha\psi
    -{\bf e}_\alpha\wedge\iota_\beta\psi+\iota_\alpha\iota_\beta\psi.
\end{align*}
Now it it is well--known that $\iota_\alpha\iota_\beta+\iota_\beta\iota_\alpha=0$.  Therefore, symmetrization yields $(\gamma_\alpha\gamma_\beta+\gamma_\beta\gamma_\alpha)\psi=-2\eta_{\alpha\beta}\psi$.
\end{proof}

\begin{fact}\label{fact:gammametric}
$\gamma_\alpha^\dagger\hat\eta+\hat\eta\gamma_\alpha=0$.
\end{fact}

\begin{proof}
It is well--known that exterior and interior products on an exterior algebra are dual with respect to the extended metric.  That is for spinors $\psi,\varphi$ we have $({\bf e}_\alpha\wedge\psi)^\dagger\hat\eta\varphi=\psi^\dagger\hat\eta\iota_\alpha\varphi$ and $(\iota_\alpha\psi)^\dagger\hat\eta\varphi=\psi^\dagger\hat\eta({\bf e}_\alpha\wedge\varphi)$.  Consequently,
\begin{align*}
  (\gamma_\alpha\psi)^\dagger\hat\eta\varphi
  &=({\bf e}_\alpha\wedge\psi-\iota_\alpha\psi)^\dagger\hat\eta\varphi
   =({\bf e}_\alpha\wedge\psi)^\dagger\hat\eta\varphi
    -(\iota_\alpha\psi)^\dagger\hat\eta\varphi\\
  &=\psi^\dagger\hat\eta\iota_\alpha\varphi
    -\psi^\dagger\hat\eta({\bf e}_\alpha\wedge\varphi)
   =\psi^\dagger\hat\eta(\iota_\alpha\varphi-{\bf e}_\alpha\wedge\varphi)
   =-\psi^\dagger\hat\eta\gamma_\alpha\varphi.\qedhere
\end{align*}
\end{proof}

Let $I=\alpha_1\cdots\alpha_k$ be a (unique) multi--index.  If $k=0$, set $\gamma_I=\gamma_\emptyset\doteq{\mathcal I}$.  And for $k\neq 0$, define $\gamma_I\doteq\gamma_{\alpha_1}\cdots\gamma_{\alpha_k}$.

\begin{fact}\label{fact:index2gamma}
If $\eta$ is diagonal, then for any (unique) multi--index $I$, $\gamma_I{\bf e}_\emptyset={\bf e}_I$.
\end{fact}

\begin{proof}
First observe that because $\eta$ is diagonal, $\iota_\alpha{\bf e}_\beta=\eta_{\alpha\beta}{\bf e}_\emptyset=0$ if $\alpha\neq\beta$.  It follows that $\iota_\alpha{\bf e}_J=0$ if $\alpha$ is not an index in $J$.  We induct on $|I|$.  Observe that $\gamma_I{\bf e}_\emptyset={\bf e}_I$ is true by definition if $|I|=0$.  So suppose that ${\bf e}_I={\bf e}_\alpha\wedge{\bf e}_J$, where $\alpha\not\in J$ (that is $\alpha$ is not an index in $J$), and we inductively assume that $\gamma_J{\bf e}_\emptyset={\bf e}_J$.  Therefore,
$$\gamma_I{\bf e}_\emptyset
   =\gamma_\alpha\gamma_J{\bf e}_\emptyset
   =\gamma_\alpha{\bf e}_J
   ={\bf e}_\alpha\wedge{\bf e}_J-\iota_\alpha{\bf e}_J
   ={\bf e}_\alpha\wedge{\bf e}_J
   ={\bf e}_I.\qedhere
$$
\end{proof}

We remark that Fact \ref{fact:index2gamma} is false if $I$ is not unique.  Indeed ${\bf e}_I=0$ in this case, while $\gamma_I$ is not necessarily 0.

\begin{fact}\label{fact:gammametricdiag}
If $\eta$ is diagonal, then $\gamma_I^\dagger\hat\eta=s_I\hat\eta\gamma_I$, where $s_I\doteq 1$ if $|I|=0,3,4$ and $s_I\doteq -1$ if $|I|=1,2$.
\end{fact}

\begin{proof}
Set $q(k)\doteq s_I$ for $|I|=k$.  Observe that $q$ satisfies the recurrence relation $q(0)=1$ and $q(k+1)=(-1)^{k+1}q(k)$.  Induct on $|I|$.  If $|I|=0$, $\gamma_\emptyset^\dagger\hat\eta=\hat\eta=s_\emptyset\hat\eta\gamma_\emptyset$.  So suppose that ${\bf e}_I={\bf e}_\alpha\wedge{\bf e}_J$ where $\alpha\not\in J$, and assume that $\gamma_J^\dagger\hat\eta=s_J\hat\eta\gamma_J$.  Setting $k\doteq|J|$, we then have
\begin{align*}
  \gamma_I^\dagger\hat\eta
  &=(\gamma_\alpha\gamma_J)^\dagger\hat\eta
   =\gamma_J^\dagger\gamma_\alpha^\dagger\hat\eta
   =-\gamma_J^\dagger\hat\eta\gamma_\alpha
   =-s_J\hat\eta\gamma_J\gamma_\alpha
   =-(-1)^{|J|}s_J\hat\eta\gamma_\alpha\gamma_J\\
  &=(-1)^{|J|+1}s_J\hat\eta\gamma_I
   =(-1)^{k+1}q(k)\hat\eta\gamma_I
   =q(k+1)\hat\eta\gamma_I
   =s_I\hat\eta\gamma_I.
\end{align*}
courtesy of Fact \ref{fact:gammametric}, and Fact \ref{fact:gammaclifford} coupled with the diagonality of $\eta$ (so that $\gamma_J\gamma_\alpha=(-1)^{|J|}\gamma_\alpha\gamma_J$).
\end{proof}

\begin{fact}
Suppose $L=({L^\alpha}_\beta)$ is a Lorentz algebra element, and let $L$ act on extended Minkowski space via the Leibniz rule:
$$L\cdot{\bf e}_\alpha={L^\beta}_\alpha{\bf e}_\beta
  \quad\text{and}\quad
  L\cdot(\psi\wedge\varphi)
  =(L\cdot\psi)\wedge\varphi+\psi\wedge(L\cdot\varphi)
$$
for spinors $\psi,\varphi$.  This defines a Lie algebra representation, and we let $\hat{L}$ denote the extension of $L$.  That is, $\hat{L}\psi\doteq L\cdot\psi$.
\end{fact}

\begin{proof}
We need to show that $[\hat{L},\hat{M}]=\widehat{[L,M]}$ for Lorentz algebra elements $L,M$, or equivalently, $[\hat{L},\hat{M}]{\bf e}_I=\widehat{[L,M]}{\bf e}_I$ for all multi--indices $I$.  Observe that this is trivially true if $|I|=0$.  We induct on the length of $I$.   Write ${\bf e}_I={\bf e}_\alpha\wedge{\bf e}_J$, and compute
\begin{align*}
  \hat{L}\hat{M}({\bf e}_\alpha\wedge{\bf e}_J)
  &=\hat{L}({M^\beta}_\alpha{\bf e}_\beta\wedge{\bf e}_J+{\bf e}_\alpha\wedge\hat{M}{\bf e}_J)\\
  &={L^\rho}_\beta{M^\beta}_\alpha{\bf e}_\rho\wedge{\bf e}_J
    +{M^\beta}_\alpha{\bf e}_\beta\wedge\hat{L}{\bf e}_J\\
  &\quad\quad
    +{L^\beta}_\alpha{\bf e}_\beta\wedge\hat{M}{\bf e}_J
    +{\bf e}_\alpha\wedge\hat{L}\hat{M}{\bf e}_J.
\end{align*}
It follows that
\begin{align*}
  [\hat{L},\hat{M}]({\bf e}_\alpha\wedge{\bf e}_J)
  &=\hat{L}\hat{M}({\bf e}_\alpha\wedge{\bf e}_J)
    -\hat{M}\hat{L}({\bf e}_\alpha\wedge{\bf e}_J)\\
  &=({L^\rho}_\beta{M^\beta}_\alpha-{M^\rho}_\beta{L^\beta}_\alpha){\bf e}_\rho\wedge{\bf e}_J
    +{\bf e}_\alpha\wedge(\hat{L}\hat{M}-\hat{M}\hat{L}){\bf e}_J\\
  &={[L,M]^\beta}_\alpha{\bf e}_\rho\wedge{\bf e}_J
    +{\bf e}_\alpha\wedge[\hat{L},\hat{M}]{\bf e}_J.
\end{align*}
Inductively, we can assume that $[\hat{L},\hat{M}]{\bf e}_J=\widehat{[L,M]}{\bf e}_J$.  Whence
\begin{align*}
  [\hat{L},\hat{M}]{\bf e}_I
  &=[\hat{L},\hat{M}]({\bf e}_\alpha\wedge{\bf e}_J)
   ={[L,M]^\beta}_\alpha{\bf e}_\rho\wedge{\bf e}_J
    +{\bf e}_\alpha\wedge[\hat{L},\hat{M}]{\bf e}_J\\
  &={[L,M]^\beta}_\alpha{\bf e}_\rho\wedge{\bf e}_J
    +{\bf e}_\alpha\wedge\widehat{[L,M]}{\bf e}_J
   =\widehat{[L,M]}({\bf e}_\alpha\wedge{\bf e}_J)\\
  &=\widehat{[L,M]}{\bf e}_I.\qedhere
\end{align*}
\end{proof}

%::::::::::::::::::::::::::::::::::::::::::::::::::::::::::::::::
\subsection{Theta matrices}

For any multi--index $I$, the $16\times 16$ matrix $\Theta_I$ is defined by the rules (i) $\Theta_I$ commutes with gamma matrices, and (ii) $\Theta_I{\bf e}_\emptyset={\bf e}_I$.  In particular, if $I$ is not unique, then $\Theta_I=0$.

\begin{fact}\label{fact:thetawell}
Theta matrices are well--defined.
\end{fact}

\begin{proof}
We must show that the definition is basis independent.  For a change of basis on Minkowski space, ${\bf e}_\alpha'={(B^{-1})^\beta}_\alpha{\bf e}_\beta$.  This extends a change of basis on $\bigwedge{\mathbb M}$ via the mutliplicative rule $\hat{B}^{-1}(\psi\wedge\varphi)=(\hat{B}^{-1}\psi)\wedge(\hat{B}^{-1}\varphi)$.  Note that this is extension is necessarily different from the definition of the extended representation: the extended representation is a Lie {\em algebra} representation, while a change of basis is a Lie {\em group} extension.  Now by general principles, we have the transformation rule $\Theta_I'={(\hat{B}^{-1})^J}_I\hat{B}\Theta_J\hat{B}^{-1}$.  On the other hand, the transformation rule for gamma matrices is $\gamma_\alpha'={(B^{-1})^\beta}_\alpha\hat{B}\gamma_\beta\hat{B}^{-1}$.  We first show that $\Theta_I'\gamma_\alpha'=\gamma_\alpha'\Theta_I'$.  Because $\gamma_\alpha$ and $\Theta_I$ commute, we have
\begin{align*}
  \Theta_I'\gamma_\alpha'
  &={(\hat{B}^{-1})^K}_I\hat{B}\Theta_K\hat{B}^{-1}
    {(B^{-1})^\beta}_\alpha\hat{B}\gamma_\beta\hat{B}^{-1}
   ={(\hat{B}^{-1})^K}_I{(B^{-1})^\beta}_\alpha\hat{B}\Theta_K\gamma_\beta\hat{B}^{-1}\\
  &={(\hat{B}^{-1})^K}_I{(B^{-1})^\beta}_\alpha\hat{B}\gamma_\beta\Theta_K\hat{B}^{-1}
   ={(B^{-1})^\beta}_\alpha\hat{B}\gamma_\beta\hat{B}^{-1}
    {(\hat{B}^{-1})^K}_I\hat{B}\Theta_K\hat{B}^{-1}\\
  &=\gamma_\alpha'\Theta_I'.
\end{align*}
Second, we show that $\Theta_I'{\bf e}_\emptyset={\bf e}_I$.  Note that $\hat{B}{\bf e}_\emptyset={\bf e}_\emptyset$, so that
\begin{align*}
  \Theta_I'{\bf e}_\emptyset
  &={(\hat{B}^{-1})^K}_I\hat{B}\Theta_K\hat{B}^{-1}{\bf e}_\emptyset
   ={(\hat{B}^{-1})^K}_I\hat{B}\Theta_K{\bf e}_\emptyset
   ={(\hat{B}^{-1})^K}_I\hat{B}{\bf e}_K\\
  &=\hat{B}\hat{B}^{-1}{\bf e}_I
   ={\bf e}_I.\qedhere
\end{align*}
\end{proof}

\begin{fact}\label{fact:thetaunique}
Any $16\times 16$ matrix $M$ that commutes with gamma matrices is completely determined by the value of $M{\bf e}_\emptyset$.  Indeed if $\eta$ is diagonal, then $M{\bf e}_J=\gamma_JM{\bf e}_\emptyset$.  And in particular, $\Theta_I{\bf e}_J=\gamma_J{\bf e}_I$.
\end{fact}

\begin{proof}
The second and third statements follows from Fact \ref{fact:index2gamma}.  The first statement then follows by making a change of basis so that $\eta$ is diagonal.
\end{proof}

\begin{fact}
Let $\Theta_\ast$ be the set of all $16\times 16$ matrices that commute with gamma matrices.  It is a sixteen--dimensional complex vector space spanned by the matrices $\Theta_I$ for all multi--indicies $I$.  Moreover, $\Theta_\ast$ is also a matrix ring with unit $\Theta_\emptyset={\mathcal I}$.
\end{fact}

\begin{proof}
We readily have that $\Theta_\ast$ is closed under both multipication, addition, and scalar multiplication.  Also ${\mathcal I}\in\Theta_\ast$.  So $\Theta_\ast$ is a matrix ring.  Note that $\Theta_\emptyset{\bf e}_\emptyset={\bf e}_\emptyset={\mathcal I}{\bf e}_\emptyset$, so that $\Theta_\emptyset={\mathcal I}$ by Fact \ref{fact:thetaunique}.  Now suppose that $M\in\Theta_\ast$.  Write $M{\bf e}_\emptyset=a^I{\bf e}_I=a^I\Theta_I{\bf e}_\emptyset$.  Then $M=a^I\Theta_I$, again by Fact \ref{fact:thetaunique}.
\end{proof}

\begin{fact}
Single--index theta matrices satisfy $\Theta_\alpha\Theta_\beta+\Theta_\beta\Theta_\alpha=-2\eta_{\alpha\beta}{\mathcal I}$, so that $\Theta_\ast$ is a Clifford algebra representation.
\end{fact}

\begin{proof}
By Fact \ref{fact:thetaunique}, we only need show the identity holds when evaluated at ${\bf e}_\emptyset$.  We compute
\begin{align*}
  \Theta_\alpha\Theta_\beta{\bf e}_\emptyset
  &=\Theta_\alpha{\bf e}_\beta
   =\gamma_\beta{\bf e}_\alpha
   ={\bf e}_{\beta\alpha}-\eta_{\beta\alpha}{\bf e}_\emptyset.
\end{align*}
It follows that $(\Theta_\alpha\Theta_\beta+\Theta_\beta\Theta_\alpha){\bf e}_\emptyset=-2\eta_{\alpha\beta}{\bf e}_\emptyset$, as ${\bf e}_{\beta\alpha}=-{\bf e}_{\alpha\beta}$.
\end{proof}

\begin{fact}
Single--index theta matrices are similar to gamma matrices.  Explicity if $\eta$ is diagonal, then $\Theta_\alpha=S\gamma_\alpha S^{-1}$, where $S$ is diagonal with ${S^I}_I=1$ if $|I|=0,1,4$ and ${S^I}_I=-1$ if $|I|=2,3$.
\end{fact}

\begin{proof}
Observe that ${S^I}_I=z(|I|)$, where $z(k)$ satisfies the recurrence relation $z(0)=1$ and $z(k+1)=(-1)^kz(k)$.  It suffices to show that $(\Theta_\alpha S-S\gamma_\alpha){\bf e}_I=0$ for all multi--indices $I$.  Now if $\alpha\not\in I$ and $|I|=k$, then
\begin{align*}
  (\Theta_\alpha S-S\gamma_\alpha){\bf e}_I
  &=z(k)\Theta_\alpha{\bf e}_I-S({\bf e}_\alpha\wedge{\bf e}_I)
   =z(k)\gamma_I{\bf e}_\alpha-z(k+1){\bf e}_\alpha\wedge{\bf e}_I\\
  &=z(k){\bf e}_I\wedge{\bf e}_\alpha-z(k+1){\bf e}_\alpha\wedge{\bf e}_I\\
  &=(-1)^kz(k){\bf e}_\alpha\wedge{\bf e}_I-z(k+1){\bf e}_\alpha\wedge{\bf e}_I
   =0.
\end{align*}
On the other hand, suppose $\alpha\in I$.  Without loss of generality, we may write ${\bf e}_I={\bf e}_\alpha\wedge{\bf e}_J$ with $\alpha\not\in J$ and $|J|=k$.  In this case,
\begin{align*}
  (\Theta_\alpha S-S\gamma_\alpha){\bf e}_I
  &=z(k+1)\Theta_\alpha({\bf e}_\alpha\wedge{\bf e}_J)
    -S\gamma_\alpha({\bf e}_\alpha\wedge{\bf e}_J)\\
  &=z(k+1)\gamma_\alpha\gamma_J{\bf e}_\alpha
    -S(-\eta_{\alpha\alpha}){\bf e}_J\\
  &=z(k+1)\gamma_\alpha({\bf e}_J\wedge{\bf e}_\alpha)
    +z(k)\eta_{\alpha\alpha}{\bf e}_J\\
  &=(-1)^kz(k+1)\gamma_\alpha({\bf e}_\alpha\wedge{\bf e}_J)
    -z(k)\gamma_\alpha({\bf e}_\alpha\wedge{\bf e}_J)\\
  &=(-1)^k[z(k+1)-(-1)^kz(k)]\gamma_\alpha({\bf e}_\alpha\wedge{\bf e}_J)
   =0.\qedhere
\end{align*}
\end{proof}

\begin{fact}\label{fact:thetametricdiag}
If $\eta$ is diagonal, then $\Theta_I^\dagger\hat\eta=s_I\hat\eta\Theta_I$, where $s_I$ is as in Fact \ref{fact:gammametricdiag}.
\end{fact}

\begin{proof}
First we show that ${\bf e}_K^\dagger(\Theta_I^\dagger\hat\eta-s_I\hat\eta\Theta_I){\bf e}_\emptyset=0$.  Indeed, using Fact \ref{fact:gammametricdiag} and Fact \ref{fact:thetaunique} we have
\begin{align*}
  {\bf e}_K^\dagger(\Theta_I^\dagger\hat\eta-s_I\hat\eta\Theta_I){\bf e}_\emptyset
  &=(\Theta_I{\bf e}_K)^\dagger\hat\eta{\bf e}_\emptyset
    -s_I{\bf e}_K^\dagger\hat\eta{\bf e}_I
   =(\gamma_K{\bf e}_I)^\dagger\hat\eta{\bf e}_\emptyset-s_I\hat\eta_{KI}\\
  &={\bf e}_I^\dagger\gamma_K^\dagger\hat\eta{\bf e}_\emptyset-s_I\hat\eta_{KI}
   =s_K{\bf e}_I^\dagger\hat\eta\gamma_K{\bf e}_\emptyset-s_I\hat\eta_{KI}\\
  &=s_K{\bf e}_I^\dagger\hat\eta{\bf e}_K-s_I\hat\eta_{KI}
   =s_K\hat\eta_{IK}-s_I\hat\eta_{KI}
   =0
\end{align*}
because $\hat\eta$ is diagonal ($\hat\eta_{IK}=0$ unless $I=K$).  More generally, we compute
\begin{align*}
  {\bf e}_L^\dagger(\Theta_I^\dagger\hat\eta-s_I\hat\eta\Theta_I){\bf e}_J
  &={\bf e}_L^\dagger(\Theta_I^\dagger\hat\eta-s_I\hat\eta\Theta_I)\gamma_J{\bf e}_\emptyset
   ={\bf e}_L^\dagger(s_J\Theta_I^\dagger\gamma_J^\dagger\hat\eta
                      -s_I\hat\eta\gamma_J\Theta_I){\bf e}_\emptyset\\
  &={\bf e}_L^\dagger(s_J\Theta_I^\dagger\gamma_J^\dagger\hat\eta
                      -s_Is_J\gamma_J^\dagger\hat\eta\Theta_I){\bf e}_\emptyset\\
  &=s_J{\bf e}_L^\dagger\gamma_J^\dagger(\Theta_I^\dagger\hat\eta
                       -s_I\hat\eta\Theta_I){\bf e}_\emptyset\\
  &=s_J(\gamma_J{\bf e}_L)^\dagger(\Theta_I^\dagger\hat\eta
                       -s_I\hat\eta\Theta_I){\bf e}_\emptyset.
\end{align*}
As $\gamma_J{\bf e}_L={(\gamma_J)^K}_L{\bf e}_K$, this last expression is equal to 0 by the previous computation.  Whence $\Theta_I^\dagger\hat\eta-s_I\hat\eta\Theta_I=0$.
\end{proof}

For the remainder of this section, we focus on rectangular coordinates $(x^\alpha)=(t,x,y,z)$ on Minkowski space, where $\eta={\it diag}(1,-1,-1,-1)$.  The multiplication table for theta matrices is given in Appendix \ref{sec:appendixA}.  Observe in particular that the theta matrix ${\mathcal J}\doteq\Theta_{txyz}$ is such that ${\mathcal J}^2=-{\mathcal I}$.  It should be noted that all theta matrices are traceless except for $\Theta_\emptyset={\mathcal I}$.  That is, ${\it tr}\,\Theta_I=16\,\delta_{I\emptyset}$.

\begin{fact}\label{fact:Jproj}
${\mathcal J}$ has eigenvalues $\pm i$.  Moreover, $P_\epsilon\doteq\tfrac{1}{2}({\mathcal I}-i\epsilon{\mathcal J})$ is projection onto the $i\epsilon$ eigenspace ${\mathcal J}$, where $\epsilon=\pm 1$.  Each eigenspace is eight--dimensional.
\end{fact}

\begin{proof}
$P_\epsilon$ is indeed a projection:
$$[\tfrac{1}{2}({\mathcal I}-i\epsilon{\mathcal J})]^2
     =\tfrac{1}{4}({\mathcal I}-2i\epsilon J-\epsilon^2{\mathcal J}^2)
     =\tfrac{1}{4}({\mathcal I}-2i\epsilon J+{\mathcal I})
     =\tfrac{1}{2}({\mathcal I}-i\epsilon{\mathcal J}),$$
so that $P_\epsilon^2=P_\epsilon$.  And we have
$$P_\epsilon{\mathcal J}=\tfrac{1}{2}({\mathcal I}-i\epsilon{\mathcal J}){\mathcal J}
               =\tfrac{1}{2}({\mathcal J}-i\epsilon{\mathcal J}^2)
               =\tfrac{1}{2}({\mathcal J}+i\epsilon{\mathcal I})
               =\tfrac{i\epsilon}{2}({\mathcal I}-i\epsilon{\mathcal J})
               =i\epsilon P_\epsilon.$$
The trace of ${\mathcal J}$ is 0, so ${\it tr}\,P_\epsilon=\tfrac{1}{2}{\it tr}\,{\mathcal I}=8$.  Whence the image of $P_\epsilon$ is eight--dimensional.
\end{proof}

For the vector $\vec{v}=(v^x,v^y,v^z)$, define the linear combination of 2--index theta matrices
\begin{equation}\label{thetavec}
  \Theta(\vec{v})
  \doteq v^x\Theta_{yz}-v^y\Theta_{xz}+v^z\Theta_{xy}.
\end{equation}

\begin{fact}\label{fact:thetavec}
For any complex vector $\vec{v}$, $\Theta(\vec{v})^\dagger\hat\eta=-\hat\eta\Theta(\vec{v}^\ast)$, $\Theta(\vec{v})^2=-(\vec{v}\cdot\vec{v}){\mathcal I}$, and $\Theta(\vec{v})$ commutes with ${\mathcal J}$.
\end{fact}

\begin{proof}
The first statement follows from Fact \ref{fact:thetametricdiag}.  Using the theta matrix multiplication table, we have
\begin{align*}
  \Theta(\vec{v})^2
  &=v_x^2\Theta_{yz}^2+v_y^2\Theta_{xz}+v_z^2\Theta_{xy}^2
    -v^xv^y(\Theta_{yz}\Theta_{xz}+\Theta_{xz}\Theta_{yz})\\
  &\quad\quad
    +v^xv^z(\Theta_{yz}\Theta_{xy}+\Theta_{xy}\Theta_{yz})
    -v^yv^z(\Theta_{xz}\Theta_{xy}+\Theta_{xy}\Theta_{xz})\\
  &=-(v_x^2+v_y^2+v_z^2){\mathcal I}
\end{align*}
as the $\Theta_{jk}$ anti--commute.  This gives the second statement.  The third statement also follows by direct computation:
\begin{align*}
  {\mathcal J}\Theta(\vec{v})
  &=\Theta_{txyz}(v^x\Theta_{yz}-v^y\Theta_{xz}+v^z\Theta_{xy})
   =-v^x\Theta_{tx}-v^y\Theta_{ty}-v^z\Theta_{tz}\\
  &=(v^x\Theta_{yz}-v^y\Theta_{xz}+v^z\Theta_{xy})\Theta_{txyz}
   =\Theta(\vec{v}){\mathcal J}.\qedhere
\end{align*}
\end{proof}

%::::::::::::::::::::::::::::::::::::::::::::::::::::::::::::::::
\subsection{Projection onto an irreducible summand}

Because it must commute with gamma matrices, a projection from extended Minkowski space to an irreducible summand necessarily has the form
\begin{equation}\label{eq:genproj}
  \pi=\tfrac{1}{4}A^I\Theta_I.
\end{equation}
An irreducible summand has complex dimension 4, so that we also must have ${\it tr}\,\pi=4$.  Now, with the single exception of $\Theta_\emptyset={\mathcal I}$, the theta matrices are traceless, which implies that $A^\emptyset=1$.

\begin{fact}\label{fact:Qproj}
Let $\vec{v}=(v^x,v^y,v^z)$ be a complex vector with $\vec{v}\cdot\vec{v}=1$.  The matrix $Q_{\vec{v}}\doteq\tfrac{1}{2}\bigl({\mathcal I}-{\mathcal J}\Theta(\vec{v})\bigr)$ is a projection onto a eight--dimensional subspace of extended Minkowski space.  $Q_{\vec{v}}$ commutes with ${\mathcal J}$.
\end{fact}

\begin{proof}
From fact \ref{fact:thetavec}, $Q_{\vec{v}}$ commutes with ${\mathcal J}$, and is a projection:
\begin{align*}
  Q_{\vec{v}}^2
  &=\tfrac{1}{4}\bigl({\mathcal I}-2{\mathcal J}\Theta(\vec{v})+{\mathcal J}^2\Theta(\vec{v})^2\bigr)
   =\tfrac{1}{4}\bigl({\mathcal I}-2{\mathcal J}\Theta(\vec{v})+(\vec{v}\cdot\vec{v}){\mathcal I}\bigr)
   =Q_{\vec{v}}.
\end{align*}
As ${\it tr}\,Q_{\vec{v}}=8$, the image of $Q_{\vec{v}}$ is a subspace of dimension eight.
\end{proof}

\begin{fact}\label{fact:irredproj}
If $\pi$ is a projection onto an irreducible summand of $\bigwedge{\mathbb M}$, and commutes with ${\mathcal J}$, then $\pi$ is equal to
$$\pi_{\vec{v}}^\epsilon
  \doteq P_\epsilon Q_{\vec{v}}
  =\tfrac{1}{4}({\mathcal I}-i\epsilon{\mathcal J})
   \bigl({\mathcal I}-{\mathcal J}\Theta(\vec{v})\bigr)
  =\tfrac{1}{4}({\mathcal I}-i\epsilon{\mathcal J})
   \bigl({\mathcal I}-i\epsilon\Theta(\vec{v})\bigr)
$$
for some $\epsilon=\pm 1$ and complex vector $\vec{v}$ with $\vec{v}\cdot\vec{v}=1$.
\end{fact}

\begin{proof}[Sketch of proof]
From the theta matrix multiplication table, $\Theta_I$ commutes with ${\mathcal J}$ only if $|I|$ is even.  It follows that if we require $\pi{\mathcal J}={\mathcal J}\pi$, then only the indices $I$ of even length can appear in equation \eqref{eq:genproj}.  That is,
$$\pi=\tfrac{1}{4}({\mathcal I}+A^{txyz}{\mathcal J}
                    +A^{tx}\Theta_{tx}+A^{ty}\Theta_{ty}+A^{tz}\Theta_{tz}
                    +A^{xy}\Theta_{xy}+A^{xz}\Theta_{xz}+A^{yz}\Theta_{yz}).
$$
On the other hand, one computes using the theta matrix multiplication table:
\begin{align*}
  \pi^2
  &=\tfrac{1}{16}[(1-A_{txyz}^2+A_{tx}^2+A_{ty}^2+A_{tz}^2
                    -A_{xy}^2-A_{xz}^2-A_{yz}^2){\mathcal I}\\
  &\quad\quad\quad
                   +2(A^{ty}A^{xz}-A^{tz}A^{xy}-A^{tx}A^{yz}-A^{txyz}){\mathcal J}\\
  &\quad\quad\quad
                   +2(A^{tx}-A^{txyz}A^{xy})\Theta_{tx}
                   +2(A^{ty}+A^{txyz}A^{xz})\Theta_{ty}\\
  &\quad\quad\quad
                   +2(A^{tz}-A^{txyz}A^{xy})\Theta_{tz}
                   +2(A^{xy}+A^{txyz}A^{tz})\Theta_{xy}\\
  &\quad\quad\quad
                   +2(A^{xz}-A^{txyz}A^{ty})\Theta_{xz}
                   +2(A^{yz}+A^{txyz}A^{tx})\Theta_{yz}]
\end{align*}  
Observe that ${\it tr}(\Theta_I\Theta_J)=0$ unless $I=J$.  Thus
$${\it tr}(\pi\Theta_{tx})
  =\tfrac{1}{4}A^{tx}{\it tr}(\Theta_{tx}^2)
  =\tfrac{1}{4}A^{tx}{\it tr}\,{\mathcal I}
  =4A^{tx}.
$$
Moreover,
$${\it tr}(\pi^2\Theta_{tx})
  =\tfrac{1}{16}\cdot 2(A^{tx}-A^{txyz}A^{yz}){\it tr}\,\Theta_{tx}^2
  =2(A^{tx}-A^{txyz}A^{yz}).
$$
Combining this computation with the previous, we get ${\it tr}(\pi^2\Theta_{tx}-\pi\Theta_{tx})=-2(A^{tx}+A^{txyz}A^{yz})$.  For $\pi$ to be a projection, we must have $\pi^2=\pi$,  so that $A^{tx}+A^{txyz}A^{yz}=0$.  Making similar computations with ${\it tr}(\pi^2\Theta_J-\pi\Theta_J)$ for $J=ty,tz,xy,xz,yz$, we arrive at the constraints
\begin{align*}
    A^{tx}+A^{txyz}A^{yz} &=0
  & A^{ty}-A^{txyz}A^{xz} &=0
  & A^{tz}+A^{txyz}A^{xy} &=0\\
    A^{xy}-A^{txyz}A^{tz} &=0
  & A^{xz}+A^{txyz}A^{ty} &=0
  & A^{yz}-A^{txyz}A^{tx} &=0.
\end{align*}
From which it follows that $A^{txyz}=-i\epsilon$, $A^{xy}=-i\epsilon A^{tz}$, $A^{xz}=i\epsilon A^{ty}$, and $A^{yz}=-i\epsilon A^{tx}$.  Whence
\begin{align*}
  \pi&=\tfrac{1}{4}[{\mathcal I}-i\epsilon{\mathcal J}+v^x(\Theta_{tx}-i\epsilon\Theta_{yz})
                    +v^y(\Theta_{ty}+i\epsilon\Theta_{xz})+v^z(\Theta_{tz}-i\epsilon\Theta_{xy})]\\
     &=\tfrac{1}{4}({\mathcal I}-i\epsilon{\mathcal J})({\mathcal I}+v^x\Theta_{tx}
                    +v^y\Theta_{ty}+v^z\Theta_{tz})
      =\tfrac{1}{4}({\mathcal I}-i\epsilon{\mathcal J})
          \bigl({\mathcal I}-{\mathcal J}\Theta(\vec{v})\bigr)
\end{align*}
and
\begin{align*}
  \pi^2
  &=\tfrac{1}{8}[(1+v_x^2+v_y^2+v_z^2){\mathcal I}-i\epsilon(1+v_x^2+v_y^2+v_z^2){\mathcal J}\\
  &\quad\quad
                 +2v^x(\Theta_{tx}-i\epsilon\Theta_{yz})+2v^y(\Theta_{ty}+i\epsilon\Theta_{xz})
                 +2v^z(\Theta_{tz}-i\epsilon\Theta_{xy})]
\end{align*}
where we have set $v^x\doteq A^{tx}$, $v^y\doteq A^{ty}$, and $v^z\doteq A^{tz}$.  Consequently, $\pi^2=\pi$ if and only if $\vec{v}\cdot\vec{v}=v_x^2+v_y^2+v_z^2=1$.
\end{proof}

\begin{fact}
${\mathcal J}\pi_{\vec{v}}^\epsilon=i\epsilon\pi_{\vec{v}}^\epsilon$.
\end{fact}

\begin{proof}
This follows from facts \ref{fact:Jproj}, \ref{fact:Qproj}, and \ref{fact:irredproj}.
\end{proof}

\begin{fact}
The projections $\pi_{\vec{v}}^+$ and $\pi_{\vec{v}'}^-$ commute.  In fact, $\pi_{\vec{v}}^+\pi_{\vec{v}'}^-=0=\pi_{\vec{v}'}^-\pi_{\vec{v}}^+$.
\end{fact}

\begin{proof}
Note that $P_+P_-=0=P_-P_+$.  As $P_\epsilon$ commutes with $Q_{\vec{v}}$, we have $\pi_{\vec{v}}^+\pi_{\vec{v}'}^-=P_+Q_{\vec{v}}P_-Q_{\vec{v}'}=P_+P_-Q_{\vec{v}}Q_{\vec{v}'}=0$.
\end{proof}

\begin{fact}
$(\pi_{\vec{v}}^\epsilon)^\dagger\hat\eta\pi_{\vec{v}'}^\epsilon=0$.
\end{fact}

\begin{proof}
First observe that $P_\epsilon^\dagger\hat\eta P_\epsilon=0$.  Indeed,
\begin{align*}
  P_\epsilon^\dagger\hat\eta P_\epsilon
  &=\tfrac{1}{4}({\mathcal I}-i\epsilon{\mathcal J})^\dagger
      \hat\eta({\mathcal I}-i\epsilon{\mathcal J})
   =\tfrac{1}{4}({\mathcal I}+i\epsilon{\mathcal J}^\dagger)
      \hat\eta({\mathcal I}-i\epsilon{\mathcal J})\\
  &=\tfrac{1}{4}(\hat\eta-i\epsilon\hat\eta{\mathcal J}
                 +i\epsilon{\mathcal J}^\dagger\hat\eta
                 +{\mathcal J}^\dagger\hat\eta{\mathcal J})
   =\tfrac{1}{4}(\hat\eta+\hat\eta{\mathcal J}^2)
   =0.
\end{align*}
Then
$$(\pi_{\vec{v}}^\epsilon)^\dagger\hat\eta\pi_{\vec{v}'}^\epsilon
  =(P_\epsilon Q_{\vec{v}})^\dagger\hat\eta P_\epsilon Q_{\vec{v}'}
  =Q_{\vec{v}}^\dagger P_\epsilon^\dagger\hat\eta P_\epsilon Q_{\vec{v}'}
  =0.\qedhere
$$
\end{proof}

In the remainder of the article, we will arbitrarily choose the projection $\pi_z\doteq\pi_{(0,0,1)}^+$.  Note that
$$\pi_z=\tfrac{1}{4}({\mathcal I}-i{\mathcal J})({\mathcal I}+\Theta_{tx})
     =\tfrac{1}{4}({\mathcal I}-i{\mathcal J}+\Theta_{tx}-i\Theta_{yz}).
$$
The extended Minkowski space elements
\begin{equation}\label{eq:summandbasis}
\begin{aligned}
  {\bf f}_0&=\tfrac{1}{2\sqrt{2}}({\bf e}_\emptyset+{\bf e}_{tz}-i{\bf e}_{xy}-i{\bf e}_{txyz}
               -i{\bf e}_t+i{\bf e}_z-{\bf e}_{txy}+{\bf e}_{xyz})\\
  {\bf f}_1&=\tfrac{1}{2\sqrt{2}}({\bf e}_{tx}-i{\bf e}_{ty}+{\bf e}_{xz}-i{\bf e}_{yz}
               +i{\bf e}_x+{\bf e}_y-i{\bf e}_{txz}-{\bf e}_{tyz})\\
  {\bf f}_2&=\tfrac{1}{2\sqrt{2}}({\bf e}_\emptyset+{\bf e}_{tz}-i{\bf e}_{xy}-i{\bf e}_{txyz}
               +i{\bf e}_t-i{\bf e}_z+{\bf e}_{txy}-{\bf e}_{xyz})\\
  {\bf f}_3&=\tfrac{1}{2\sqrt{2}}({\bf e}_{tx}-i{\bf e}_{ty}+{\bf e}_{xz}-i{\bf e}_{yz}
               -i{\bf e}_x-{\bf e}_y+i{\bf e}_{txz}+{\bf e}_{tyz}).
\end{aligned}
\end{equation}
are seen to be fixed by $\pi_z$.  That is, $\pi_z{\bf f}_\alpha={\bf f}_\alpha$.  In addition, they are mutually orthonormal with respect to the standard Hermitian metric: ${{\bf f}_\alpha}^\dagger{\bf f}_\beta=\delta_{\alpha\beta}$.  Because the gamma matrices commute with $\pi_z$, the span of basis \eqref{eq:summandbasis} is acted on by the gamma matrices.  For example,
\begin{align*}
  \gamma_x{\bf f}_0
  &=\tfrac{1}{2\sqrt{2}}
    \gamma_x({\bf e}_\emptyset+{\bf e}_{tz}-i{\bf e}_{xy}-i{\bf e}_{txyz}
             -i{\bf e}_t+i{\bf e}_z-{\bf e}_{txy}+{\bf e}_{xyz})\\
  &=\tfrac{1}{2\sqrt{2}}({\bf e}_x-{\bf e}_{txz}-i{\bf e}_y+i{\bf e}_{tyz}
                         +i{\bf e}_{tx}+i{\bf e}_{xz}+{\bf e}_{ty}+{\bf e}_{yz})
   =i{\bf f}_3
\end{align*}
Similarly, one computes $\gamma_x{\bf f}_1=i{\bf f}_2$, $\gamma_x{\bf f}_2=-i{\bf f}_1$, and $\gamma_x{\bf f}_3=-i{\bf f}_0$.  So if we identify an element in the span of basis \eqref{eq:summandbasis} with a four--dimensional column vector, then $\gamma_x$ acts as the $4\times 4$ matrix
$$(\pi_z)_\ast\gamma_x
  \doteq
  \left(\begin{smallmatrix} 0 & 0 &  0 & -i\\
                            0 & 0 & -i &  0\\
                            0 & i &  0 &  0\\
                            i & 0 &  0 &  0\end{smallmatrix}\right)
$$
when applied to that element.  I.e.,
$${\bf v}=a{\bf f}_0+b{\bf f}_1+c{\bf f}_2+d{\bf f}_3
   \mapsto
   \left(\begin{smallmatrix} a\\ b\\ c\\ d\end{smallmatrix}\right)
$$
and
$$\gamma_x{\bf v}=a(i{\bf f}_3)+b(i{\bf f}_2)+c(-i{\bf f}_1)+d(-i{\bf f}_0)
  \mapsto
  \left(\begin{smallmatrix} -id\\ -ic\\ ib\\ ia\end{smallmatrix}\right)
  =\left(\begin{smallmatrix} 0 & 0 &  0 & -i\\
                            0 & 0 & -i &  0\\
                            0 & i &  0 &  0\\
                            i & 0 &  0 &  0
   \end{smallmatrix}\right)
   \left(\begin{smallmatrix} a\\ b\\ c\\ d\end{smallmatrix}\right).
$$
Performing similar computations with the other gamma matrices, we find that when restricted to the basis elements \eqref{eq:summandbasis},
\begin{align*}
   (\pi_z)_\ast\gamma_t
    &=i\left(\begin{smallmatrix} 1 & 0 & 0 & 0\\ 0 & 1 & 0 & 0\\
            0 & 0 & -1 & 0\\ 0 & 0 & 0 & -1\end{smallmatrix}\right)
  & (\pi_z)_\ast\gamma_x
    &=-i\left(\begin{smallmatrix} 0 & 0 &  0 & 1\\ 0 & 0 & 1 &  0\\
             0 & -1 &  0 &  0\\ -1 & 0 &  0 &  0\end{smallmatrix}\right)\\
  (\pi_z)_\ast\gamma_y
    &=-i\left(\begin{smallmatrix} 0 & 0 & 0 & -i\\ 0 & 0 & i & 0\\
             0 & i & 0 & 0\\ -i & 0 & 0 & 0\end{smallmatrix}\right)
  & (\pi_z)_\ast\gamma_z
    &=-i\left(\begin{smallmatrix} 0 & 0 & 1 & 0\\ 0 & 0 & 0 & -1\\
             -1 & 0 & 0 & 0\\ 0 & 1 & 0 & 0\end{smallmatrix}\right).
\end{align*}
This gives the Dirac representation for gamma matrices in the standard formulation of the Dirac equation, with the factor of $i$ due to our choice of sign convention for the Clifford algebra relation, equation \eqref{eq:clifford}.

%%%%%%%%%%%%%%%%%%%%%%%%%%%%%%%%%%%%%%%%%%%%%%%%%%%%%%%%%%%%%%%%
\section{Plane waves}

In rectangular coordinates $(x^\alpha)=(t,x,y,z)$, the Dirac equation is
$$D\psi=m\psi,
  \quad\text{where}\quad
  D\doteq\gamma^\alpha\partial_\alpha.
$$
For the four--momentum ${\bf p}=(p^t,p^x,p^y,p^z)$, define the matrices
$$K_{\bf p}\doteq m{\mathcal I}+ip^\alpha\gamma_\alpha
  \quad\text{and}\quad
  \Pi_{\bf p}\doteq\tfrac{1}{2m}(m{\mathcal I}-ip^\alpha\gamma_\alpha).
$$

\begin{fact}
If $p^\alpha p_\alpha=m^2$, then $\Pi_{\bf p}$ is projection onto the kernel of $K_{\bf p}$.
\end{fact}

\begin{proof}
Consider the matrix $\Pi_{\bf p}'\doteq\tfrac{1}{2m}K_{\bf p}$.  By virtue of equation \eqref{eq:clifford},
\begin{align*}
  (\Pi_{\bf p}')^2
  &=\tfrac{1}{4m^2}(m{\mathcal I}+ip^\alpha\gamma_\alpha)^2
   =\tfrac{1}{4m^2}(m^2{\mathcal I}+2imp^\alpha\gamma_\alpha
                    -p^\alpha p^\beta\gamma_\alpha\gamma_\beta)\\
  &=\tfrac{1}{4m^2}(m^2{\mathcal I}+2imp^\alpha\gamma_\alpha
                    +p^\alpha p_\alpha{\mathcal I})
   =\tfrac{1}{2m}(m{\mathcal I}+ip^\alpha\gamma_\alpha)
   =\Pi_{\bf p}'.
\end{align*}
That is, $\Pi_{\bf p}'$ is a projection whose kernel is equal to that of $K_{\bf p}$.  It follows that the image of its complement ${\mathcal I}-\Pi_{\bf p}'=\Pi_{\bf p}$ is equal to the kernel of $K_{\bf p}$.
\end{proof}

\begin{fact}
The plane wave $\psi=e^{-ip_\alpha x^\alpha}\chi$ solves the Dirac equation iff and only $p^\alpha p_\alpha=m^2$ and $\chi$ is in the kernel of $K_{\bf p}$.
\end{fact}

\begin{proof}
For the plane wave $\psi$, the square of the Dirac operator is
$$D^2\psi=\gamma^\alpha(-ip_\alpha)\gamma^\beta(-ip_\beta)\psi
         =-p^\alpha p^\beta\gamma_\alpha\gamma_\beta\psi
         =p^\alpha p_\alpha\psi
$$
by equation \eqref{eq:clifford}, so that $D^2\psi-m^2\psi=(p^\alpha p_\alpha-m^2)\psi$.  On the other hand,
\begin{align*}
  D\psi-m\psi
  &=\gamma^\alpha(-ip_\alpha)\psi-m\psi
   =-e^{-ip_\beta x^\beta}(ip^\alpha\gamma_\alpha+m{\mathcal I})\chi
   =-e^{-ip_\alpha x^\alpha}K_{\bf p}\chi.
\end{align*}
If $\psi$ solves the Dirac equation, it must also solve the square of the Dirac equation, $D^2\psi=m^2\psi$.  Whence $p^\alpha p_\alpha=m^2$ and $K_{\bf p}\chi=0$.  Conversely, the latter of these two conditions is enough to guarantee that $\psi$ solves the Dirac equation.
\end{proof}

In the remainder, we assume that $p^\alpha p_\alpha=m^2$.  Define the spinors
\begin{align*}
  \chi_0^+({\bf p})
    &\doteq\tfrac{1}{\sqrt{2}m}(m{\bf e}_\emptyset-ip^\alpha{\bf e}_\alpha)
     =\tfrac{1}{\sqrt{2}m}(m{\bf e}_\emptyset-ip^t{\bf e}_t-ip^x{\bf e}_x
                           -ip^y{\bf e}_y-ip^z{\bf e}_z)\\
  \chi_1^+({\bf p})
    &\doteq\Theta_{yz}\chi_0^+({\bf p})
     =\tfrac{1}{\sqrt{2}m}(m{\bf e}_{yz}-ip^t{\bf e}_{tyz}-ip^x{\bf e}_{xyz}
                           -ip^y{\bf e}_z+ip^z{\bf e}_y)\\
  \chi_2^+({\bf p})
    &\doteq\Theta_{xz}\chi_0^+({\bf p})
     =\tfrac{1}{\sqrt{2}m}(m{\bf e}_{xz}-ip^t{\bf e}_{txz}-ip^x{\bf e}_z
                           +ip^y{\bf e}_{xyz}+ip^z{\bf e}_x)\\
  \chi_3^+({\bf p})
    &\doteq\Theta_{xy}\chi_0^+({\bf p})
     =\tfrac{1}{\sqrt{2}m}(m{\bf e}_{xy}-ip^t{\bf e}_{txy}-ip^x{\bf e}_y
                           +ip^y{\bf e}_x-ip^z{\bf e}_{xyz})
\end{align*}
and
\begin{align*}
  \chi_0^-({\bf p})
    &\doteq{\mathcal J}\chi_0^+({\bf p})
     =\tfrac{1}{\sqrt{2}m}(m{\bf e}_{txyz}+ip^t{\bf e}_{xyz}+ip^x{\bf e}_{tyz}
                           -ip^y{\bf e}_{txz}+ip^z{\bf e}_{txy})\\
  \chi_1^-({\bf p})
    &\doteq{\mathcal J}\chi_1^+({\bf p})
     =\tfrac{1}{\sqrt{2}m}(-m{\bf e}_{tx}-ip^t{\bf e}_x-ip^x{\bf e}_t
                           +ip^y{\bf e}_{txy}+ip^z{\bf e}_{txz})\\
  \chi_2^-({\bf p})
    &\doteq{\mathcal J}\chi_2^+({\bf p})
     =\tfrac{1}{\sqrt{2}m}(m{\bf e}_{ty}+ip^t{\bf e}_y+ip^x{\bf e}_{txy}
                           +ip^y{\bf e}_t-ip^z{\bf e}_{tyz})\\
  \chi_3^-({\bf p})
    &\doteq{\mathcal J}\chi_3^+({\bf p})
     =\tfrac{1}{\sqrt{2}m}(-m{\bf e}_{tz}-ip^t{\bf e}_z-ip^x{\bf e}_{txz}
                           -ip^y{\bf e}_{tyz}-ip^z{\bf e}_t)
\end{align*}

In general, $\Theta_I\chi_0^+({\bf p})$ is necessarily in the kernel of $K_{\bf p}$ (equivalently, in the image of $\Pi_{\bf p}$) for any multi--index $I$.  However, these spinors are not all independent.

\begin{fact}\label{fact:spindep}
$p^\alpha\Theta_\alpha\chi_0^+({\bf p})=im\chi_0^+({\bf p})$.
\end{fact}

\begin{proof}
Compute:
\begin{align*}
  p^\alpha\Theta_\alpha\chi_0^+({\bf p})
  &=\tfrac{1}{\sqrt{2}m}p^\alpha\Theta_\alpha(m{\bf e}_\emptyset-ip^\beta{\bf e}_\beta)
   =\tfrac{1}{\sqrt{2}m}p^\alpha(m{\bf e}_\alpha-ip^\beta\gamma_\beta{\bf e}_\alpha)\\
  &=\tfrac{1}{\sqrt{2}m}(mp^\alpha{\bf e}_\alpha
              -ip^\beta p^\alpha \gamma_\beta\gamma_\alpha{\bf e}_\emptyset)
   =\tfrac{1}{\sqrt{2}m}(mp^\alpha{\bf e}_\alpha
              +ip^\beta p^\alpha\eta_{\beta\alpha}{\bf e}_\emptyset)\\
  &=\tfrac{1}{\sqrt{2}m}(mp^\alpha{\bf e}_\alpha
              +im^2{\bf e}_\emptyset)
   =\tfrac{1}{\sqrt{2}m}im(-ip^\alpha{\bf e}_\alpha
              +m{\bf e}_\emptyset)\\
  &=im\chi_0^+({\bf p})\qedhere
\end{align*}
\end{proof}

\begin{fact}\label{fact:chiplus}
The spinors $\chi_\alpha^+({\bf p})$ for $\alpha=0,1,2,3$ are mutually orthogonal, and with length 1, with respect to the extended metric.
\end{fact}

\begin{proof}
For $I=xy,xz,yz$, we have $\Theta_I^\dagger\hat\eta\Theta_I=-\hat\eta\Theta_I^2=\hat\eta$.  It follows that $\chi_\alpha^+({\bf p})^\dagger\hat\eta\chi_\alpha^+({\bf p})=\chi_0^+({\bf p})^\dagger\hat\eta\chi_0^+({\bf p})$.  Moroever,
$$\chi_0^+({\bf p})^\dagger\hat\eta\chi_0^+({\bf p})
  =\tfrac{1}{2m^2}(m{\bf e}_\emptyset+ip^\alpha{\bf e}_\alpha)
   \hat\eta(m{\bf e}_\emptyset-ip^\beta{\bf e}_\beta)
  =\tfrac{1}{2m^2}(m^2+p^\alpha p^\beta\eta_{\alpha\beta})
  =1.
$$
So $\chi_\alpha^+({\bf p})$ has length 1.  The fact that $\chi_\alpha^+({\bf p})^\dagger\hat\eta\chi_\beta^+({\bf p})=0$ for $\alpha\neq\beta$ follows by direct computation.  For instance,
$$\chi_2^+({\bf p})^\dagger\hat\eta\chi_3^+({\bf p})
  =\tfrac{1}{2m^2}(-p^yp^z\eta_{xx}\eta_{yy}\eta_{zz}+p^zp^y\eta_{xx})
   =\tfrac{1}{2m^2}(p^yp^z-p^zp^y)
   =0.\qedhere
$$
\end{proof}

\begin{fact}
The spinors $\chi_\alpha^-({\bf p})$ for $\alpha=0,1,2,3$ are mutually orthogonal, and with length $-1$, with respect to the extended metric.
\end{fact}

\begin{proof}
This follows from Fact \ref{fact:chiplus} and ${\mathcal J}^\dagger\hat\eta{\mathcal J}=\hat\eta{\mathcal J}^2=-\hat\eta$.
\end{proof}

\begin{fact}
The spinors $\chi_\alpha^+({\bf p})$ and $\chi_\alpha^-({\bf p})$ for $\alpha=0,1,2,3$ form a basis of the kernel of $K_{\bf p}$, orthogonal with respect to the extended metric.
\end{fact}

\begin{proof}
First, observe that $\chi_0^+({\bf p})$ belongs to the kernel of $K_{\bf p}$:
\begin{align*}
  K_{\bf p}\chi_0^+({\bf p})
  &=\tfrac{1}{\sqrt{2}m}(m{\mathcal I}+ip^\alpha\gamma_\alpha)
                        (m{\bf e}_\emptyset-ip^\beta{\bf e}_\beta)\\
  &=\tfrac{1}{\sqrt{2}m}(m^2{\bf e}_\emptyset-imp^\beta{\bf e}_\beta
              +imp^\alpha\gamma_\alpha{\bf e}_\emptyset
              +p^\alpha p^\beta\gamma_\alpha{\bf e}_\beta)\\
  &=\tfrac{1}{\sqrt{2}m}[m^2{\bf e}_\emptyset-imp^\beta{\bf e}_\beta
              +imp^\alpha{\bf e}_\alpha
              +p^\alpha p^\beta({\bf e}_{\alpha\beta}-\eta_{\alpha\beta}{\bf e}_\emptyset)]\\
  &=\tfrac{1}{\sqrt{2}m}(m^2-p^\alpha p_\alpha){\bf e}_\emptyset
   =0.
\end{align*}
As gamma and theta matrices commute, theta matrices commute with $K_{\bf p}$.  We then have that $\chi_\alpha^{\pm}({\bf p})$ is also in the kernel of $K_{\bf p}$.  Moreover, ${\it tr}\,\Pi_{\bf p}=\tfrac{1}{2m}(m\,{\it tr}\,{\mathcal I})=8$, so the kernel of $K_{\bf p}$ has dimension 8.  It remains to verify that $\chi_\alpha^+({\bf p})^\dagger\hat\eta\chi_\beta^-({\bf p})=0$.  This follows by direct computation.  E.g.,
\begin{align*}
  \chi_1^+({\bf p})^\dagger\hat\eta\chi_0^-({\bf p})
  &=\tfrac{1}{2m^2}(-p^tp^x\eta_{tt}\eta_{yy}\eta_{zz}-p^xp^t\eta_{xx}\eta_{yy}\eta_{zz})\\
  &=\tfrac{1}{2m^2}(-p^tp^x+p^xp^t)
   =0.\qedhere
\end{align*}
\end{proof}

In {\it Forces on an exterior bundle,} it is shown that if we assume that the Lagrangian density for an electromagnetic field $A^\alpha$ is
$$L=\tfrac{1}{2}\psi^\dagger\hat\eta D\psi+c.c.
    -m\psi^\dagger\hat\eta\psi
    +\tau{\it tr}(\Omega_{\alpha\beta}\Omega^{\alpha\beta}),
$$
where $D\psi=\gamma^\alpha(\partial_\alpha+iA_\alpha)\psi$ and $\Omega_{\alpha\beta}=\partial_\alpha A_\beta-\partial_\beta A_\alpha$, then the resulting field equation is
$$\partial_\beta\partial^\beta-\partial_\alpha\partial_\beta A^\beta
  =-ie\psi^\dagger\hat\eta\gamma_\alpha\psi
$$
for some constant $e$.  So in the context of extended Minkowsk space, it is natural to take
$$j^\alpha=-ie\psi^\dagger\hat\eta\gamma^\alpha\psi$$
as the charge current of the spinor field $\psi$.

\begin{fact}
The charge current of $\psi=e^{-ip_\alpha x^\alpha}\chi_\alpha^\epsilon({\bf p})$ is $j^\alpha=\epsilon ep^\alpha/m$, for $\epsilon=\pm 1$ and $\alpha=0,1,2,3$.
\end{fact}

\begin{proof}
For $I=xy,xz,yz$, we have $\Theta_I^\dagger\hat\eta\gamma^\alpha\Theta_I=-\hat\eta\gamma^\alpha\Theta_I^2=\hat\eta\gamma^\alpha$.  Moreover, ${\mathcal J}^\dagger\hat\eta\gamma^\alpha{\mathcal J}=\hat\eta\gamma^\alpha{\mathcal J}^2=-\hat\eta\gamma^\alpha$.  Therefore it suffices to compute the quantity $-ie\chi_0^+({\bf p})^\dagger\hat\eta\gamma^\alpha\chi_0^+({\bf p})$.
\begin{align*}
  -ie\chi_0^+({\bf p})^\dagger\hat\eta\gamma^\alpha\chi_0^+({\bf p})
  &=\tfrac{e\eta^{\alpha\beta}}{i2m^2}(m{\bf e}_\emptyset+ip^\mu{\bf e}_\mu)
    \hat\eta\gamma_\beta(m{\bf e}_\emptyset-ip^\nu{\bf e}_\nu)\\
  &=\tfrac{e\eta^{\alpha\beta}}{i2m^2}(m{\bf e}_\emptyset+ip^\mu{\bf e}_\mu)
    \hat\eta[m{\bf e}_\beta-ip^\nu({\bf e}_{\beta\nu}-\eta_{\beta\nu}{\bf e}_\emptyset)]\\
  &=\tfrac{e\eta^{\alpha\beta}}{i2m^2}(imp^\nu\eta_{\beta\nu}
                                +imp^\mu\eta_{\mu\beta})
   =\tfrac{ep^\alpha}{m}.\qedhere
\end{align*}
\end{proof}

\begin{fact}
A null spinor has zero charge current.
\end{fact}

\begin{proof}[Sketch of proof]
A null spinor has the form $\chi({\bf p})=B^\alpha\chi_\alpha^+({\bf p})+C^\alpha\chi_\alpha^-({\bf p})$, where $\bar{B}^\alpha B_\alpha=\bar{C}^\alpha C_\alpha$.  One shows that (i) $-i\chi_\alpha^+({\bf p})^\dagger\hat\eta\gamma^\mu\chi_\beta^+({\bf p})=\delta_{\alpha\beta}\tfrac{p^\mu}{m}$ by direct computation on a case by case basis.  For example, if $\alpha=1$ and $\beta=3$, we first observe that
\begin{align*}
  \chi_1^+({\bf p})\hat\eta\gamma_\mu\chi_3^+({\bf p})
  &=\chi_0^+({\bf p})^\dagger\Theta_{yz}^\dagger\hat\eta\gamma_\mu\Theta_{xy}\chi_0^+({\bf p})
  =-\chi_0^+({\bf p})^\dagger\hat\eta\gamma_\mu\Theta_{yz}\Theta_{xy}\chi_0^+({\bf p})\\
  &=-\chi_0^+({\bf p})^\dagger\hat\eta\gamma_\mu\Theta_{xz}\chi_0^+({\bf p}).
\end{align*}
Next we compute
\begin{align*}
  \gamma_\mu\Theta_{xz}{\bf e}_\emptyset
  &=\gamma_\mu{\bf e}_{xz}
   =-\eta_{\mu x}{\bf e}_z+\eta_{\mu z}{\bf e}_x+\cdots\\
  \gamma_\mu\Theta_{xz}{\bf e}_\alpha
  &=\gamma_\mu\gamma_\alpha{\bf e}_{xz}
   =\gamma_\mu({\bf e}_{\alpha xz}-\eta_{\alpha x}{\bf e}_z+\eta_{\alpha z}{\bf e}_x)
   =(\eta_{\alpha x}\eta_{\mu z}-\eta_{\alpha z}\eta_{\mu x}){\bf e}_\emptyset
    +\cdots
\end{align*}
where $\cdots$ indicates terms of degree at least 2.  So that
\begin{align*}
  &\chi_0^+({\bf p})^\dagger\hat\eta\gamma_\mu\Theta_{xz}\chi_0^+({\bf p})
   =\tfrac{1}{2m^2}(m{\bf e}_\emptyset+ip^\beta{\bf e}_\beta)
     \hat\eta\gamma_\mu\Theta_{xz}(m{\bf e}_\emptyset-ip^\alpha{\bf e}_\alpha)\\
  &\quad\quad
   =\tfrac{1}{2m^2}[-imp^\alpha(\eta_{\alpha x}\eta_{\mu z}-\eta_{\alpha z}\eta_{\mu x})
                    +imp^\beta(-\eta_{\mu x}\eta_{\beta z}+\eta_{\mu z}\eta_{\beta x})]
   =0.
\end{align*}
Other cases are similar.  Note that we then have (ii) $-i\chi_\alpha^-({\bf p})^\dagger\hat\eta\gamma^\mu\chi_\beta^-({\bf p})=-\delta_{\alpha\beta}\tfrac{p^\mu}{m}$, as $\chi_\alpha^-({\bf p})={\mathcal J}\chi_\alpha^-({\bf p})$.  Similarly, one also shows on a case by case basis that (iii) $-ie\chi_\alpha^+({\bf p})^\dagger\hat\eta\gamma^\mu\chi_\beta^-({\bf p})=0$.  From (i), (ii), and (iii) it follows that
$$-i\chi({\bf p})^\dagger\hat\eta\gamma^\mu\chi({\bf p})
  =(\bar{B}^\alpha B_\alpha-\bar{C}^\alpha C_\alpha)\tfrac{p^\mu}{m}
  =0.\qedhere
$$
\end{proof}

%::::::::::::::::::::::::::::::::::::::::::::::::::::::::::::::::
\subsection{Helicity operators}

\begin{lemma}\label{lem:interiorext}
For any Lorentz algebra element $L$, $\iota_\rho\hat{L}-\hat{L}\iota_\rho=-{L^\sigma}_\rho\iota_\sigma$.
\end{lemma}

\begin{proof}
We show that ($\ast$) $(\iota_\rho\hat{L}-\hat{L}\iota_\rho+{L^\sigma}_\rho\iota_\sigma)\psi=0$ by induction on the degree of $\psi$.  Note that $(\iota_\rho\hat{L}-\hat{L}\iota_\rho+{L^\sigma}_\rho\iota_\sigma){\bf e}_\emptyset=0$.  So ($\ast$) is true in degree 0.  Now assume that ($\ast$) holds for degree $k\geq 0$.  Then for $\varphi$ of degree $k$, we compute
\begin{align*}
  (\iota_\rho\hat{L}-\hat{L}\iota_\rho)({\bf e}_\alpha\wedge\varphi)
  &=\iota_\rho[{L^\beta}_\alpha{\bf e}_\beta\wedge\varphi+{\bf e}_\alpha\wedge\hat{L}\varphi)]
    -\hat{L}[\eta_{\rho\alpha}\varphi-{\bf e}_\alpha\wedge\iota_\rho\varphi]\\
  &={L^\beta}_\alpha\eta_{\rho\beta}\varphi-{L^\beta}_\alpha{\bf e}_\beta\wedge\iota_\rho\varphi
    +\eta_{\rho\alpha}\hat{L}\varphi-{\bf e}_\alpha\wedge\iota_\rho\hat{L}\varphi\\
  &\quad\quad
    -[\eta_{\rho\alpha}\hat{L}\varphi-{L^\beta}_\alpha{\bf e}_\beta\wedge\iota_\rho\varphi
      -{\bf e}_\alpha\wedge\hat{L}\iota_\rho\varphi]\\
  &=L_{\rho\alpha}\varphi-{\bf e}_\alpha\wedge(\iota_\rho\hat{L}-\hat{L}\iota_\rho)\varphi
   =-L_{\alpha\rho}\varphi+{\bf e}_\alpha\wedge({L^\sigma}_\rho\iota_\sigma\varphi)\\
  &=-{L^\sigma}_\rho(\eta_{\sigma\alpha}\varphi-{\bf e}_\alpha\wedge\iota_\sigma\varphi)
   =-{L^\sigma}_\rho\iota_\sigma({\bf e}_\alpha\wedge\varphi)
\end{align*}
via the inductive assumption.  And by linearity, ($\ast$) holds in degree $k+1$.
\end{proof}

\begin{lemma}\label{lem:gammaext}
For any Lorentz algebra element $L$, $\gamma_\rho\hat{L}-\hat{L}\gamma_\rho=-{L^\sigma}_\rho\gamma_\sigma$.
\end{lemma}

\begin{proof}
Using lemma \ref{lem:interiorext}, we compute for any spinor $\psi$,
\begin{align*}
  (\gamma_\rho\hat{L}-\hat{L}\gamma_\rho)\psi
    &=\gamma_\rho\hat{L}\psi-\hat{L}({\bf e}_\rho\wedge\psi-\iota_\rho\psi)\\
    &=\gamma_\rho\hat{L}\psi-{L^\sigma}_\rho{\bf e}_\sigma\wedge\psi
      -{\bf e}_\rho\wedge\hat{L}\psi+\hat{L}\iota_\rho\psi\\
    &=\gamma_\rho\hat{L}\psi-{L^\sigma}_\rho{\bf e}_\sigma\wedge\psi
      -{\bf e}_\rho\wedge\hat{L}\psi+\iota_\rho\hat{L}\psi+{L^\sigma}_\rho\iota_\sigma\psi\\
    &=\gamma_\rho\hat{L}\psi-{L^\sigma}_\rho\gamma_\sigma\psi-\gamma_\rho\hat{L}\psi
     =-{L^\sigma}_\rho\gamma_\sigma\psi.\qedhere
\end{align*}
\end{proof}

\begin{fact}\label{fact:spinext}
The spin and extended representations commute.  Moroever, both commute with ${\mathcal J}$.
\end{fact}

\begin{proof}
Let $L$ be a Lorentz algebra element.  From lemma \ref{lem:gammaext}, we have
\begin{align*}
  [\sigma(L)\hat{L}-\hat{L}\sigma(L)]
  &=-\tfrac{1}{4}L^{\rho\sigma}(\gamma_\rho\gamma_\sigma\hat{L}\psi
                                -\hat{L}\gamma_\rho\gamma_\sigma)\\
  &=-\tfrac{1}{4}L^{\rho\sigma}[\gamma_\rho(\hat{L}\gamma_\sigma-{L^\beta}_\sigma\gamma_\beta)
                               -(\gamma_\rho\hat{L}+{L^\beta}_\rho\gamma_\beta)\gamma_\sigma]\\
  &=\tfrac{1}{4}(L^{\rho\sigma}{L^\beta}_\sigma\gamma_\rho\gamma_\beta
                 +L^{\rho\sigma}{L^\beta}_\rho\gamma_\beta\gamma_\sigma)\\
  &=\tfrac{1}{4}(-{L^\rho}_\sigma L^{\sigma\beta}\gamma_\rho\gamma_\beta
                 +{L^\beta}_\rho L^{\rho\sigma}\gamma_\beta\gamma_\sigma)
   =0.
\end{align*}
For commutation with ${\mathcal J}$, we only need to demonstrate that the commutator $[\hat{L},{\mathcal J}]$ is zero.  First observe that $[\hat{L},{\mathcal J}]{\bf e}_\emptyset=-\hat{L}{\bf e}_{txyz}=-{\it tr}(L){\bf e}_{txyz}=0$.  So for the multi--index $I$, $[\hat{L},{\mathcal J}]{\bf e}_I=0$ if $|I|=0$.  On the other hand if ${\bf e}_I={\bf e}_\alpha\wedge{\bf e}_J$ (so that $\alpha$ is not contained in $J$), we have
\begin{align*}
  [\hat{L},{\mathcal J}]{\bf e}_I
  &=[\hat{L},{\mathcal J}]\gamma_\alpha{\bf e}_J
   =(\hat{L}{\mathcal J}\gamma_\alpha-{\mathcal J}\hat{L}\gamma_\alpha){\bf e}_J\\
  &=\{\hat{L}\gamma_\alpha{\mathcal J}-{\mathcal J}(\gamma_\alpha\hat{L}
                                          +{L^\beta}_\alpha\gamma_\beta)\}{\bf e}_J\\
  &=\{(\gamma_\alpha\hat{L}+{L^\beta}_\alpha\gamma_\beta){\mathcal J}
      -\gamma_\alpha{\mathcal J}\hat{L}-{L^\beta}_\alpha{\mathcal J}\gamma_\beta\}{\bf e}_J
   =\gamma_\alpha[\hat{L},{\mathcal J}]{\bf e}_J.
\end{align*}
Therefore, induction on the length of $I$ implies that $[\hat{L},{\mathcal J}]{\bf e}_I=0$.
\end{proof}

Now for the unit vector $\vec{u}=\vec{p}/|\vec{p}|=(u^x,u^y,u^z)$, consider the Lorentz algebra element
$$L_{\vec{u}}=\left(\begin{smallmatrix} 0 & 0 & 0 & 0\\ 0 & 0 & -u^z & u^y\\
                    0 & u^z & 0 & -u^x\\ 0 & -u^y & u^x & 0\end{smallmatrix}\right)
             =\left(\begin{smallmatrix} 0 & \vec{0}\\
                     {\vec{0}\,}^T & \Omega_{\vec{u}}\end{smallmatrix}\right)
$$
which is the generator of rotations about the axis $\vec{u}$.  In terms of components, ${(\Omega_{\vec{u}})^j}_k=\eta^{jp}\epsilon_{qpk}u^q$.  The spin representation of $L_{\vec{u}}$ is
\begin{align*}
  \sigma(L_{\vec u})
  &=-\tfrac{1}{4}L^{\alpha\beta}\gamma_\alpha\gamma_\beta
   =-\tfrac{1}{2}(L^{xy}\gamma_{xy}+L^{xz}\gamma_{xz}+L^{yz}\gamma_{yz})\\
  &=-\tfrac{1}{2}(-{L^x}_y\gamma_{xy}-{L^x}_z\gamma_{xz}-{L^y}_z\gamma_{yz})
   =-\tfrac{1}{2}(u^z\gamma_{xy}-u^y\gamma_{xz}+u^x\gamma_{yz}).
\end{align*}
We will call this the {\em spin helicity operator.}

\begin{fact}
The eigenvalues of $\sigma(L_{\vec{u}})$ are $\pm\tfrac{i}{2}$.  Moreover for $\epsilon=\pm 1$, the operator $\pi_{i\epsilon/2}\doteq\tfrac{1}{2}{\mathcal I}-i\epsilon\sigma(L_{\vec{u}})$ is projection onto the $\tfrac{i\epsilon}{2}$--eigenspace.
\end{fact}

\begin{proof}
First we compute $\sigma(L_{\vec{u}})^2=-\tfrac{1}{4}{\mathcal I}$.  Indeed,
\begin{align*}
  \sigma(L_{\vec{u}})^2
  &=\tfrac{1}{4}(u^z\gamma_{xy}-u^y\gamma_{xz}+u^x\gamma_{yz})^2\\
  &=\tfrac{1}{4}[u_z^2\gamma_{xy}^2-u^zu^y(\gamma_{xy}\gamma_{xz}+\gamma_{xz}\gamma_{xy})
      +u^zu^x(\gamma_{xy}\gamma_{yz}+\gamma_{yz}\gamma_{xy})\\
  &\quad\quad
      +u_y^2\gamma_{xz}^2+u^yu^x(\gamma_{yz}\gamma_{yz}+\gamma_{yz}\gamma_{xz})
      +u_x^2\gamma_{yz}^2]\\
  &=\tfrac{1}{4}[-u_z^2\gamma_x^2\gamma_y^2+u^zu^y\gamma_x^2(\gamma_{xz}+\gamma_{zy})
      +u^zu^x\gamma_y^2(\gamma_{xz}+\gamma_{zx})\\
  &\quad\quad
      -u_y^2\gamma_x^2\gamma_z^2-u^yu^x\gamma_z^2(\gamma_{xy}+\gamma_{yx})
      -u_x^2\gamma_y^2\gamma_z^2]
   =-\tfrac{1}{4}{\mathcal I}
\end{align*}
via the Clifford relation and the fact that $\vec{u}$ has unit length.  We then have
$$\pi_{i\epsilon/2}^2
  =\tfrac{1}{4}{\mathcal I}-i\epsilon\sigma(L_{\vec{u}})-\sigma(L_{\vec{u}})^2
  =\tfrac{1}{2}{\mathcal I}-i\epsilon\sigma(L_{\vec{u}})
  =\pi_{i\epsilon/2}.
$$
Whence $\pi_{i\epsilon/2}$ is a projection.  And,
$$\sigma(L_{\vec{u}})\pi_{i\epsilon/2}
  =\tfrac{1}{2}\sigma(L_{\vec{u}})-i\epsilon\sigma(L_{\vec{u}})^2
  =\tfrac{1}{2}\sigma(L_{\vec{u}})+\tfrac{i\epsilon}{4}{\mathcal I}
  =\tfrac{i\epsilon}{2}\pi_{i\epsilon/2}
$$
so that the image of $\pi_{i\epsilon/2}$ is contained in the $\tfrac{i\epsilon}{2}$--eigenspace of $\sigma(L_{\vec{u}})$.  Moreover, $\pi_{i/2}+\pi_{-i/2}={\mathcal I}$.  Hence, the image of $\pi_{i\epsilon/2}$ is the entire eigenspace.
\end{proof}

One the other hand, we can extend $L_{\vec{u}}$ to an operator $\hat{L}_{\vec{u}}$ on extended Minkowski space.  We will call this the {\em extended helicity operator.}

\begin{fact}
$\hat{L}_{\vec{u}}$ has eigenvalues $0,\pm i$.  Furthermore, the operators $\pi_0\doteq\hat{L}_{\vec{u}}^2+{\mathcal I}$ and $\pi_{i\epsilon}\doteq-\tfrac{1}{2}(\hat{L}_{\vec{u}}^2+i\epsilon\hat{L}_{\vec{u}})$ are projections onto the the $0$--eigenspace and $i\epsilon$--eigenspaces for $\epsilon=\pm 1$.
\end{fact}

\begin{proof}
We claim that ($\ast$) $\hat{L}_{\vec{u}}^3=-\hat{L}_{\vec{u}}$.  Note that $L_{\vec{u}}{\bf e}_t=0$, so it suffices to compute the effect on basis elements ${\bf e}_I$ where the multi--index $I$ only contains $x,y,z$.  E.g., $\hat{L}_{\vec{u}}{\bf e}_{txy}=(L_{\vec{u}}{\bf e}_t)\wedge{\bf e}_{xy}+{\bf e}_t\wedge(\hat{L}_{\vec{u}}{\bf e}_{xy})={\bf e}_t\wedge(\hat{L}_{\vec{u}}{\bf e}_{xy})$.  Now
\begin{equation}\label{eq:Ldeg1}
  L_{\vec{u}}{\bf e}_x=u^z{\bf e}_y-u^y{\bf e}_z,\quad
  L_{\vec{u}}{\bf e}_y=-u^z{\bf e}_x+u^x{\bf e}_z,\quad
  L_{\vec{u}}{\bf e}_z=u^y{\bf e}_x-u^x{\bf e}_y.
\end{equation}
Observe $L_{\vec{u}}^2{\bf e}_x=u^z(-u^z{\bf e}_x+u^x{\bf e}_z)-u^y(u^y{\bf e}_x-u^x{\bf e}_y)=u^x\vec{u}-(u_x^2+u_y^2+u_z^2){\bf e}_x=u^x\vec{u}-{\bf e}_x$.  Here we have identified $\vec{u}=u^x{\bf e}_x+u^y{\bf e}_y+u^z{\bf e}_z$.  And similary, one computes
$$L_{\vec{u}}^2{\bf e}_x=u^x\vec{u}-{\bf e}_x,\quad
  L_{\vec{u}}^2{\bf e}_y=u^y\vec{u}-{\bf e}_y,\quad
  L_{\vec{u}}^2{\bf e}_z=u^z\vec{u}-{\bf e}_z.
$$
Noting that $L_{\vec{u}}\vec{u}=0$, we then have $L_{\vec{u}}^3{\bf e}_\alpha=-L_{\vec{u}}{\bf e}_\alpha$.  For basis elements of degree two, we have
$$\hat{L}_{\vec{u}}{\bf e}_{xy}=\vec{u}\wedge{\bf e}_z,\quad
  \hat{L}_{\vec{u}}{\bf e}_{xz}=-\vec{u}\wedge{\bf e}_y,\quad
  \hat{L}_{\vec{u}}{\bf e}_{yz}=\vec{u}\wedge{\bf e}_x.
$$
For example, $\hat{L}_{\vec{u}}{\bf e}_{xy}=(L_{\vec{u}}{\bf e}_x)\wedge{\bf e}_y+{\bf e}_x\wedge(L_{\vec{u}}{\bf e}_y)=(-u^y{\bf e}_z)\wedge{\bf e}_y+{\bf e}_x\wedge(u^x{\bf e}_z)=(u^x{\bf e}_x+u^y{\bf e}_y)\wedge{\bf e}_z=\vec{u}\wedge{\bf e}_z$.  Therefore,
$$\hat{L}_{\vec{u}}^3{\bf e}_{xy}
  =\vec{u}\wedge(L_{\vec u}^2{\bf e}_z)
  =\vec{u}\wedge(u^z\vec{u}-{\bf e}_z)
  =-\vec{u}\wedge{\bf e}_z
  =-\hat{L}_{\vec{u}}{\bf e}_{xy}.
$$
Likewise, one computes $\hat{L}_{\vec{u}}^3{\bf e}_I=-\hat{L}_{\vec{u}}{\bf e}_I$ for $I=xy,xz,yz$.  Finally, observe that $\hat{L}_{\vec{u}}{\bf e}_{xyz}=(L_{\vec{u}}{\bf e}_x)\wedge{\bf e}_{yz}=0$.  So that ($\ast$) holds trivially when applied to ${\bf e}_{xyz}$.  This exhausts all possibilities, whence $\hat{L}_{\vec{u}}^3{\bf e}_I=-\hat{L}_{\vec{u}}{\bf e}_I$ for any $I$.

Using the veracity of ($\ast$), we show that $\pi_0$ and $\pi_{i\epsilon}$ are in fact projections onto the stated eigenspaces.  Now,
\begin{align*}
  \pi_0^2
    &=(\hat{L}_{\vec{u}}^2+{\mathcal I})^2
     =\hat{L}_{\vec{u}}^4+2\hat{L}_{\vec{u}}^2+{\mathcal I}
     =-\hat{L}_{\vec{u}}^2+2\hat{L}_{\vec{u}}^2+{\mathcal I}
     =\pi_0\\
  \pi_{i\epsilon}^2
    &=[-\tfrac{1}{2}(\hat{L}_{\vec{u}}^2+i\epsilon\hat{L}_{\vec{u}})]^2
     =\tfrac{1}{4}(\hat{L}_{\vec{u}}^4+2i\epsilon\hat{L}_{\vec{u}}^3-\hat{L}_{\vec{u}}^2)
     =\tfrac{1}{4}(-\hat{L}_{\vec{u}}^2-2i\epsilon\hat{L}_{\vec{u}}-\hat{L}_{\vec{u}}^2)
     =\pi_{i\epsilon}
\end{align*}
so that $\pi_0$ and $\pi_{i\epsilon}$ are both projections.  Moreover,
\begin{align*}
  \hat{L}_{\vec u}\pi_0
    &=\hat{L}_{\vec u}(\hat{L}_{\vec{u}}^2+{\mathcal I})
     =\hat{L}_{\vec u}^3+\hat{L}_{\vec u}
     =-\hat{L}_{\vec u}+\hat{L}_{\vec u}
     =0\\
  \hat{L}_{\vec u}\pi_{i\epsilon}
    &=-\tfrac{1}{2}\hat{L}_{\vec u}(\hat{L}_{\vec{u}}^2+i\epsilon\hat{L}_{\vec{u}})
     =-\tfrac{1}{2}(\hat{L}_{\vec{u}}^3+i\epsilon\hat{L}_{\vec{u}}^2)
     =-\tfrac{1}{2}(-\hat{L}_{\vec{u}}+i\epsilon\hat{L}_{\vec{u}}^2)
     =i\epsilon\pi_{i\epsilon}
\end{align*}
Therefore the images of $\pi_0$ and $\pi_{i\epsilon}$ are contained in $0$ and $i\epsilon$--eigenspaces of $\hat{L}_{\vec{u}}$.  The images are actually equal to the stated eigenspaces, as
$$\pi_0+\pi_i+\pi_{-i}
  =(\hat{L}_{\vec{u}}^2+{\mathcal I})
   -\tfrac{1}{2}(\hat{L}_{\vec{u}}^2+i\hat{L}_{\vec{u}})
   -\tfrac{1}{2}(\hat{L}_{\vec{u}}^2-i\hat{L}_{\vec{u}})
  ={\mathcal I}.\qedhere
$$
\end{proof}

\begin{fact}\label{fact:projcommute}
The projections $\Pi_{\bf p}$, $\pi_{i/2}$, $\pi_{-i/2}$, $\pi_0$, $\pi_i$, $\pi_{-i}$ commute.  Moreover, they all commute with ${\mathcal J}$.
\end{fact}

\begin{proof}
First observe that the spin helicity operator commutes with $p^\alpha\gamma_\alpha$:
\begin{align*}
  p^\alpha[\gamma_\alpha\sigma(L_{\vec{u}})-\sigma(L_{\vec{u}})\gamma_\alpha]
  &=-\tfrac{1}{4}p^\alpha(L_{\vec{u}})^{\rho\sigma}(\gamma_\alpha\gamma_\rho\gamma_\sigma
                                         -\gamma_\rho\gamma_\sigma\gamma_\alpha)\\
  &=-\tfrac{1}{4}p^\alpha(L_{\vec{u}})^{\rho\sigma}[
       (-\gamma_\rho\gamma_\alpha-2\eta_{\alpha\rho}{\mathcal I})\gamma_\sigma
       -\gamma_\rho(-\gamma_\alpha\gamma_\sigma-2\eta_{\sigma\alpha}{\mathcal I})]\\
  &=\tfrac{1}{2}(L_{\vec{u}})^{\rho\sigma}(p_\rho\gamma_\sigma-p_\sigma\gamma_\rho)
   =-{(L_{\vec{u}})^\rho}_\sigma p^\sigma\gamma_\rho
   =-{(\Omega_{\vec{u}})^j}_kp^k\gamma_j\\
  &=-\eta^{jr}\epsilon_{srk}u^sp^k\gamma_j
   =0
\end{align*}
as $u^k=p^k/|\vec{p}|$.  It follows from this observation that $\Pi_{\bf p}$ and $\pi_{i\epsilon/2}$ commute.  Now note that $p^\alpha{(L_{\vec{u}})^j}_\alpha=p^k{(\Omega_{\vec{u}})^j}_k=(u^k/|\vec{u}|)\eta^{jq}\epsilon_{rqk}u^r=0$.  Therefore using Lemma \ref{lem:gammaext}, we have that $p^\alpha\gamma_\alpha$ and $\hat{L}_{\vec{u}}$ commute:
$$p^\alpha\gamma_\alpha\hat{L}_{\vec{u}}-\hat{L}_{\vec{u}}p^\alpha\gamma_\alpha
  =-p^\alpha{(L_{\vec{u}})^\beta}_\alpha\gamma_\beta
   =0.
$$
And it follows that $\Pi_{\bf p}$ commutes with both $\pi_0$ and $\pi_{i\epsilon}$.  The remaining projections commute courtesty of Fact \ref{fact:spinext}.  Commutation with ${\mathcal J}$ follows from Fact \ref{fact:spinext}.
\end{proof}

\begin{lemma}
$\Theta(\vec{u})$ and $\hat{L}_{\vec{u}}$ commute.
\end{lemma}

\begin{proof}
We first claim that $\hat{L}_{\vec{u}}\Theta(\vec{u}){\bf e}_\emptyset=0$.  Indeed, from equation \eqref{eq:Ldeg1},
\begin{align*}
  \hat{L}_{\vec{u}}\Theta(\vec{u}){\bf e}_\emptyset
  &=\hat{L}_{\vec{u}}(u^x{\bf e}_{yz}-u^y{\bf e}_{xz}+u^z{\bf e}_{xy})\\
  &=u^x(-u^z{\bf e}_{xz}+u^y{\bf e}_{yx})-u^y(u^z{\bf e}_{yz}-u^x{\bf e}_{xy})\\
  &\quad\quad
    +u^z(-u^y{\bf e}_{zy}+u^x{\bf e}_{xz})=0.
\end{align*}
Second, we claim that the commutator $[\hat{L},\gamma_I]$ commutes with theta matrices for any Lorentz algebra element $L$ and any multi--index $I$.  If $|I|=0$, this is trivially true.  By virtue of Lemma \ref{lem:gammaext},
\begin{align*}
  [\hat{L},\gamma_\alpha\gamma_J]
  &=\hat{L}\gamma_\alpha\gamma_J-\gamma_\alpha\gamma_J\hat{L}
   =(\gamma_\alpha\hat{L}+{L^\beta}_\alpha\gamma_\beta)\gamma_J
     -\gamma_\alpha\gamma_J\hat{L}\\
  &=\gamma_\alpha[\hat{L},\gamma_J]+{L^\beta}_\alpha\gamma_\beta\gamma_J.
\end{align*}
So by induction on the length of $I$, the commutator commutes with any theta matrix.  Finally, for any multi--index $I$ we have
\begin{align*}
  \bigl(\hat{L}_{\vec{u}}\Theta(\vec{u})-\Theta(\vec{u})\hat{L}_{\vec{u}}\bigr){\bf e}_I
  &=\bigl(\hat{L}_{\vec{u}}\Theta(\vec{u})-\Theta(\vec{u})\hat{L}_{\vec{u}}\bigr)
           \gamma_I{\bf e}_\emptyset\\
  &=\{\hat{L}_{\vec{u}}\gamma_I\Theta(\vec{u})
      -\Theta(\vec{u})[\hat{L}_{\vec{u}},\gamma_I]\}{\bf e}_\emptyset
      -\Theta(\vec{u})\gamma_I\hat{L}_{\vec{u}}{\bf e}_\emptyset\\
  &=\gamma_I\hat{L}_{\vec{u}}\Theta(\vec{u}){\bf e}_\emptyset
     +\{[\hat{L}_{\vec{u}},\gamma_I]\Theta(\vec{u})
        -\Theta(\vec{u})[\hat{L}_{\vec{u}},\gamma_I]\}{\bf e}_\emptyset
   =0
\end{align*}
(recall that $\hat{L}_{\vec{u}}{\bf e}_\emptyset=0$) by the previous two claims.
\end{proof}

\begin{fact}\label{fact:halfeigen}
$\varphi_\epsilon^+({\bf p})\doteq\tfrac{1}{\sqrt{2}}\bigl({\mathcal I}-i\epsilon\Theta(\vec{u})\bigr)\chi_0^+({\bf p})$ is a unit positive spinor that is a simultaneous eigenvector of both the spin and extended helicity operators, with respective eigenvaues $i\epsilon/2$ and $0$.  Moroever, $\varphi_\epsilon^-({\bf p})\doteq{\mathcal J}\varphi_\epsilon^+({\bf p})$ is a unit negative spinor that is also a simultaneous eigenvector with the same eigenvalues.
\end{fact}

\begin{proof}
$\varphi_\epsilon^+({\bf p})$ is a sum of the plane waves $\chi_\alpha^+({\bf p})$, so is positive.  Its extended length is
\begin{align*}
  \varphi_\epsilon^+({\bf p})^\dagger\hat\eta\varphi_\epsilon^+({\bf p})
    &=\tfrac{1}{2}\chi_0^+({\bf p})^\dagger\bigl({\mathcal I}+i\epsilon\Theta(\vec{u})^\dagger\bigr)
       \hat\eta\bigl({\mathcal I}-i\epsilon\Theta(\vec{u})\bigr)\chi_0^+({\bf p})\\
    &=\tfrac{1}{2}\chi_0^+({\bf p})^\dagger
        \hat\eta\bigl({\mathcal I}-i\epsilon\Theta(\vec{u})\bigr)^2\chi_0^+({\bf p})\\
    &=\tfrac{1}{2}\chi_0^+({\bf p})^\dagger
        \hat\eta\bigl({\mathcal I}-2i\epsilon\Theta(\vec{u})+{\mathcal I}\bigr)\chi_0^+({\bf p})
     =\chi_0^+({\bf p})^\dagger\hat\eta\chi_0^+({\bf p})
     =1
\end{align*}
by the definition and orthogonality of the $\chi_\alpha^+({\bf p})$, as well as the fact $\Theta(\vec{u})^2=-{\mathcal I}$.  Note that $\pi_{i\epsilon/2}{\bf e}_\emptyset=[\tfrac{1}{2}{\mathcal I}-i\epsilon\sigma(L_{\vec{u}})]{\bf e}_\emptyset=\tfrac{1}{2}\bigl({\mathcal I}-i\epsilon\Theta(\vec{u})\bigr){\bf e}_\emptyset$ and $\Pi_{\bf p}{\bf e}_\emptyset=\tfrac{1}{\sqrt{2}}\chi_0^+({\bf p})$.  Therefore,
\begin{align*}
  \varphi_\epsilon^+({\bf p})
  &=\tfrac{1}{\sqrt{2}}\bigl({\mathcal I}-i\epsilon\Theta(\vec{u})\bigr)\chi_0^+({\bf p})
   =\bigl({\mathcal I}-i\epsilon\Theta(\vec{u})\bigr)\Pi_{\bf p}{\bf e}_\emptyset\\
  &=\Pi_{\bf p}\bigl({\mathcal I}-i\epsilon\Theta(\vec{u})\bigr){\bf e}_\emptyset
   =2\Pi_{\bf p}\pi_{i\epsilon/2}{\bf e}_\emptyset.
\end{align*}
Moreover, $\pi_0{\bf e}_\emptyset=(\hat{L}_{\vec{u}}^2+{\mathcal I}){\bf e}_\emptyset={\bf e}_\emptyset$.  Hence $\varphi_\epsilon^+({\bf p})=2\Pi_{\bf p}\pi_{i\epsilon/2}\pi_0{\bf e}_\emptyset$.  It follows that $\varphi_\epsilon^+({\bf p})$ is in the $i\epsilon/2$--eigenspace of the spin helicity operator, and in the $0$--eigenspace of the extended helicity operator.  As ${\mathcal J}$ commutes with the projection operators, the analogous statements for $\varphi_\epsilon^-({\bf p})={\mathcal J}\varphi_\epsilon^+({\bf p})$ follow as well.
\end{proof}

\begin{fact}
Let $\vec{v}=(v^x,v^y,v^z)$ be a complex vector such that $\vec{u}\times\vec{v}=i\vec{v}$ and $\vec{v}^\ast\cdot\vec{v}=1$.  Then $\psi_\epsilon^+({\bf p})$, where $\psi_+^+({\bf p})\doteq\Theta(\vec{v})\chi_0^+({\bf p})$ and $\psi_-^+({\bf p})\doteq\Theta(\vec{v}^\ast)\chi_0^+({\bf p})$, is a unit positive spinor that is a simultaneous eigenvector of the both the spin and extended helicity operators, with respective eigenvalues $i\epsilon/2$ and $i\epsilon$.  Moreover, $\psi_\epsilon^-({\bf p})\doteq{\mathcal J}\psi_\epsilon^+({\bf p})$ is a unit negative spinor that is also a simultaneous eigenvector with the same eigenvalues.
\end{fact}

\begin{proof}
Note that $\vec{u}\cdot\vec{v}=0=\vec{v}\cdot\vec{v}$.  Using the multiplication table for theta matrices, we compute
\begin{align*}
  \Theta(\vec{u})\Theta(\vec{v})
  &=(u^x\Theta_{yz}-u^y\Theta_{xz}+u^z\Theta_{xy})
    (v^x\Theta_{yz}-v^y\Theta_{xz}+v^z\Theta_{xy})\\
  &=-(u^xv^x+u^yv^y+u^zv^z){\mathcal I}
    +(u^xv^y-u^yv^x)\Theta_{xy}\\
  &\quad\quad\quad\quad
    -(u^zv^x-u^xv^z)\Theta_{xz}
    +(u^yv^z-u^zv^y)\Theta_{yz}\\
  &=-(\vec{u}\cdot\vec{v}){\mathcal I}
    -\Theta_{\vec{u}\times\vec{v}}.
\end{align*}
So $\Theta(\vec{u})\Theta(\vec{v})=-i\Theta(\vec{v})$ and $\Theta(\vec{v})\Theta(\vec{u})=i\Theta(\vec{v})$.  Observe that if we replace $\vec{u}$ with the complex conjugate $\vec{v}^\ast$ of $\vec{v}$ in the above computation, then the diagonal elements of $\Theta(\vec{v}^\ast)\Theta(\vec{v})$ are all equal to $-\vec{v}^\ast\cdot\vec{v}=-1$.

We only need to verify the statements for $\psi_+^+({\bf p})$.  The computations for $\psi_-^+({\bf p})$ are the same, but with $\vec{v}$ replaced with its complex conjugate $\vec{v}^\ast$.  Evidently, $\psi_+^+({\bf p})$ is a positive spinor.  And by the previous observation, its length is
$$\psi_+^+({\bf p})^\dagger\hat\eta\psi_+^+({\bf p})
   =\chi_0^+({\bf p})^\dagger\Theta(\vec{v}^\ast)^\dagger
    \hat\eta\Theta(\vec{v})\chi_0^+({\bf p})
   =-\chi_0^+({\bf p})^\dagger\hat\eta
       \Theta(\vec{v}^\ast)\Theta(\vec{v})\chi_0^+({\bf p})
   =1.
$$ 

Now we show that ($\ast$) $\Theta(\vec{v}){\bf e}_\emptyset=v^x{\bf e}_{yz}-v^y{\bf e}_{xz}+v^z{\bf e}_{xy}$ is an eigenvector of $\sigma(L_{\vec{u}})$ with eigenvalue $i/2$.  Indeed, note that $\sigma(L_{\vec{u}}){\bf e}_\emptyset=\tfrac{1}{2}\Theta(\vec{u}){\bf e}_\emptyset$.  Thus
$$\sigma(L_{\vec{u}})\Theta(\vec{v}){\bf e}_\emptyset
  =\Theta(\vec{v})\sigma(L_{\vec{u}}){\bf e}_\emptyset
  =\tfrac{1}{2}\Theta(\vec{v})\Theta(\vec{u}){\bf e}_\emptyset
  =\tfrac{i}{2}\Theta(\vec{v}){\bf e}_\emptyset.
$$

Next we show that ($\ast\ast$) $\Theta(\vec{v}){\bf e}_\emptyset$ is also an eigenvector of $\hat{L}_{\vec{u}}$ with eigenvalue $i$.  Using equation \eqref{eq:Ldeg1},
\begin{align*}
  \hat{L}_{\vec{u}}\Theta(\vec{v}){\bf e}_\emptyset
  &=\hat{L}_{\vec{u}}(v^x{\bf e}_{yz}-v^y{\bf e}_{xz}+v^z{\bf e}_{xy})\\
  &=v^x(-u^z{\bf e}_{xz}+u^y{\bf e}_{yx})-v^y(u^z{\bf e}_{yz}-u^x{\bf e}_{xy})
     +v^z(-u^y{\bf e}_{zy}+u^x{\bf e}_{xz})\\
  &=(u^yv^z-u^zv^y){\bf e}_{yz}+(u^xv^z-u^zv^x){\bf e}_{xz}+(u^xv^y-u^yv^x){\bf e}_{xy}\\
  &=(\vec{u}\times\vec{v})^x{\bf e}_{yz}-(\vec{u}\times\vec{v})^y{\bf e}_{xz}
    +(\vec{u}\times\vec{v})^z{\bf e}_{xy}\\
  &=iv^x{\bf e}_{yz}-iv^y{\bf e}_{xz}+iv^z{\bf e}_{xy}
   =i\Theta(\vec{v}){\bf e}_\emptyset.
\end{align*}

Finally, we show that $\psi_+^+({\bf p})$ is in the image of $\Pi_{\bf p}\pi_{i/2}\pi_i$.  Indeed, by ($\ast$), we have
$$\pi_{i/2}\Theta(\vec{v}){\bf e}_\emptyset
  =[\tfrac{1}{2}{\mathcal I}-i\sigma(L_{\vec{u}})]\Theta(\vec{v}){\bf e}_\emptyset
  =[\tfrac{1}{2}-i(\tfrac{i}{2})]\Theta(\vec{v}){\bf e}_\emptyset
  =\Theta(\vec{v}){\bf e}_\emptyset.
$$
And by ($\ast\ast$),
$$\pi_i\Theta(\vec{v}){\bf e}_\emptyset
  =-\tfrac{1}{2}(\hat{L}_{\vec{u}}^2+i\hat{L}_{\vec{u}})\Theta(\vec{v}){\bf e}_\emptyset
  =-\tfrac{1}{2}[(i)^2+i(i)]\Theta(\vec{v}){\bf e}_\emptyset
  =\Theta(\vec{v}){\bf e}_\emptyset.
$$
Consequently,
\begin{align*}
  \psi_+^+({\bf p})
  &=\Theta(\vec{v})\chi_0^+({\bf p})
   =\sqrt{2}\,\Theta(\vec{v})\Pi_{\bf p}{\bf e}_\emptyset
   =\sqrt{2}\,\Pi_{\bf p}\Theta(\vec{v}){\bf e}_\emptyset\\
  &=\sqrt{2}\,\Pi_{\bf p}\pi_{i/2}\pi_i\Theta(\vec{v}){\bf e}_\emptyset.\qedhere
\end{align*}
\end{proof}

%%%%%%%%%%%%%%%%%%%%%%%%%%%%%%%%%%%%%%%%%%%%%%%%%%%%%%%%%%%%%%%%
\subsection{Projections}

We apply the projection $\pi_z$ to the spinors in the previous section.  Recall that
$$\pi_z=\pi_{(0,0,1)}^+
       =\tfrac{1}{4}({\mathcal I}-i{\mathcal J})({\mathcal I}-{\mathcal J}\Theta_{xy})
       =\tfrac{1}{4}({\mathcal I}-i{\mathcal J}+\Theta_{tz}-i\Theta_{xy}).
$$

\begin{lemma}\label{lem:projzchip}
$\pi_z\Theta_{xz}=-i\pi_z\Theta_{yz}$ and $\pi_z\Theta_{xy}=i\pi_z$.
\end{lemma}

\begin{proof}
Using the theta matrix multiplication table, we compute
\begin{align*}
  \pi_z\Theta_{xz}
    &=\tfrac{1}{4}({\mathcal I}-i{\mathcal J}+\Theta_{tz}-i\Theta_{xy})\Theta_{xz}
     =\tfrac{1}{2}(\Theta_{xz}-i\Theta_{ty}+\Theta_{tx}-i\Theta_{yz})\\
  \pi_z\Theta_{yz}
    &=\tfrac{1}{4}({\mathcal I}-i{\mathcal J}+\Theta_{tz}-i\Theta_{xy})\Theta_{yz}
     =\tfrac{1}{4}(\Theta_{yz}+i\Theta_{tx}+\Theta_{ty}+i\Theta_{xz})\\
  \pi_z\Theta_{xy}
    &=\tfrac{1}{4}({\mathcal I}-i{\mathcal J}+\Theta_{tz}-i\Theta_{xy})\Theta_{xy}
     =\tfrac{1}{4}(\Theta_{xy}+i\Theta_{tz}+{\mathcal J}+i{\mathcal I}).
  \qedhere
\end{align*}
\end{proof}

\begin{fact}
$\pi_z\chi_2^+({\bf p})=-i\pi_z\chi_1^+({\bf p})$ and $\pi_z\chi_3^+({\bf p})=i\pi_z\chi_0^+({\bf p})$.
\end{fact}

\begin{proof}
Apply lemma \ref{lem:projzchip} to $\chi_2^+({\bf p})=\Theta_{xz}\chi_0^+({\bf p})$ and $\chi_3^+({\bf p})=\Theta_{xy}\chi_0^+({\bf p})$.
\end{proof}

\begin{fact}\label{fact:halfeigenexpl}
Define $\zeta\doteq u^z{\bf e}_{xy}-u^y{\bf e}_{xz}+u^x{\bf e}_{yz}$.  Then for ${\bf p}=(p^\alpha)=(E,\vec{p})$,
\begin{align*}
  \varphi_\epsilon ^+({\bf p})
    &=\tfrac{1}{2m}(m{\bf e}_\emptyset+i\epsilon m\zeta-ip^\alpha{\bf e}_\alpha
                    +\epsilon E{\bf e}_t\wedge\zeta+\epsilon|\vec{p}|{\bf e}_{xyz}).
\end{align*}
\end{fact}

\begin{proof}
Note that $\Theta(\vec{u}){\bf e}_\emptyset=u^x{\bf e}_{yz}-u^y{\bf e}_{xz}+u^x{\bf e}_{xy}=\zeta$.  Then $\Theta(\vec{u}){\bf e}_t=\gamma_t\Theta(\vec{u}){\bf e}_\emptyset=\gamma_t\zeta={\bf e}_t\wedge\zeta-\iota_t\zeta={\bf e}_t\wedge\zeta$.  Moreover,
\begin{align*}
  p^k\Theta(\vec{u}){\bf e}_k
  &=p^k\gamma_k\zeta
   =p^k\gamma_k(u^x{\bf e}_{yz}-u^y{\bf e}_{xz}+u^z{\bf e}_{xy})\\
  &=p^k[u^x({\bf e}_{kyz}-\eta_{ky}{\bf e}_z+\eta_{kz}{\bf e}_y)
        -u^y({\bf e}_{kxz}-\eta_{kx}{\bf e}_z+\eta_{kz}{\bf e}_x)\\
  &\quad\quad\quad\quad
        +u^z({\bf e}_{kxy}-\eta_{kx}{\bf e}_y+\eta_{ky}{\bf e}_x)]\\
  &=u^x(p^x{\bf e}_{xyz}-p_y{\bf e}_z+p_z{\bf e}_y)
    -u^y(p^y{\bf e}_{yxz}-p_x{\bf e}_z+p_z{\bf e}_x)\\
  &\quad\quad\quad\quad
    +u^z(p^z{\bf e}_{zxy}-p_x{\bf e}_y+p_y{\bf e}_x)\\
  &=(u^xp^x+u^yp^y+u^zp^z){\bf e}_{xyz}
   =|\vec{p}|{\bf e}_{xyz}
\end{align*}
because $p^k=|\vec{p}|u^k$ and $u_k=-u^k$.  The formulas for $\varphi_\epsilon^+({\bf p})$ now follow from Fact \ref{fact:halfeigen}.
\end{proof}

\begin{fact}\label{fact:projhalfeigen}
Up to complex scalar multiple, the projection $\pi_z\varphi_\epsilon^+({\bf p})$ is given by
\begin{align*}
  \pi_z\varphi_\epsilon^+({\bf p})
  &=(E+m+\epsilon|\vec{p}|)(|\vec{p}|-\epsilon p^z){\bf f}_0
    -\epsilon(E+m+\epsilon|\vec{p}|)(p^x+ip^y){\bf f}_1\\
  &\quad\quad
   -(E-m+\epsilon|\vec{p}|)(|\vec{p}|-\epsilon p^z){\bf f}_2
   +\epsilon(E-m+\epsilon|\vec{p}|)(p^x+ip^y){\bf f}_3.
\end{align*}
\end{fact}

\begin{proof}
We use Fact \ref{fact:halfeigenexpl} and the orthogonality relation $({\bf f}_\alpha)^\dagger{\bf f}_\beta=\delta_{\alpha\beta}$ of the basis for the image of $\pi_z$ given in equation \eqref{eq:summandbasis} to compute.  Set $\kappa\doteq\tfrac{1}{4\sqrt{2}m}$.
\begin{align*}
  ({\bf f}_0)^\dagger\varphi_\epsilon^+({\bf p})
  &=\kappa(m-\epsilon mu^z+E-p^z-\epsilon Eu^z+\epsilon|\vec{p}|)\\
  &=\kappa(E+m+\epsilon|\vec{p}|)(1-\epsilon u^z)\\
  ({\bf f}_1)^\dagger\varphi_\epsilon^+({\bf p})
  &=\kappa(-i\epsilon mu^y-\epsilon mu^x-p^x-ip^y-i\epsilon Eu^y-\epsilon Eu^x)\\
  &=-\kappa\epsilon(E+m+\epsilon|\vec{p}|)(u^x+iu^y)\\
  ({\bf f}_2)^\dagger\varphi_\epsilon^+({\bf p})
  &=\kappa(m-\epsilon mu^z-E+p^z+\epsilon Eu^z-\epsilon|\vec{p}|)\\
  &=-\kappa(E-m+\epsilon|\vec{p}|)(1-\epsilon u^z)\\
  ({\bf f}_3)^\dagger\varphi_\epsilon^+({\bf p})
  &=\kappa(-i\epsilon mu^y-\epsilon mu^x+p^x+ip^y+i\epsilon Eu^y+\epsilon Eu^x)\\
  &=\kappa\epsilon(E-m+\epsilon|\vec{p}|)(u^x+iu^y).
\end{align*}
The formula stated in the fact then follows, provided we take the scalar multiple to be $\kappa/|\vec{p}|$.
\end{proof}

\begin{fact}\label{fact:projhalfeigennorm}
$[\pi_z\varphi_\epsilon^+({\bf p})]^\dagger\pi_z\varphi_\epsilon^+({\bf p})=8E|\vec{p}|(E+\epsilon|\vec{p}|)(|\vec{p}|-\epsilon p^z)$.
\end{fact}

\begin{proof}
Compute using Fact \ref{fact:halfeigen}:
\begin{align*}
  [\pi_z\varphi_\epsilon^+({\bf p})]^\dagger\pi_z\varphi_\epsilon^+({\bf p})
  &=(E+m+\epsilon|\vec{p}|)^2(|\vec{p}|-\epsilon p^z)^2
    +(E+m+\epsilon|\vec{p}|)^2(p_x^2+p_y^2)\\
  &
    +(E-m+\epsilon|\vec{p}|)^2(|\vec{p}|-\epsilon p^z)^2
    +(E-m+\epsilon|\vec{p}|)^2(p_x^2+p_y^2)\\
  &=2|\vec{p}|(|\vec{p}|-\epsilon p^z)
    [(E+m+\epsilon|\vec{p}|)^2+(E-m+\epsilon|\vec{p}|)^2]\\
  &=4|\vec{p}|(|\vec{p}|-\epsilon p^z)
    [(E+\epsilon|\vec{p}|)^2+m^2]\\
  &=4|\vec{p}|(|\vec{p}|-\epsilon p^z)
    [(E+\epsilon|\vec{p}|)^2+E^2-|\vec{p}|^2]\\
  &=4|\vec{p}|(|\vec{p}|-\epsilon p^z)(E+\epsilon|\vec{p}|)(2E).\qedhere
\end{align*}
\end{proof}

Observe that $\pi_z\varphi_0^+({\bf p})=0$ if $\vec{p}=(0,0,p^z)$ with $p^z>0$.  This is a consequence of the choice of projection $\pi_z$.  We can resolve this pathology without choosing a different projection (and basis for the image of the projection), by multiplying by an appropriate phase factor and normalizing.

\begin{lemma}
$p^x+ip^y=e^{i\theta}\sqrt{p_x^2+p_y^2}$ for some angle $\theta$.  And for such an angle, we have $\tfrac{p^x+ip^y}{\sqrt{|\vec{p}|-\epsilon p^z}}=e^{i\theta}\sqrt{|\vec{p}|+\epsilon p^z}$.
\end{lemma}

\begin{proof}
The existence of $\theta$ follows from the fact that $|p^x+ip^y|=\sqrt{p_x^2+p_y^2}$.  Now, $|\vec{p}|^2=p_x^2+p_y^2+p_z^2$ implies that $p_x^2+p_y^2=(|\vec{p}|+\epsilon p^z)(|\vec{p}|-\epsilon p^z)$.  Thus
$$\tfrac{p^x+ip^y}{\sqrt{|\vec{p}|-\epsilon p^z}}
  =\tfrac{e^{i\theta}\sqrt{p_x^2+p_y^2}}{\sqrt{|\vec{p}|-\epsilon p^z}}
  =\tfrac{e^{i\theta}\sqrt{(|\vec{p}|+\epsilon p^z)(|\vec{p}|-\epsilon p^z)}}
         {\sqrt{|\vec{p}|-\epsilon p^z}}
  =e^{i\theta}\sqrt{|\vec{p}|+\epsilon p^z}.\qedhere
$$
\end{proof}

\begin{fact}\label{fact:projhalfeigennormed}
Let $\tilde\varphi_\epsilon({\bf p})$ denote $\pi_z\varphi_\epsilon^+({\bf p})$ normalized to $\sqrt{2E}$.  That is, such that $\tilde\varphi_\epsilon({\bf p})^\dagger\tilde\varphi_\epsilon({\bf p})=2E$.  Then
\begin{align*}
  \tilde\varphi_\epsilon({\bf p})
  &=\tfrac{(E+\epsilon|\vec{p}|+m)\sqrt{|\vec{p}|-\epsilon p^z}}
          {2\sqrt{|\vec{p}|(E+\epsilon|\vec{p}|)}}\,{\bf f}_0
     -\tfrac{\epsilon e^{i\theta}(E+\epsilon|\vec{p}|+m)\sqrt{|\vec{p}+\epsilon p^z}}
            {2\sqrt{|\vec{p}|(E+\epsilon|\vec{p}|)}}\,{\bf f}_1\\
  &\quad\quad
     -\tfrac{(E+\epsilon|\vec{p}|-m)\sqrt{|\vec{p}|-\epsilon p^z}}
          {2\sqrt{|\vec{p}|(E+\epsilon|\vec{p}|)}}\,{\bf f}_2
     +\tfrac{\epsilon e^{i\theta}(E+\epsilon|\vec{p}|-m)\sqrt{|\vec{p}+\epsilon p^z}}
            {2\sqrt{|\vec{p}|(E+\epsilon|\vec{p}|)}}\,{\bf f}_3
\end{align*}
where $\theta$ is such that $\tfrac{p^x+ip^y}{\sqrt{|\vec{p}|-\epsilon p^z}}=e^{i\theta}\sqrt{|\vec{p}|+\epsilon p^z}$.
\end{fact}

\begin{proof}
This follows from Facts \ref{fact:projhalfeigen} and \ref{fact:projhalfeigennorm}, and the previous lemma.
\end{proof}

We remark that we can rewrite the quotients $\tfrac{E+\epsilon|\vec{p}|\pm m}{\sqrt{E+\epsilon|\vec{p}|}}$ in Fact \ref{fact:projhalfeigennormed} using the following lemma.

\begin{lemma}
$\tfrac{E+\epsilon|\vec{p}|\pm m}{\sqrt{E+\epsilon|\vec{p}|}}
   =\pm\bigl(\sqrt{E+|\vec{p}|}-\sqrt{E-|\vec{p}|}\bigr)$.
\end{lemma}

\begin{proof}
As $m^2=E^2-|\vec{p}|^2$, we have $m=\sqrt{(E+\epsilon|\vec{p}|)(E-\epsilon|\vec{p}|)}$.  Hence
\begin{align*}
  \tfrac{E+\epsilon|\vec{p}|\pm m}
        {\sqrt{E+\epsilon|\vec{p}|}}
  &=\tfrac{E+\epsilon|\vec{p}|\pm\sqrt{(E+\epsilon|\vec{p}|)(E-\epsilon|\vec{p}|)}}
          {\sqrt{E+\epsilon|\vec{p}|}}
   =\sqrt{E+\epsilon|\vec{p}|}\pm\sqrt{E-\epsilon|\vec{p}|}.
\end{align*}
And the latter is equal to $\pm\bigl(\sqrt{E+|\vec{p}|}-\sqrt{E-|\vec{p}|}\bigr)$ for $\epsilon=\pm 1$.
\end{proof}

%%%%%%%%%%%%%%%%%%%%%%%%%%%%%%%%%%%%%%%%%%%%%%%%%%%%%%%%%%%%%%%%%
\section{Circular beams}

Using cylindrical coordinates $(x^\alpha)=(t,s,\xi,z)$ for ${\mathbb M}$, where $x=s\cos\xi$ and $y=s\sin\xi$, the Minkowski metric is
$$\eta={\it diag}(1,-1,-s^2,-1).$$

\begin{fact}\label{fact:circrect}
${\bf e}_s=\tfrac{x}{s}{\bf e}_x+\tfrac{y}{s}{\bf e}_y$ and ${\bf e}_\xi=-y{\bf e}_x+x{\bf e}_y$.
\end{fact}

\begin{proof}
For any function $f=f(x,y,z)$, we have
$$\partial_sf=\tfrac{\partial f}{\partial x}\tfrac{\partial x}{\partial s}
              +\tfrac{\partial f}{\partial y}\tfrac{\partial y}{\partial s}
              +\tfrac{\partial f}{\partial z}\tfrac{\partial z}{\partial s}
             =\cos\xi\,\partial_xf+\sin\xi\,\partial_yf
             =\tfrac{x}{s}\partial_xf+\tfrac{y}{s}\partial_yf.
$$
Because we are identifying ${\bf e}_\alpha$ with $\partial_\alpha$, the first identity follows.  Similarly,
$$\partial_\xi f=\tfrac{\partial f}{\partial x}\tfrac{\partial x}{\partial\xi}
              +\tfrac{\partial f}{\partial y}\tfrac{\partial y}{\partial\xi}
              +\tfrac{\partial f}{\partial z}\tfrac{\partial z}{\partial\xi}
             =-s\sin\xi\,\partial_xf+s\cos\xi\,\partial_yf
             =-y\partial_xf+x\partial_yf.\qedhere
$$
\end{proof}

%::::::::::::::::::::::::::::::::::::::::::::::::::::::::::::::::
\subsection{Dirac equation}

\begin{lemma}\label{lem:christoffel}
${\Gamma^s}_{\xi\xi}=-s$ and ${\Gamma^\xi}_{s\xi}={\Gamma^\xi}_{\xi s}=\tfrac{1}{s}$.  All other Christoffel symbols are trivial.
\end{lemma}

\begin{proof}
The only nontrivial metric derivative is $\eta_{\xi\xi s}=\partial_s\eta_{\xi\xi}=-2s$.  It follows that $\Gamma_{s\xi\xi}=-\tfrac{1}{2}\eta_{\xi\xi s}=s$ and $\Gamma_{\xi s\xi}=\tfrac{1}{2}\eta_{\xi\xi s}=-s$.  Hence ${\Gamma^s}_{\xi\xi}=\eta^{ss}\Gamma_{s\xi\xi}=-s$ and ${\Gamma^\xi}_{s\xi}=\eta^{\xi\xi}\Gamma_{\xi s\xi}=\tfrac{1}{s}$.
\end{proof}

\begin{fact}\label{fact:circconn}
The nontrivial effects of the extended connection on rank one basis elements are $\hat\Gamma_s{\bf e}_\xi=\tfrac{1}{s}{\bf e}_\xi$, $\hat\Gamma_\xi{\bf e}_s=\tfrac{1}{s}{\bf e}_\xi$, and $\hat\Gamma_\xi{\bf e}_\xi=-s{\bf e}_s$.
\end{fact}

\begin{proof}
On rank one elements, the extended connection is just the usual connection.  So, this just a restatement of Lemma \ref{lem:christoffel}.  E.g., $\Gamma_s{\bf e}_\xi={\Gamma^\alpha}_{s\xi}{\bf e}_\alpha={\Gamma^\xi}_{s\xi}{\bf e}_\xi=\tfrac{1}{s}{\bf e}_\xi$.
\end{proof}

The Dirac operator is $D=\gamma^\alpha\partial_\alpha+\gamma^s\hat\Gamma_s+\gamma^\xi\hat\Gamma_\xi$, as we have $\hat\Gamma_t=0$ and $\hat\Gamma_z=0$.  So if we are looking for circular beam spinors, that is spinors $\psi$ of the form
\begin{equation}\label{eq:circbeam}
  \psi=e^{-i(Et-kz)}\varphi
\end{equation}
where $\varphi=\varphi^I{\bf e}_I$ is a function of $s$ only, then the Dirac equation $D\psi=m\psi$ becomes
\begin{equation}\label{eq:scircdirac}
  \gamma^s\partial_s\varphi-iE\gamma^t\varphi+ik\gamma^z\varphi
  +\gamma^s\hat\Gamma_s\varphi+\gamma^\xi\hat\Gamma_\xi\varphi
  =m\varphi.
\end{equation}

%::::::::::::::::::::::::::::::::::::::::::::::::::::::::::::::::
\subsection{Solutions}

\begin{fact}
The square of the Dirac equation $D^2\psi=m^2\psi$ implies that
\begin{equation}\label{eq:circdirac}
  \partial_s^2\varphi
           +A\partial_s\varphi
           +B\varphi
  =m^2\varphi,
\end{equation}
where $A\doteq2\hat\Gamma_s+\tfrac{1}{s}{\mathcal I}$ and $B\doteq(E^2-k^2){\mathcal I}+\tfrac{1}{s}\hat\Gamma_s+\tfrac{1}{s^2}\hat\Gamma_\xi^2$ are diagonal operators.
\end{fact}

\begin{proof}
The general form of the square of the Dirac operator is computed in Appendix B.  Using equation \eqref{eq:circbeam}, the second order derivative term of $D^2\psi$ is
$$-\eta^{\rho\sigma}\partial_{\rho\sigma}\psi
  =-\eta^{tt}(iE)^2\psi-\eta^{ss}\partial_s^2\psi-\eta^{zz}(ik)^2\psi
  =(E^2+\partial_s^2-k^2)\psi.
$$
The first order derivative term is
$$(\eta^{\rho\sigma}{\Gamma^\alpha}_{\rho\sigma}{\mathcal I}
    -2\hat\Gamma^\alpha)\partial_\alpha\psi
  =(\eta^{\xi\xi}{\Gamma^s}_{\xi\xi}{\mathcal I}-2\eta^{ss}\hat\Gamma_s)\partial_s\psi
   =(\tfrac{1}{s}{\mathcal I}+2\hat\Gamma_s)\psi.
$$
The metric $\eta$ is flat.  That is, the curvature of the connection is trivial: $\Omega_{\alpha\beta}=0$ (in rectangular coordinates the Christoffel symbols are trivial, so the curvature matrix must be trivial in any coordinate system --- we can also verify this directly for the metric in cylindrical coordinates).  The derivative--free term of $D^2\psi$ is then
\begin{align*}
  (\eta^{\rho\sigma}{\Gamma^\beta}_{\rho\sigma}\hat\Gamma_\beta
   -\partial^\beta\hat\Gamma_\beta-\hat\Gamma^\beta\hat\Gamma_\beta)\psi
  &=(\eta^{\xi\xi}{\Gamma^s}_{\xi\xi}\hat\Gamma_s
     -\eta^{ss}\partial_s\hat\Gamma_s-\eta^{ss}\hat\Gamma_s^2
     -\eta^{\xi\xi}\hat\Gamma_\xi^2)\psi\\
  &=(\tfrac{1}{s}\hat\Gamma_s-\tfrac{1}{s}\hat\Gamma_s
     +\tfrac{1}{s}\hat\Gamma_s+\tfrac{1}{s^2}\hat\Gamma_\xi^2)\psi\\
  &=(\tfrac{1}{s}\hat\Gamma_s+\tfrac{1}{s^2}\hat\Gamma_\xi^2)\psi
\end{align*}
because $\partial_s\hat\Gamma_s=-\tfrac{1}{s}\hat\Gamma_s$ and $\hat\Gamma_s^2=\tfrac{1}{s}\hat\Gamma_s$, courtesy of Fact \ref{fact:circconn}.  That $\hat\Gamma_s$ and $\hat\Gamma_\xi^2$ are diagonal also follows from Fact \ref{fact:circconn}.
\end{proof}

\begin{fact}\label{fact:circdirac2soln}
The equation \eqref{eq:circdirac} has nonsingular solutions $\varphi=\varphi^I{\bf e}_I$ with
\begin{enumerate}
\item $\varphi^I=A^IJ_0(ps)$ for $I=\emptyset,tz,t,z$
\item $\varphi^I=A^IJ_1(ps)$ for $I=ts,sz,s,tsz$
\item $\varphi^I=A^I\tfrac{J_1(ps)}{ps}$ for $I=t\xi,\xi z,\xi,t\xi z$
\item $\varphi^I=0$ for $I=s\xi,ts\xi z,ts\xi,s\xi z$,
\end{enumerate}
where $p\doteq\sqrt{E^2-k^2-m^2}$, $A^I$ are constants, and $J_0$ and $J_1$ are the zeroth and first order Bessel functions of the first kind.
\end{fact}

\begin{proof}
For $I=\emptyset,tz,t,z$, equation \eqref{eq:circdirac} restricts to $\partial_s^2\varphi^I+\tfrac{1}{s}\partial_s\varphi^I+p^2\varphi^I=0$, for which $J_0(ps)$ is the only nonsingular solution.

If $I=ts,sz,s,tsz$, we have $\hat\Gamma_s{\bf e}_I=0$ and $\hat\Gamma_\xi^2{\bf e}_I=-{\bf e}_I$.  So that equation \eqref{eq:circdirac} restricts to $\partial_s^2\varphi^I+\tfrac{1}{s}\partial_s\varphi^I+(p^2-\tfrac{1}{s^2})\varphi^I=0$, whose only nonsingular solution is $J_1(ps)$.

For $I=t\xi,\xi z,\xi,t\xi z$, $\hat\Gamma_s{\bf e}_I=\tfrac{1}{s}{\bf e}_I$ and $\hat\Gamma_\xi^2{\bf e}_I=-{\bf e}_I$.  And equation \eqref{eq:circdirac} becomes $\partial_s^2\varphi^I+\tfrac{3}{s}\partial_s\varphi^I+p^2\varphi^I=0$.  If we write $\varphi^I=\tfrac{1}{ps}f$, then
\begin{align*}
  0&=\partial_s^2\varphi^I+\tfrac{3}{s}\partial_s\varphi^I+p^2\varphi^I\\
   &=(\tfrac{1}{ps}\partial_s^2f-\tfrac{2}{ps^2}\partial_sf+\tfrac{2}{ps^3}f)
     +\tfrac{3}{s}(\tfrac{1}{ps}\partial_sf-\tfrac{1}{ps^2}f)+\tfrac{p}{s}f\\
   &=\tfrac{1}{ps}[\partial_s^2f+\tfrac{1}{s}\partial_sf+(p^2-\tfrac{1}{s^2})f].
\end{align*}
I.e., $\partial_s^2f+\tfrac{1}{s}\partial_sf+(p^2-\tfrac{1}{s^2})f=0$, whose only nonsingular solution is $f=J_1(ps)$.  So that $\varphi^I=\tfrac{1}{ps}J_1(ps)$.

Finally for $I=s\xi,ts\xi z,ts\xi,s\xi z$, $\hat\Gamma_s{\bf e}_I=\tfrac{1}{s}{\bf e}_I$.  On the other hand, $\hat\Gamma_\xi{\bf e}_{s\xi}=(\tfrac{1}{s}{\bf e}_\xi)\wedge{\bf e}_\xi+{\bf e}_s\wedge(-s{\bf e}_s)=0$.  Hence $\hat\Gamma_\xi^2{\bf e}_I=0$.  Equation \eqref{eq:circdirac} then reduces to $\partial_s^2\varphi^I+\tfrac{3}{s}\partial_s\varphi^I+(p^2+\tfrac{1}{s^2})\varphi^I=0$.  Again writing $\varphi^I=\tfrac{1}{ps}f$ and calculating as above, we get $\partial_s^2f+\tfrac{1}{s}\partial_sf+p^2f=0$.  And the only nonsingular solution is $f=J_0(ps)$.  However $J_0(0)=1$, so that $\varphi^I=\tfrac{1}{ps}J_0(ps)$ is nonsingular.
\end{proof}

\begin{fact}\label{fact:tz-comm}
$\Theta_t$ and $\Theta_z$ commute with the Dirac operator.
\end{fact}

\begin{proof}
The metric component $\eta_{\xi\xi}=-s^2$ is the only component that is not constant.  So the only gamma matrix that is not constant is $\gamma_\xi$.  And only theta matrices $\Theta_I$ with $\xi\in I$ can be non--constant.  So we only need to show that $\Theta_t$ and $\Theta_z$ commute with $\hat\Gamma_s$ and $\hat\Gamma_\xi$.

Suppose that $I$ is a multi--index with $\xi\not\in I$.  Then $\Theta_t{\bf e}_I=\gamma_I{\bf e}_t=k{\bf e}_K$, with $\xi\not\in K$ ($k$ a constant).  So $\hat\Gamma_s\Theta_t{\bf e}_I=0=\Theta_t\hat\Gamma_s{\bf e}_I$.  Moreover, $\Theta_t({\bf e}_\xi\wedge{\bf e}_I)=\gamma_\xi\gamma_I{\bf e}_t=k{\bf e}_\xi\wedge{\bf e}_K$.  So we have $\hat\Gamma_s\Theta_t({\bf e}_\xi\wedge{\bf e}_I)=\tfrac{k}{s}{\bf e}_\xi\wedge{\bf e}_K=\Theta_t\hat\Gamma_s({\bf e}_\xi\wedge{\bf e}_I)$.  Hence $\Theta_t$ commutes with $\hat\Gamma_s$.

Now suppose that $I$ contains neither $s$ nor $\xi$.  In this case $\Theta_t{\bf e}_I=k{\bf e}_K$ with $s,\xi\not\in K$.  Therefore $\hat\Gamma_\xi\Theta_t{\bf e}_I=0=\Theta_t\hat\Gamma_\xi{\bf e}_I$.  And $\hat\Gamma_\xi\Theta_t({\bf e}_s\wedge{\bf e}_I)=\tfrac{k}{s}{\bf e}_\xi\wedge{\bf e}_I=\Theta_t\hat\Gamma_\xi({\bf e}_s\wedge{\bf e}_I)$.  Furthermore, $\hat\Gamma_\xi\Theta_t({\bf e}_\xi\wedge{\bf e}_I)=-ks{\bf e}_s\wedge{\bf e}_I=\Theta_t\hat\Gamma_\xi({\bf e}_\xi\wedge{\bf e}_I)$.  On the other hand, $\hat\Gamma_\xi{\bf e}_{s\xi}=(\tfrac{1}{s}{\bf e}_\xi)\wedge{\bf e}_\xi+{\bf e}_s\wedge(-s{\bf e}_s)=0$.  Thus $\hat\Gamma_\xi\Theta_t({\bf e}_{s\xi}\wedge{\bf e}_I)=k\hat\Gamma_\xi\gamma_{s\xi}{\bf e}_K=k\hat\Gamma_\xi({\bf e}_{s\xi}\wedge{\bf e}_I)=0=\Theta_t\hat\Gamma_\xi({\bf e}_{s\xi}\wedge{\bf e}_I)$.  Consequently, $\Theta_t$ commutes with $\hat\Gamma_\xi$.

The same line of reasoning shows that $\Theta_z$ commutes with $\hat\Gamma_s$ and $\hat\Gamma_\xi$.
\end{proof}

\begin{fact}
Let $D_s$ be the operator on the left hand side of equation \eqref{eq:scircdirac}.  The restriction of the operator $\pi_c\doteq\tfrac{1}{2m}(m{\mathcal I}+D_s)$ to the solution space of equation \eqref{eq:circdirac} is projection onto the solution space of equation \eqref{eq:scircdirac}.
\end{fact}

\begin{proof}
Suppose that $\varphi$ solves equation \eqref{eq:circdirac}, so that $D_s^2\varphi=m^2\varphi$.  Then we have:
\begin{align*}
  \pi_c^2\varphi
  &=\tfrac{1}{4m^2}(m^2{\mathcal I}+2mD_s+D_s^2)\varphi
   =\tfrac{1}{4m^2}(m^2\varphi+2mD_s\varphi+D_s^2\varphi)\\
  &=\tfrac{1}{4m^2}(m^2\varphi+2mD_s\varphi+m^2\varphi)
   =\tfrac{1}{2m}(m\varphi+D_s\varphi)
   =\pi_c\varphi.
\end{align*}
So that the restriction of $\pi_c$ is a projection operator.  Moreover if $\varphi'=\pi_c\varphi$, then $D_s\varphi'=m\varphi'$.  Indeed, we compute
\begin{align*}
  D_s\varphi'
  &=D_s\pi_c\varphi
   =D_s\tfrac{1}{2m}(m\varphi+D_s\varphi)
   =\tfrac{1}{2m}(mD_s\varphi+D_s^2\varphi)\\
  &=\tfrac{1}{2m}(mD_s\varphi+m^2\varphi)
   =m\pi_c\varphi
   =m\varphi'
\end{align*}
Finally, $D_s$ has no diagonal elements, so the trace of $\pi_c$ is 8, which is the dimension of the solution space of the extended Dirac equation.
\end{proof}

\begin{fact}\label{fact:cnonsingsol}
The space of nonsingular solutions of equation \eqref{eq:scircdirac} is spanned by
\begin{align*}
  \varphi_0(E,k)
    &\doteq\tfrac{1}{\sqrt{2}m}[mJ_0(ps){\bf e}_\emptyset-iEJ_0(ps){\bf e}_t
                                +pJ_1(ps){\bf e}_s-ikJ_0(ps){\bf e}_z]\\
  \varphi_1(E,k)
    &=\tfrac{1}{\sqrt{2}m}[mJ_1(ps){\bf e}_{sz}+ikJ_1(ps){\bf e}_s
                           -pJ_0(ps){\bf e}_z-iEJ_1(ps){\bf e}_{tsz}]\\
  \Theta_{tz}\varphi_0(E,k)
    &=\tfrac{1}{\sqrt{2}m}[mJ_0(ps){\bf e}_{tz}+ikJ_0(ps){\bf e}_t
                           +iEJ_0(ps){\bf e}_z-pJ_1(ps){\bf e}_{tsz}]\\
  \Theta_{tz}\varphi_1(E,k)
    &\doteq\tfrac{1}{\sqrt{2}m}[mJ_1(ps){\bf e}_{ts}+pJ_0(ps){\bf e}_t
                               +iEJ_1(ps){\bf e}_s-ikJ_1(ps){\bf e}_{tsz}].
\end{align*}
\end{fact}

\begin{proof}
Using the indentity $\partial_uJ_0(u)=-J_1(u)$, we readily compute:
\begin{align*}
  D_sJ_0(ps){\bf e}_\emptyset
  &=-\gamma_s[-pJ_1(ps)]{\bf e}_\emptyset
    -iJ_0(ps)(E\gamma_t+k\gamma_z){\bf e}_\emptyset\\
  &=pJ_1(ps){\bf e}_s-iEJ_0(ps){\bf e}_t-ikJ_0(ps){\bf e}_z.
\end{align*}
It follows that
$$\pi_cJ_0(ps){\bf e}_\emptyset
  =\tfrac{1}{2m}[mJ_0(ps){\bf e}_\emptyset+pJ_1(ps){\bf e}_s
                -iEJ_0(ps){\bf e}_t-ikJ_0(ps){\bf e}_z].
$$
I.e., $\pi_cJ_0(Ps){\bf e}_\emptyset=\tfrac{1}{\sqrt{2}}\varphi_0(E,k)$.  Moreover, $\partial_uJ_1(u)=J_0(u)-\tfrac{1}{u}J_1(u)$, $\hat\Gamma_s{\bf e}_{sz}=0$, and $\hat\Gamma_\xi{\bf e}_{sz}=\tfrac{1}{s}{\bf e}_{\xi z}$, so that
\begin{align*}
  D_sJ_1(ps){\bf e}_{sz}
  &=-\gamma_s[pJ_0(ps)-\tfrac{1}{s}J_1(ps)]{\bf e}_{sz}
    -iJ_1(ps)(E\gamma_t+k\gamma_z){\bf e}_{sz}
    -\tfrac{1}{s^3}J_1(ps)\gamma_\xi{\bf e}_{\xi z}\\
  &=-[pJ_0(ps)-\tfrac{1}{s}J_1(ps)]{\bf e}_z
   -iEJ_1(ps){\bf e}_{tsz}+ikJ_1(ps){\bf e}_s
   -\tfrac{1}{s}J_1(ps){\bf e}_z\\
  &=-pJ_0(ps){\bf e}_z+ikJ_1(ps){\bf e}_s-iEJ_1(ps){\bf e}_{tsz}.
\end{align*}
And we have $\pi_cJ_1(ps){\bf e}_{sz}=\tfrac{1}{\sqrt{2}}\varphi_1(E,k)$.

Let ${\mathcal S}$ denote the space of solutions to equation \eqref{eq:circdirac}.  By Fact \ref{fact:tz-comm}, $\Theta_t$ and $\Theta_z$ commute with $\pi_c$.  Whence, $\varphi_0(E,k)$, $\varphi_1(E,k)$, $\Theta_{tz}\varphi_0(E,k)$, and $\Theta_{tz}\varphi_1(E,k)$ are all in the image of $\pi_c|{\mathcal S}$.  Note that these four functions are independent.

It remains to show that any function in the image of $\pi_c|{\mathcal S}$ that is not spanned by these four functions is singular.  Compute
\begin{align*}
  D_s\tfrac{J_1(ps)}{ps}{\bf e}_\xi
  &=-\gamma_s\partial_s\tfrac{J_1(ps)}{ps}{\bf e}_\xi
    +(-iE\gamma_t-ik\gamma_z-\gamma_s\hat\Gamma_s-\tfrac{1}{s^2}\gamma_\xi\hat\Gamma_\xi)
      \tfrac{J_1(ps)}{ps}{\bf e}_\xi\\
  &=-\gamma_s[\tfrac{1}{s}J_0(ps)-\tfrac{2}{ps^2}J_1(ps)]{\bf e}_\xi
    +\tfrac{J_1(ps)}{ps}(-iE{\bf e}_{t\xi}+ik{\bf e}_{\xi z}
                         -\tfrac{1}{s}\gamma_s{\bf e}_\xi+\tfrac{1}{s}\gamma_\xi{\bf e}_s)\\
  &=-[\tfrac{1}{s}J_0(ps)-\tfrac{2}{ps^2}J_1(ps)]{\bf e}_{s\xi}
    +\tfrac{J_1(ps)}{ps}(-iE{\bf e}_{t\xi}+ik{\bf e}_{\xi z}
                         -\tfrac{1}{s}{\bf e}_{s\xi}-\tfrac{1}{s}{\bf e}_{s\xi})\\
  &=-\tfrac{1}{s}J_0(ps){\bf e}_{s\xi}
    +\tfrac{J_1(ps)}{ps}(-iE{\bf e}_{t\xi}+ik{\bf e}_{\xi z}).
\end{align*}
So that $\pi_c\tfrac{J_1(ps)}{ps}{\bf e}_\xi=-\tfrac{iE}{ps}J_1(ps){\bf e}_{t\xi}-\tfrac{1}{s}J_0(ps){\bf e}_{s\xi}+\tfrac{ik}{ps}J_1(ps){\bf e}_{\xi z}+\tfrac{m}{ps}J_1(ps){\bf e}_\xi$.  Observe that the ${\bf e}_{s\xi}$--term is singular.  Applying $\Theta_t$, $\Theta_z$, and $\Theta_{tz}$ to this singular solution of equation \eqref{eq:scircdirac} produces other independent singular solutions.  That is, $\pi_c|{\mathcal S}$ is spanned by the four non--singular solutions in the statement of the fact and these four singular solutions.
\end{proof}

\begin{fact}\label{fact:cnonsing}
The nonsingular solutions of equation \eqref{eq:scircdirac} are of the form
$$(A\,{\mathcal I}+B\Theta_t+C\Theta_z+D\Theta_{tz})\varphi_0(E,k)$$
for complex scalars $A,B,C,D$.  In particular,
\begin{equation}\label{eq:csol1id}
  \varphi_1(E,k)=\tfrac{1}{p}(ik\,{\mathcal I}+m\Theta_z+iE\Theta_{tz})\varphi_0(E,k).
\end{equation}
\end{fact}

\begin{proof}
From the previous fact, we know that $\varphi_0(E,k)$ is a nonsingular solution.  The first statement follows, provided that equation \eqref{eq:csol1id} holds.  This identity follows by routine computation:
\begin{align*}
  &(ik\,{\mathcal I}+m\Theta_z+iE\Theta_{tz})\varphi_0(E,k)\\
  &=\tfrac{1}{\sqrt{2}m}[imkJ_0(ps){\bf e}_\emptyset+kEJ_0(ps){\bf e}_t
                         +ikpJ_1(ps){\bf e}_s+k^2J_0(ps){\bf e}_z\\
  &\quad\quad\quad
                         +m^2J_0(ps){\bf e}_z-imEJ_0(ps){\bf e}_{tz}
                         +mpJ_1(ps){\bf e}_{sz}-imkJ_0(ps){\bf e}_\emptyset\\
  &\quad\quad\quad
                         +iEmJ_0(ps){\bf e}_{tz}-E^2J_0(ps){\bf e}_z
                         -iEpJ_1(ps){\bf e}_{tsz}-EkJ_1(ps){\bf e}_t]\\
  &=\tfrac{p}{\sqrt{2}m}[mJ_1(ps){\bf e}_{sz}+ikJ_0(ps){\bf e}_s
                         -pJ_0(ps){\bf e}_z-iEJ_1(ps){\bf e}_{tsz}]\\
  &=p\varphi_1(E,k).\qedhere
\end{align*}
\end{proof}

\begin{fact}
$-i\varphi_0(E,k)^\dagger\hat\eta\gamma^\beta\varphi_0(E,k)=0$ for $\beta=s,\xi$.  Moreover,
$$-i\varphi_0(E,k)^\dagger\hat\eta\gamma^t\varphi_0(E,k)
  =\tfrac{E}{m}J_0^2(ps)$$
and
$$-i\varphi_0(E,k)^\dagger\hat\eta\gamma^z\varphi_0(E,k)
  =\tfrac{k}{m}J_0^2(ps).$$
\end{fact}

\begin{proof}
Observe that because $\hat\eta$ is diagonal and $\varphi_0(E,k)$ only has nonzero components ${\bf e}_I$ with $I=\emptyset,t,s,z$, we have that
$${\bf e}_s^\dagger\hat\eta\gamma_t\varphi_0(E,k)
  =-{\bf e}_s^\dagger\gamma_t^\dagger\hat\eta\varphi_0(E,k)
  =-(\gamma_t{\bf e}_s)^\dagger\hat\eta\varphi_0(E,k)
  =-{\bf e}_{ts}^\dagger\hat\eta\varphi_0(E,k)
  =0.$$
Similarly, ${\bf e}_z^\dagger\hat\eta\gamma_t\varphi_0(E,k)=0$.  Thus,
\begin{align*}
  \varphi_0(E,k)^\dagger\hat\eta\gamma^t\varphi_0(E,k)
  &=\tfrac{J_0^2(ps)}{2m^2}(m{\bf e}_\emptyset-iE{\bf e}_t)^\dagger
           \hat\eta\gamma_t(m{\bf e}_\emptyset-iE{\bf e}_t)\\
  &=\tfrac{J_0^2(ps)}{2m^2}(m{\bf e}_\emptyset-iE{\bf e}_t)^\dagger
           \hat\eta(m{\bf e}_t+iE{\bf e}_\emptyset)\\
  &=\tfrac{J_0^2(ps)}{2m^2}(iEm+iEm)
   =\tfrac{iEJ_0^2(ps)}{m}.
\end{align*}
Likewise, ${\bf e}_t^\dagger\hat\eta\gamma_s\varphi_0(E,k)=0$ and ${\bf e}_z^\dagger\hat\eta\gamma_s\varphi_0(E,k)=0$.  And so,
\begin{align*}
  \varphi_0(E,k)^\dagger\hat\eta\gamma^s\varphi_0(E,k)
  &=-\tfrac{1}{2m^2}[mJ_0(ps){\bf e}_\emptyset+pJ_1(ps){\bf e}_s]^\dagger
           \hat\eta\gamma_s[mJ_0(ps){\bf e}_\emptyset+pJ_1(ps){\bf e}_s]\\
  &=-\tfrac{1}{2m^2}[mJ_0(ps){\bf e}_\emptyset+pJ_1(ps){\bf e}_s]^\dagger
           \hat\eta[mJ_0(ps){\bf e}_s+pJ_1(ps){\bf e}_\emptyset]\\
  &=-\tfrac{1}{2m^2}[mpJ_0(ps)J_1(ps)-mpJ_0(ps)J_1(ps)]
   =0.
\end{align*}
We also have
\begin{align*}
  \varphi_0(E,k)^\dagger\hat\eta\gamma^z\varphi_0(E,k)
  &=-\tfrac{J_0^2(ps)}{2m^2}(m{\bf e}_\emptyset-ik{\bf e}_z)^\dagger
            \hat\eta\gamma_z(m{\bf e}_\emptyset-ik{\bf e}_z)\\
  &=-\tfrac{J_0^2(ps)}{2m^2}(m{\bf e}_\emptyset-ik{\bf e}_z)^\dagger
           \hat\eta(m{\bf e}_z-ik{\bf e}_\emptyset)\\
  &=-\tfrac{J_0^2(ps)}{2m^2}(-imk-imk)
   =\tfrac{ikJ_0^2(ps)}{m}.
\end{align*}
Moreover, we see that ${\bf e}_I^\dagger\hat\eta\gamma_\xi\varphi_0(E,k)=0$ for $I=\emptyset,t,s,z$.
\end{proof}

\begin{fact}
$-i\varphi_1(E,k)^\dagger\hat\eta\gamma^\beta\varphi_1(E,k)=0$ for $\beta=s,\xi$.  Moreover,
$$-i\varphi_1(E,k)^\dagger\hat\eta\gamma^t\varphi_1(E,k)
  =\tfrac{E}{m}J_1^2(ps)$$
and
$$-i\varphi_1(E,k)^\dagger\hat\eta\gamma^z\varphi_1(E,k)
  =\tfrac{k}{m}J_1^2(ps).$$
\end{fact}

\begin{proof}
Note that $\varphi_1(E,k)$ as nonzero components ${\bf e}_I$ where $I=sz,s,z,tsz$.  It follows that ${\bf e}_I\hat\eta\gamma_\xi\varphi_1(E,k)=0$.  And we have
\begin{align*}
  \varphi_1(E,k)^\dagger\hat\eta\gamma^t\varphi_1(E,k)
  &=\tfrac{J_1^2(ps)}{2m^2}(m{\bf e}_{sz}-iE{\bf e}_{tsz})^\dagger
      \hat\eta\gamma_t(m{\bf e}_{sz}-iE{\bf e}_{tsz})\\
  &=\tfrac{J_1^2(ps)}{2m^2}(m{\bf e}_{sz}-iE{\bf e}_{tsz})^\dagger
      \hat\eta(m{\bf e}_{tsz}+iE{\bf e}_{sz})\\
  &=\tfrac{J_1^2(ps)}{2m^2}(imE+imE)
   =\tfrac{iE}{m}J_1(ps)^2,
\end{align*}
\begin{align*}
  &\varphi_1(E,k)^\dagger\hat\eta\gamma^s\varphi_1(E,k)\\
  &=-\tfrac{1}{2m^2}[mJ_1(ps){\bf e}_{sz}-pJ_0(ps){\bf e}_z]^\dagger
     \hat\eta\gamma_s[mJ_1(ps){\bf e}_{sz}-pJ_0(ps){\bf e}_z]\\
  &=-\tfrac{1}{2m^2}[mJ_1(ps){\bf e}_{sz}-pJ_0(ps){\bf e}_z]^\dagger
     \hat\eta[mJ_1(ps){\bf e}_z-pJ_0(ps){\bf e}_{sz}]\\
  &=-\tfrac{1}{2m^2}[-mpJ_1(ps)J_0(ps)+mpJ_0(ps)J_1(ps)]
   =0,
\end{align*}
\begin{align*}
  \varphi_1(E,k)^\dagger\hat\eta\gamma^z\varphi_1(E,k)
  &=-\tfrac{J_1^2(ps)}{2m^2}(m{\bf e}_{sz}+ik{\bf e}_s)^\dagger
     \hat\eta\gamma_z(m{\bf e}_{sz}+ik{\bf e}_s)\\
  &=-\tfrac{J_1^2(ps)}{2m^2}(m{\bf e}_{sz}+ik{\bf e}_s)^\dagger
     \hat\eta(-m{\bf e}_s-ik{\bf e}_{sz})\\
  &=-\tfrac{J_1^2(ps)}{2m^2}(-imk-imk)
   =\tfrac{ik}{m}J_1^2(ps).\qedhere
\end{align*}
\end{proof}

\begin{fact}
For $\varphi_2(E,k)\doteq\Theta_{tz}\varphi_0(E,k)$, we have
$$\varphi_2(E,k)^\dagger\hat\eta\gamma^\alpha\varphi_2(E,k)
  =-\varphi_0(E,k)^\dagger\hat\eta\gamma^\alpha\varphi_0(E,k).$$
And for $\varphi_3(E,k)\doteq\Theta_{tz}\varphi_1(E,k)$, we have
$$\varphi_2(E,k)^\dagger\hat\eta\gamma^\alpha\varphi_2(E,k)
  =-\varphi_1(E,k)^\dagger\hat\eta\gamma^\alpha\varphi_1(E,k).$$
\end{fact}

\begin{proof}
Both of these statements follow from: $\Theta_{tz}^\dagger\hat\eta\Theta_{tz}=-\hat\eta\Theta_{tz}^2=\hat\eta$.
\end{proof}

%::::::::::::::::::::::::::::::::::::::::::::::::::::::::::::::::
\subsection{Projections}

Recall that $\pi_z=\pi_{(0,0,1)}^+=\tfrac{1}{4}({\mathcal I}-i{\mathcal J}+\Theta_{tz}-i\Theta_{xy})$.

\begin{fact}
$\pi_z\Theta_{tz}\psi_0(E,k)=\pi_z\psi_0(E,k)$ and $\pi_z\Theta_{tz}\psi_1(E,k)=\pi_z\psi_1(E,k)$.
\end{fact}

\begin{proof}
This follows from the observation
$$\pi_z\Theta_{tz}
  =\tfrac{1}{4}({\mathcal I}-i{\mathcal J}+\Theta_{tz}-i\Theta_{xy})\Theta_{tz}
  =\tfrac{1}{4}(\Theta_{tz}-i\Theta_{xy}+{\mathcal I}-i{\mathcal J})
  =\pi_z.\qedhere
$$
\end{proof}

\begin{fact}
$\pi_z\varphi_0(E,k)=\tfrac{1}{\sqrt{2}m}A^\alpha\,{\bf f}_\alpha$, where $A^0=(E-k+m)J_0(ps)$, $A^1=-ip\,\tfrac{x+iy}{s}J_1(ps)$, $A^2=-(E-k-m)J_0(ps)$, and $A^3=-A^1$.
\end{fact}

\begin{proof}
From Fact \ref{fact:cnonsingsol} and the definition of the basis ${\bf f}_\alpha$ in equation \eqref{eq:summandbasis}, we compute in rectangular coordinates (using fact \ref{fact:circrect}):
\begin{align*}
  {\bf f}_0^\dagger\varphi_0(E,k)
    &=\tfrac{1}{\sqrt{2}m}[mJ_0(ps)+EJ_0(ps)-kJ_0(ps)]
     =\tfrac{1}{\sqrt{2}m}(E-k+m)J_0(ps)\\
  {\bf f}_1^\dagger\varphi_0(E,k)
    &=\tfrac{1}{\sqrt{2}m}[-ip\tfrac{x}{s}J_1(ps)+p\tfrac{y}{s}J_1(ps)]
     =-\tfrac{ip}{\sqrt{2}m}\tfrac{x+iy}{s}\,J_1(ps)\\
  {\bf f}_2^\dagger\varphi_0(E,k)
    &=\tfrac{1}{\sqrt{2}m}[mJ_0(ps)-EJ_0(ps)+kJ_0(ps)]
     =-\tfrac{1}{\sqrt{2}m}(E-k-m)J_0(ps)\\
  {\bf f}_3^\dagger\varphi_0(E,k)
    &=\tfrac{1}{\sqrt{2}m}[ip\tfrac{x}{s}J_1(ps)-p\tfrac{y}{s}J_1(ps)]
     =\tfrac{ip}{\sqrt{2}m}\tfrac{x+iy}{s}\,J_1(ps).\qedhere
\end{align*}
\end{proof}

\begin{fact}
$\pi_z\varphi_1(E,k)=\tfrac{1}{\sqrt{2}m}B^\alpha\,{\bf f}_\alpha$, where $B^0=ipJ_0(ps)$, $B^1=(E+k+m)\,\tfrac{x+iy}{s}J_1(ps)$, $B^2=-B^0$, and $B^3=-(E+k-m)\,\tfrac{x+iy}{s}J_1(ps)$.
\end{fact}

\begin{proof}
From Fact \ref{fact:cnonsingsol} and the definition of the basis ${\bf f}_\alpha$ in equation \eqref{eq:summandbasis}, we compute in rectangular coordinates (using fact \ref{fact:circrect}):
\begin{align*}
  {\bf f}_0^\dagger\varphi_1(E,k)
    &=\tfrac{1}{\sqrt{2}m}ipJ_0(ps)\\
  {\bf f}_1^\dagger\varphi_1(E,k)
    &=\tfrac{1}{\sqrt{2}m}[m\tfrac{x}{s}J_1(ps)+im\tfrac{y}{s}J_1(ps)+k\tfrac{x}{s}J_1(ps)\\
    &\quad\quad\quad\quad  +ik\tfrac{y}{s}J_1(ps)+E\tfrac{x}{s}J_1(ps)+iE\tfrac{y}{s}J_1(ps)]\\
    &=\tfrac{1}{\sqrt{2}m}(E+k+m)\tfrac{x+iy}{s}\,J_1(ps)\\
  {\bf f}_2^\dagger\varphi_1(E,k)
    &=\tfrac{1}{\sqrt{2}m}(-ip)J_0(ps)\\
  {\bf f}_3^\dagger\varphi_0(E,k)
    &=\tfrac{1}{\sqrt{2}m}[m\tfrac{x}{s}J_1(ps)+im\tfrac{y}{s}J_1(ps)-k\tfrac{x}{s}J_1(ps)\\
    &\quad\quad\quad\quad -ik\tfrac{y}{s}J_1(ps)-E\tfrac{x}{s}J_1(ps)-iE\tfrac{y}{s}J_1(ps)]\\
    &=-\tfrac{1}{\sqrt{2}m}(E+k-m)\tfrac{x+iy}{s}\,J_1(ps).\qedhere
\end{align*}
\end{proof}

%::::::::::::::::::::::::::::::::::::::::::::::::::::::::::::::::
\subsection{Plane wave decomposition}

For the four--vector ${\bf q}=(q^\alpha)=(q^t,q^x,q^y,q^z)$, define the distribution
$$\Psi_c(E,k,{\bf q})
  \doteq\tfrac{1}{2\pi p}\delta(q^t-E)\delta(q^z-k)\delta(q-p)
$$
where $p\doteq\sqrt{E^2-m^2-k^2}$ and $q\doteq\sqrt{q_x^2+q_y^2}$.
Recall that for any integer $n$, we have the identity
\begin{equation}\label{eq:besseln}
  J_n(u)=\tfrac{1}{2\pi}\int_{-\pi}^\pi e^{i(n\tau-u\sin\tau)}\,d\tau
\end{equation}
where $J_n(u)$ is the $n$--th order Bessel function of the first kind.

\begin{lemma}\label{lem:bessint1}
For all $\xi$, $\displaystyle\tfrac{1}{2\pi}\int_0^{2\pi}e^{ips\cos(\zeta-\xi)}\,d\zeta=J_0(ps)$.
\end{lemma}

\begin{proof}
Making the substitution $\theta\doteq\zeta-\xi-\tfrac{\pi}{2}$, we have
$$\tfrac{1}{2\pi}\int_0^{2\pi}e^{ips\cos(\zeta-\xi)}\,d\zeta
  =\tfrac{1}{2\pi}\int_{-\xi-\tfrac{\pi}{2}}^{2\pi-\xi-\tfrac{\pi}{2}}e^{-ips\sin\theta}\,d\theta
  =\tfrac{1}{2\pi}\int_{-\pi}^{\pi}e^{-ips\sin\theta}\,d\theta
$$
by the $2\pi$--periodicity of the integrand.  This is equal to $J_0(ps)$.
\end{proof}

\begin{lemma}\label{lem:bessint2}
For all $\xi$, $\displaystyle\tfrac{1}{2\pi}\int_0^{2\pi}e^{ips\cos(\zeta-\xi)}(\cos\zeta\,{\bf e}_x+\sin\zeta\,{\bf e}_y)\,d\zeta=iJ_1(ps){\bf e}_s$.
\end{lemma}

\begin{proof}
From Fact \ref{fact:circrect} and the identities $x=s\cos\xi$, $y=s\sin\xi$, we have
\begin{align*}
  \cos\zeta\,{\bf e}_x+\sin\zeta\,{\bf e}_y
  &=\cos\zeta\,(\tfrac{x}{s}{\bf e}_s-\tfrac{y}{s^2}{\bf e}_\xi)
    +\sin\zeta\,(\tfrac{y}{s}{\bf e}_s+\tfrac{x}{s^2}{\bf e}_\xi)\\
  &=\cos(\zeta-\xi){\bf e}_s+\tfrac{1}{s}\sin(\zeta-\xi){\bf e}_\xi.
\end{align*}
Observe that $\int_0^{2\pi}e^{ips\cos(\zeta-\xi)}\sin(\zeta-\xi)\,d\zeta=-\tfrac{1}{ips}e^{ips\cos(\zeta-\xi)}|_0^{2\pi}=0$.  Thus, using the substitution $\theta\doteq\zeta-\xi-\tfrac{\pi}{2}$ again,
\begin{align*}
  &\tfrac{1}{2\pi}\int_0^{2\pi}e^{ips\cos(\zeta-\xi)}(\cos\zeta\,{\bf e}_x+\sin\zeta\,{\bf e}_y)\,d\zeta
    =\tfrac{1}{2\pi}\int_0^{2\pi}e^{ips\cos(\zeta-\xi)}\cos(\zeta-\xi)\,{\bf e}_s\,d\zeta\\
  &=-\tfrac{1}{2\pi}\int_{-\xi-\tfrac{\pi}{2}}^{2\pi-\xi-\tfrac{\pi}{2}}
                       e^{-ips\sin\theta}\sin\theta\,{\bf e}_s\,d\theta
   =\tfrac{i}{4\pi}\int_{-\xi-\tfrac{\pi}{2}}^{2\pi-\xi-\tfrac{\pi}{2}}
        e^{-ips\sin\theta}(e^{i\theta}-e^{-i\theta})\,{\bf e}_s\,d\theta\\
  &=\tfrac{i}{4\pi}\int_{-\xi-\tfrac{\pi}{2}}^{2\pi-\xi-\tfrac{\pi}{2}}
      [e^{i(\theta-ps\sin\theta)}-e^{i(-\theta-ps\sin\theta)}]\,{\bf e}_s\,d\theta
   =\tfrac{i}{2}[J_1(ps)-J_{-1}(ps)]{\bf e}_s,
\end{align*}
courtesy of equation \eqref{eq:besseln}.  On the other hand, $J_{-1}(u)=-J_1(u)$.
\end{proof}

\begin{fact}\label{fact:cpsi0dist}
$\psi_0(E,k)\doteq e^{-i(Et-kz)}\varphi_0(E,k)=\displaystyle\int_{\mathbb M}\Psi_c(E,k,{\bf q})e^{-iq_\alpha x^\alpha}\chi_0^+({\bf q})\,d^4{\bf q}$.
\end{fact}

\begin{proof}
Write $q^x=q\cos\zeta$ and $q^y=q\sin\zeta$.  Then
\begin{align*}
  &\int_{\mathbb M}\Psi_c(E,k,{\bf q})e^{-iq_\alpha x^\alpha}\chi_0^+({\bf q})\,d^4{\bf q}\\
  &=\int_{q^z=-\infty}^\infty\int_{\zeta=0}^{2\pi}\int_{q=0}^\infty\int_{q^t=-\infty}^\infty
       \Psi_c(E,k,{\bf q})e^{-iq_\alpha x^\alpha}\chi_0^+({\bf q})\,q\,dq^t\,dq\,d\zeta\,dq^z\\
  &=\tfrac{1}{2\pi p}e^{-i(Et-kz)}\int_{\zeta=0}^{2\pi}\int_{q=0}^\infty\delta(q-p)e^{i(q^xx+q^yy)}
      \tfrac{1}{\sqrt{2}m}(m{\bf e}_\emptyset-iE{\bf e}_t-ik{\bf e}_z-iq^x{\bf e}_x-iq^y{\bf e}_y)\,
      q\,dq\,d\zeta\\
  &=\tfrac{1}{2\sqrt{2}\pi m}e^{-i(Et-kz)}\int_{\zeta=0}^{2\pi}e^{ip(x\cos\zeta+y\sin\zeta)}
      (m{\bf e}_\emptyset-iE{\bf e}_t-ik{\bf e}_z-ip\cos\zeta\,{\bf e}_x-ip\sin\zeta\,{\bf e}_y)
      \,d\zeta.
\end{align*}
Now $x=s\cos\xi$ and $y=s\sin\xi$, so that $x\cos\zeta+y\sin\zeta=s\cos(\zeta-\xi)$.  From Lemmas \ref{lem:bessint1} and \ref{lem:bessint2} we then have
\begin{align*}
  &\int_{\mathbb M}\Psi_c(E,k,{\bf q})e^{-iq_\alpha x^\alpha}\chi_0^+({\bf q})\,d^4{\bf q}\\
  &=\tfrac{1}{\sqrt{2}m}e^{-i(Et-kz)}[(m{\bf e}_\emptyset-iE{\bf e}_t-ik{\bf e}_z)J_0(ps)
                                      +pJ_1(ps){\bf e}_s]
   =\psi_0(E,k).\qedhere
\end{align*}
\end{proof}

\begin{fact}\label{fact:cphi01}
Set $\Theta_C\doteq\tfrac{1}{p}(ik{\mathcal I}+m\Theta_z+iE\Theta_{tz})$.  Then
\begin{align*}
  \Theta_{tz}\chi_0^+({\bf p})
    &=-\chi_3^-({\bf p})\\
  \Theta_C\chi_0^+({\bf p})
    &=\tfrac{1}{i\sqrt{p_x^2+p_y^2}}[p^y\chi_1^+({\bf p})+p^x\chi_2^+({\bf p})]\\
  \Theta_{tz}\Theta_C\chi_0^+({\bf p})
    &=\tfrac{i}{\sqrt{p_x^2+p_y^2}}[p^x\chi_1^-({\bf p})-p^y\chi_2^-({\bf p})].
\end{align*}
\end{fact}

\begin{proof}
We have $\chi_3^-({\bf p})={\mathcal J}\chi_3^+({\bf p})={\mathcal J}\Theta_{xy}\chi_0^+({\bf p})=-\Theta_{tz}\chi_0^+({\bf p})$.  Observe that
$$\Theta_z\Theta_C\chi_0^+({\bf p})
  =\tfrac{i}{p}(k\Theta_z-im{\mathcal I}+E\Theta_t)\chi_0^+({\bf p})
  =-\tfrac{i}{p}(p^x\Theta_x+p^y\Theta_t)\chi_0^+({\bf p}),
$$
by virtue of fact \ref{fact:spindep}: $(E\Theta_t+p^x\Theta_x+p^y\Theta_y+k\Theta_z)\chi_0^+({\bf p})=im\chi_0^+({\bf p})$.  Hence
$$\Theta_C\chi_0^+({\bf p})
  =\Theta_z\Theta_z\Theta_C\chi_0^+({\bf p})
  =\tfrac{1}{ip}(p^x\Theta_{xz}+p^y\Theta_{yz})\chi_0^+({\bf p})
  =\tfrac{1}{ip}[p^x\chi_2^+({\bf p})+p^y\chi_1^+({\bf p})].
$$
And
\begin{align*}
  \Theta_{tz}\Theta_C\chi_0^+({\bf p})
  &=-\Theta_t\Theta_z\chi_0^+({\bf p})
   =\tfrac{i}{p}(-p^x\Theta_{tx}-p^y\Theta_{ty})\chi_0^+({\bf p})\\
  &=\tfrac{i}{p}(p^x{\mathcal J}\Theta_{yz}-p^y{\mathcal J}\Theta_{xz})\chi_0^+({\bf p})
   =\tfrac{i}{p}[p^x\chi_1^-({\bf p})-p^y\chi_2^-({\bf p})].\qedhere
\end{align*}
\end{proof}

\begin{fact}
$\psi_1(E,k)\doteq e^{-i(Et-kz)}\varphi_1(E,k)=\displaystyle\int_{\mathbb M}\Psi_c(E,k,{\bf q})e^{-iq_\alpha x^\alpha}\chi_c^+({\bf q})\,d^4{\bf q}$, where $\chi_c^+({\bf p})\doteq\tfrac{1}{i\sqrt{p_x^2+p_y^2}}[p^y\chi_1^+({\bf p})+p^x\chi_2^+({\bf p})]$.
\end{fact}

\begin{proof}
From facts \ref{fact:cphi01}, \ref{fact:cpsi0dist}, and \ref{fact:cnonsing}, we have
\begin{align*}
  \int_{\mathbb M}\Psi_c(E,k,{\bf q})e^{-iq_\alpha x^\alpha}\chi_c^+({\bf q})\,d^4{\bf q}
  &=\Theta_C\int_{\mathbb M}\Psi_c(E,k,{\bf q})e^{-iq_\alpha x^\alpha}\chi_0^+({\bf q})\,d^4{\bf q}\\
  &=\Theta_C\psi_0(E,k)
   =\psi_1(E,k).\qedhere
\end{align*}
\end{proof}

%%%%%%%%%%%%%%%%%%%%%%%%%%%%%%%%%%%%%%%%%%%%%%%%%%%%%%%%%%%%%%%%%
\section{Spherical waves}

We use spherical polar coordinates $(x^\alpha)=(t,r,\theta\,\phi)$, where $x=r\sin\theta\cos\phi$, $y=r\sin\theta\sin\phi$, $z=r\cos\theta$.  In this case, the Minkowski metric is
$$\eta={\it diag}(1,-1,-r^2,-r^2\sin^2\theta).$$

\begin{fact}
${\bf e}_r=\tfrac{x}{r}{\bf e}_x$, ${\bf e}_\theta=\tfrac{xz}{s}{\bf e}_x+\tfrac{yz}{s}{\bf e}_y-s{\bf e}_z$, and ${\bf e}_\phi=-y{\bf e}_x+x{\bf e}_y$.  Here $r=\sqrt{x^2+y^2+z^2}$ and $s=\sqrt{x^2+y^2}$.
\end{fact}

\begin{proof}
The basis element ${\bf e}_r$ can be identified with the functional $\tfrac{\partial}{\partial r}$ (and similarly for all coordinates).  Now,
\begin{align*}
  \tfrac{\partial f}{\partial r}
  &=\tfrac{\partial f}{\partial x}\tfrac{\partial x}{\partial r}
    +\tfrac{\partial f}{\partial y}\tfrac{\partial y}{\partial r}
    +\tfrac{\partial f}{\partial z}\tfrac{\partial z}{\partial r}
   =\sin\theta\cos\phi\,\tfrac{\partial f}{\partial x}
    +\sin\theta\sin\phi\,\tfrac{\partial f}{\partial y}
    +\cos\theta\,\tfrac{\partial f}{\partial z}\\
  &=\tfrac{x}{r}\,\tfrac{\partial f}{\partial x}
    +\tfrac{y}{r}\,\tfrac{\partial f}{\partial y}
    +\tfrac{z}{r}\,\tfrac{\partial f}{\partial z}.
\end{align*}
Which yields the first identity.  The second and third are similar.  Noting that $s=r\sin\theta$, so that $\cos\phi=\tfrac{x}{s}$ and $\sin\phi=\tfrac{y}{s}$, we have
\begin{align*}
  \tfrac{\partial f}{\partial\theta}
  &=\tfrac{\partial f}{\partial x}\tfrac{\partial x}{\partial\theta}
    +\tfrac{\partial f}{\partial y}\tfrac{\partial y}{\partial\theta}
    +\tfrac{\partial f}{\partial z}\tfrac{\partial z}{\partial\theta}
   =r\cos\theta\cos\phi\,\tfrac{\partial f}{\partial x}
    +r\cos\theta\sin\phi\,\tfrac{\partial f}{\partial y}
    -r\sin\theta\,\tfrac{\partial f}{\partial z}\\
  &=\tfrac{xz}{s}\,\tfrac{\partial f}{\partial x}
    +\tfrac{yz}{s}\,\tfrac{\partial f}{\partial y}
    -s\,\tfrac{\partial f}{\partial z},\\
  \tfrac{\partial f}{\partial\phi}
  &=\tfrac{\partial f}{\partial x}\tfrac{\partial x}{\partial\phi}
    +\tfrac{\partial f}{\partial y}\tfrac{\partial y}{\partial\phi}
    +\tfrac{\partial f}{\partial z}\tfrac{\partial z}{\partial\phi}
   =-r\sin\theta\sin\phi\,\tfrac{\partial f}{\partial x}
    +r\sin\theta\cos\phi\,\tfrac{\partial f}{\partial y}\\
  &=-y\,\tfrac{\partial f}{\partial x}
    +x\,\tfrac{\partial f}{\partial y}.\qedhere
\end{align*}
\end{proof}

%::::::::::::::::::::::::::::::::::::::::::::::::::::::::::::::::
\subsection{Dirac equation}

\begin{fact}\label{fact:sconn}
The nontrivial components of the connection matrices are
\begin{align*}
  \Gamma_r{\bf e}_\theta&=\tfrac{1}{r}{\bf e}_\theta
  & \Gamma_r{\bf e}_\phi&=\tfrac{1}{r}{\bf e}_\phi
  & \Gamma_\theta{\bf e}_r&=\tfrac{1}{r}{\bf e}_\theta
  & \Gamma_\theta{\bf e}_\theta &= -r{\bf e}_r
  & \Gamma_\theta{\bf e}_\phi &= \cot\theta\,{\bf e}_\phi\\
  \Gamma_\phi{\bf e}_r &=\tfrac{1}{r}{\bf e}_\phi
  & \Gamma_\phi{\bf e}_\theta &= \cot\theta\,{\bf e}_\phi
  & \Gamma_\phi{\bf e}_\phi &= \makebox[0pt][l]{$-r\sin^2\theta\,{\bf e}_r-\sin\theta\cos\theta\,{\bf e}_\theta$.}
\end{align*}
\end{fact}

\begin{proof}
We compute the nontrivial Christoffel symbols of the metric--compatible connection:
\begin{align*}
  \Gamma_{r\theta}^\theta &=\tfrac{1}{2}\eta^{\theta\theta}\eta_{\theta\theta r}
                           =\tfrac{1}{2}(-\tfrac{1}{r^2})(-2r)
                           =\tfrac{1}{r}\\
  \Gamma_{r\phi}^\phi &=\tfrac{1}{2}\eta^{\phi\phi}\eta_{\phi\phi r}
                       =\tfrac{1}{2}(-\tfrac{1}{r^2\sin^2\theta})(-2r\sin^2\theta)
                       =\tfrac{1}{r}\\
  \Gamma_{\theta\theta}^r &=-\tfrac{1}{2}\eta^{rr}\eta_{\theta\theta r}
                           =-\tfrac{1}{2}(-1)(-2r)
                           =-r\\
  \Gamma_{\theta\phi}^\phi &=\tfrac{1}{2}\eta^{\phi\phi}\eta_{\phi\phi\theta}
                            =\tfrac{1}{2}(-\tfrac{1}{r^2\sin^2\theta})(-2r^2\sin\theta\cos\theta)
                            =\cot\theta\\
  \Gamma_{\phi\phi}^r &=-\tfrac{1}{2}\eta^{rr}\eta_{\phi\phi r}
                       =-\tfrac{1}{2}(-1)(-2r\sin^2\theta)
                       =-r\sin^2\theta\\
  \Gamma_{\phi\phi}^\theta &=-\tfrac{1}{2}\eta^{\theta\theta}\eta_{\phi\phi\theta}
                       =-\tfrac{1}{2}(-\tfrac{1}{r^2})(-2r^2\sin\theta\cos\theta)
                       =-\sin\theta\cos\theta.\qedhere
\end{align*}
\end{proof}

We seek a solution to the Dirac equation $\gamma^\alpha\hat\nabla_\alpha\psi=m\psi$ of the form
\begin{equation}\label{eq:ssepsol}
  \psi(t,r)=e^{-iEt}\varphi(r)
\end{equation}
where $\varphi=\varphi^I{\bf e}_I$.  Specifically, $\psi$ has no dependence on the angular variables $\theta$ or $\phi$.

\begin{fact}\label{fact:sdiracop}
For spinor $\psi$ as in equation \eqref{eq:ssepsol}, the effect of the Dirac operator $D=\gamma^\alpha\hat\nabla_\alpha$ is $D\psi=e^{-iEt}(-iE\gamma_t\varphi-\gamma_r\partial_r\varphi+\gamma^k\hat\Gamma_k\varphi)$, where
\begin{align*}
  \gamma^k\hat\Gamma_k\varphi
  &=(-\tfrac{2}{r}\varphi^r-\cot\theta\,\varphi^\theta){\bf e}_\emptyset
    -\cot\theta\,\varphi^{tr\theta}{\bf e}_{tr}
    +\tfrac{2}{r}\varphi^{tr\theta}{\bf e}_{t\theta}\\
  &\quad
    +(\tfrac{2}{r}\varphi^{tr\phi}+\cot\theta\,\varphi^{t\theta\phi}){\bf e}_{t\phi}
    -\tfrac{2}{r}\varphi^\theta{\bf e}_{r\theta}
    +(\cot\theta\,\varphi^{r\theta\phi}-\tfrac{2}{r}\varphi^\phi){\bf e}_{r\phi}\\
  &\quad
    -(\tfrac{2\cot\theta}{r^2}\varphi^\phi+\tfrac{2}{r}\varphi^{r\theta\phi})
       {\bf e}_{\theta\phi}
    +(\tfrac{4}{r}\varphi^{t\theta\phi}-\tfrac{2\cot\theta}{r^2}\varphi^{tr\phi})
       {\bf e}_{tr\theta\phi}\\
  &\quad
    +(\tfrac{2}{r}\varphi^{tr}+\cot\theta\,\varphi^{t\theta}){\bf e}_t
    +\cot\theta\,\varphi^{r\theta}{\bf e}_r
    -\tfrac{2}{r}\varphi^{r\theta}{\bf e}_\theta\\
  &\quad
    -(\tfrac{2}{r}\varphi^{r\phi}+\cot\theta\,\varphi^{\theta\phi}){\bf e}_\phi
    +\tfrac{2}{r}\varphi^{t\theta}{\bf e}_{tr\theta}
    +(\tfrac{2}{r}\varphi^{t\phi}-\cot\theta\,\varphi^{tr\theta\phi}){\bf e}_{tr\phi}\\
  &\quad
    +(\tfrac{2}{r}\varphi^{tr\theta\phi}+\tfrac{2\cot\theta}{r^2}\varphi^{t\phi})
       {\bf e}_{t\theta\phi}
    +(\tfrac{2\cot\theta}{r^2}\varphi^{r\phi}-\tfrac{4}{r}\varphi^{\theta\phi})
       {\bf e}_{r\theta\phi}.
\end{align*}
\end{fact}

\begin{proof}
Note that $\gamma^k\hat\Gamma_k=-\gamma_r\hat\Gamma_r-\tfrac{1}{r^2}\gamma_\theta\hat\Gamma_\theta-\tfrac{1}{r^2\sin^2\theta}\gamma_\phi\hat\Gamma_\phi$.  So $\gamma^k\hat\Gamma_k{\bf e}_t=0$.  And we compute:
\begin{align*}
  \gamma^k\hat\Gamma_k{\bf e}_r
    &=-\tfrac{1}{r^2}\gamma_\theta(\tfrac{1}{r}{\bf e}_\theta)
      -\tfrac{1}{r^2\sin^2\theta}\gamma_\phi(\tfrac{1}{r}{\bf e}_\phi)
     =-\tfrac{1}{r}{\bf e}_\emptyset-\tfrac{1}{r}{\bf e}_\emptyset
     =-\tfrac{2}{r}{\bf e}_\emptyset\\
  \gamma^k\hat\Gamma_k{\bf e}_\theta
    &=-\gamma_r(\tfrac{1}{r}{\bf e}_\theta)
      -\tfrac{1}{r^2}\gamma_\theta(-r{\bf e}_r)
      -\tfrac{1}{r^2\sin^2\theta}\gamma_\phi(\cot\theta\,{\bf e}_\phi)
     =-\tfrac{2}{r}{\bf e}_{r\theta}-\cot\theta\,{\bf e}_\emptyset\\
  \gamma^k\hat\Gamma_k{\bf e}_\phi
    &=-\gamma_r(\tfrac{1}{r}{\bf e}_\phi)
      -\tfrac{1}{r^2}\gamma_\theta(\cot\theta\,{\bf e}_\phi)
      -\tfrac{1}{r^2\sin^2\theta}\gamma_\phi
         (-r\sin^2\theta\,{\bf e}_r-\sin\theta\cos\theta\,{\bf e}_\theta)\\
    &=-\tfrac{1}{r}{\bf e}_{r\phi}-\tfrac{\cot\theta}{r^2}{\bf e}_{\theta\phi}
      -\tfrac{1}{r}{\bf e}_{r\phi}-\tfrac{\cot\theta}{r^2}{\bf e}_{\theta\phi}
     =-\tfrac{2}{r}{\bf e}_{r\phi}-\tfrac{2\cot\theta}{r^2}{\bf e}_{\theta\phi}.
\end{align*}
\begin{align*}
  \gamma^k\hat\Gamma_k{\bf e}_{tr}
    &=-\tfrac{1}{r^2}\gamma_\theta(\tfrac{1}{r}{\bf e}_{t\theta})
      -\tfrac{1}{r^2\sin^2\theta}\gamma_\phi(\tfrac{1}{r}{\bf e}_{t\phi})
     =\tfrac{1}{r}{\bf e}_t+\tfrac{1}{r}{\bf e}_t
     =\tfrac{2}{r}{\bf e}_t\\
  \gamma^k\hat\Gamma_k{\bf e}_{t\theta}
    &=-\gamma_r(\tfrac{1}{r}{\bf e}_{t\theta})
      -\tfrac{1}{r^2}\gamma_\theta(-r{\bf e}_{tr})
      -\tfrac{1}{r^2\sin^2\theta}\gamma_\phi(\cot\theta\,{\bf e}_{t\phi})
     =\tfrac{2}{r}{\bf e}_{tr\theta}+\cot\theta\,{\bf e}_t\\
  \gamma^k\hat\Gamma_k{\bf e}_{t\phi}
    &=-\gamma_r(\tfrac{1}{r}{\bf e}_{t\phi})
      -\tfrac{1}{r^2}\gamma_\theta(\cot\theta\,{\bf e}_{t\phi})
      -\tfrac{1}{r^2\sin^2\theta}\gamma_\phi
         (-r\sin^2\theta\,{\bf e}_{tr}-\sin\theta\cos\theta\,{\bf e}_{t\theta})\\
    &=\tfrac{1}{r}{\bf e}_{tr\phi}+\tfrac{\cot\theta}{r^2}{\bf e}_{t\theta\phi}
      +\tfrac{1}{r}{\bf e}_{tr\phi}+\tfrac{\cot\theta}{r^2}{\bf e}_{t\theta\phi}
     =\tfrac{2}{r}{\bf e}_{tr\phi}+\tfrac{2\cot\theta}{r^2}{\bf e}_{t\theta\phi}.
\end{align*}
\begin{align*}
  \gamma^k\hat\Gamma_k{\bf e}_{r\theta}
    &=-\gamma_r(\tfrac{1}{r}{\bf e}_{r\theta})
      -\tfrac{1}{r^2\sin^2\theta}\gamma_\phi(-\tfrac{1}{r}{\bf e}_{\theta\phi}
                             +\cot\theta\,{\bf e}_{r\phi})
     =-\tfrac{2}{r}{\bf e}_\theta+\cot\theta\,{\bf e}_r\\
  \gamma^k\hat\Gamma_k{\bf e}_{r\phi}
    &=-\gamma_r(\tfrac{1}{r}{\bf e}_{r\phi})
      -\tfrac{1}{r^2}\gamma_\theta(\tfrac{1}{r}{\bf e}_{\theta\phi}+\cot\theta\,{\bf e}_{r\phi})
      -\tfrac{1}{r^2\sin^2\theta}\gamma_\phi(-\sin\theta\cos\theta\,{\bf e}_{r\theta})\\
    &=-\tfrac{2}{r}{\bf e}_\phi+\tfrac{2\cot\theta}{r^2}{\bf e}_{r\theta\phi}\\
  \gamma^k\hat\Gamma_k{\bf e}_{\theta\phi}
    &=-\gamma_r(\tfrac{2}{r}{\bf e}_{\theta\phi})
      -\tfrac{1}{r^2}\gamma_\theta(-r{\bf e}_{r\phi}+\cot\theta\,{\bf e}_{\theta\phi})
      -\tfrac{1}{r^2\sin^2\theta}\gamma_\phi(r\sin^2\theta\,{\bf e}_{r\theta})\\
    &=-\tfrac{4}{r}{\bf e}_{r\theta\phi}-\cot\theta\,{\bf e}_\phi
\end{align*}
\begin{align*}
  \gamma^k\hat\Gamma_k{\bf e}_{tr\theta}
    &=-\gamma_r(\tfrac{1}{r}{\bf e}_{tr\theta})
      -\tfrac{1}{r^2\sin^2\theta}\gamma_\phi(-\tfrac{1}{r}{\bf e}_{t\theta\phi}
                             +\cot\theta\,{\bf e}_{tr\phi})
     =\tfrac{2}{r}{\bf e}_{t\theta}-\cot\theta\,{\bf e}_{tr}\\
  \gamma^k\hat\Gamma_k{\bf e}_{tr\phi}
    &=-\gamma_r(\tfrac{1}{r}{\bf e}_{tr\phi})
      -\tfrac{1}{r^2}\gamma_\theta(\tfrac{1}{r}{\bf e}_{t\theta\phi}+\cot\theta\,{\bf e}_{tr\phi})
      -\tfrac{1}{r^2\sin^2\theta}\gamma_\phi(-\sin\theta\cos\theta\,{\bf e}_{tr\theta})\\
    &=\tfrac{2}{r}{\bf e}_{t\phi}-\tfrac{2\cot\theta}{r^2}{\bf e}_{tr\theta\phi}\\
  \gamma^k\hat\Gamma_k{\bf e}_{t\theta\phi}
    &=-\gamma_r(\tfrac{2}{r}{\bf e}_{t\theta\phi})
      -\tfrac{1}{r^2}\gamma_\theta(-r{\bf e}_{tr\phi}+\cot\theta\,{\bf e}_{t\theta\phi})
      -\tfrac{1}{r^2\sin^2\theta}\gamma_\phi(r\sin^2\theta\,{\bf e}_{tr\theta})\\
    &=\tfrac{4}{r}{\bf e}_{tr\theta\phi}+\cot\theta\,{\bf e}_{t\phi}\\
  \gamma^k\hat\Gamma_k{\bf e}_{r\theta\phi}
    &=-\gamma_r(\tfrac{2}{r}{\bf e}_{r\theta\phi})
      -\tfrac{1}{r^2}\gamma_\theta(\cot\theta\,{\bf e}_{r\theta\phi})
     =-\tfrac{2}{r}{\bf e}_{\theta\phi}+\cot\theta\,{\bf e}_{r\phi}.\qedhere
\end{align*}
\end{proof}

\begin{fact}\label{fact:sspinorform}
A spinor $\psi$ as in equation \eqref{eq:ssepsol} solves the the Dirac equation only if $\varphi$ is of the form $\varphi=\varphi^\emptyset{\bf e}_\emptyset+\varphi^{tr}{\bf e}_{tr}+\varphi^t{\bf e}_t+\varphi^r{\bf e}_r$.
\end{fact}

\begin{proof}
Because the right--hand side of the Dirac equation $D\psi=m\psi$ does not involve either $\theta$ or $\phi$, the the left--hand side $D\psi$ cannot either.  Fact \ref{fact:sdiracop} thus implies that $\varphi^\theta=0$, $\varphi^{tr\theta}=0$, $\varphi^{t\theta\phi}=0$, $\varphi^{r\theta\phi}=0$, $\varphi^\phi=0$, $\varphi^{tr\phi}=0$, $\varphi^{t\theta}=0$, $\varphi^{r\theta}=0$, $\varphi^{\theta\phi}=0$, $\varphi^{tr\theta\phi}=0$, $\varphi^{t\phi}=0$, and $\varphi^{r\phi}=0$.
\end{proof}

\begin{fact}
A spinor $\psi$ as in equation \eqref{eq:ssepsol} solves the Dirac equation if and only if $\varphi$ is a solution of the four coupled equations
\begin{equation}\label{eq:sdirac}
\begin{aligned}
    &-\partial_r\varphi^r-\tfrac{2}{r}\varphi^r+iE\varphi^t=m\varphi^\emptyset
  & &\partial_r\varphi^t-iE\varphi^r=m\varphi^{tr}\\
    &\partial_r\varphi^{tr}+\tfrac{2}{r}\varphi^{tr}-iE\varphi^\emptyset=m\varphi^t
  & &-\partial_r\varphi^\emptyset+iE\varphi^{tr}=m\varphi^r.
\end{aligned}
\end{equation}
\end{fact}

\begin{proof}
From Facts \ref{fact:sdiracop} and \ref{fact:sspinorform}, the Dirac equation implies $-iE\gamma_t\varphi-\gamma_r\partial_r\varphi+\gamma^k\hat\Gamma_k\varphi=m\varphi$, where
$$\gamma_t\varphi
  =\gamma_t(\varphi^\emptyset{\bf e}_\emptyset+\varphi^{tr}{\bf e}_{tr}
            +\varphi^t{\bf e}_t+\varphi^r{\bf e}_r)
  =\varphi^\emptyset{\bf e}_t-\varphi^{tr}{\bf e}_r
            -\varphi^t{\bf e}_\emptyset+\varphi^r{\bf e}_{tr}
$$
$$\gamma_r\partial_r\varphi
  =\gamma_r\partial_r(\varphi^\emptyset{\bf e}_\emptyset+\varphi^{tr}{\bf e}_{tr}
                       +\varphi^t{\bf e}_t+\varphi^r{\bf e}_r)
  =\partial_r\varphi^\emptyset{\bf e}_r-\partial_r\varphi^{tr}{\bf e}_t
   -\partial_r\varphi^t{\bf e}_{tr}+\partial_r\varphi^r{\bf e}_\emptyset,
$$
and $\gamma^k\hat\Gamma_k\varphi=-\tfrac{2}{r}\varphi^r{\bf e}_\emptyset+\tfrac{2}{r}\varphi^{tr}{\bf e}_t$.  So that
\begin{align*}
  &-iE\gamma_t\varphi-\gamma_r\partial_r\varphi+\gamma^k\hat\Gamma_k\varphi
   =(iE\varphi^t-\partial_r\varphi^r-\tfrac{2}{r}\varphi^r){\bf e}_\emptyset
    +(-iE\varphi^r+\partial_r\varphi^t){\bf e}_{tr}\\
  &\quad\quad\quad
   +(-iE\varphi^\emptyset+\partial_r\varphi^{tr}+\tfrac{2}{r}\varphi^{tr}){\bf e}_t
   +(iE\varphi^{tr}-\partial_r\varphi^\emptyset){\bf e}_r.\qedhere
\end{align*}
\end{proof}

%::::::::::::::::::::::::::::::::::::::::::::::::::::::::::::::::
\subsection{Solutions}

\begin{fact}\label{fact:ssol2}
Let $D_r$ be such that $D_r\varphi=m\varphi$, as in equation \eqref{eq:sdirac}.  Then $D_r^2\varphi$ is diagonal with
\begin{align*}
  D_r^2\varphi
    &=(\partial_r^2\varphi^\emptyset+\tfrac{2}{r}\partial_r\varphi^\emptyset
       +E^2\varphi^\emptyset){\bf e}_\emptyset
      +(\partial_r^2\varphi^{tr}+\tfrac{2}{r}\partial_r\varphi^{tr}
        -\tfrac{2}{r^2}\varphi^{tr}+E^2\varphi^{tr}){\bf e}_{tr}\\
    &\quad
      +(\partial_r^2\varphi^t+\tfrac{2}{r}\partial_r\varphi^t+E^2\varphi^t){\bf e}_t
      +(\partial_r^2\varphi^r+\tfrac{2}{r}\partial_r\varphi^r
        -\tfrac{2}{r^2}\varphi^r+E^2\varphi^r){\bf e}_r.
\end{align*}
\end{fact}

\begin{proof}
Note that from Fact \ref{fact:sconn}, we have ${\Gamma^t}_{\rho\sigma}=0$, $\hat\Gamma_t=0$, $\hat\Gamma_r\psi=0$, $\partial_\phi\hat\Gamma_\phi=0$, $(\partial_\theta\hat\Gamma_\theta)\psi=0$.  And
$$\hat\Gamma_\theta^2{\bf e}_r=-{\bf e}_r
  \quad\text{and}\quad
  \hat\Gamma_\phi^2{\bf e}_r=\tfrac{1}{r}(-r\sin^2\theta{\bf e}_r
                             -\sin\theta\cos\theta\,{\bf e}_\theta)
$$
(similary for ${\bf e}_{tr}$).  Moreover, the extended curvature $\hat\Omega_{\rho\sigma}=0$ as $\eta$ is a flat metric.  Thus from Fact \ref{fact:DD} in the appendix, $D^2\psi=-\eta^{\alpha\beta}\partial_{\alpha\beta}\psi+{\mathcal A}^\alpha\partial_\alpha\psi+{\mathcal B}\psi$, where
\begin{align*}
  -\eta^{\alpha\beta}\partial_{\alpha\beta}\psi
    &=-\partial_t^2\psi+\partial_r^2\psi
     =E^2\psi+\partial_r^2\psi\\
  {\mathcal A}^\alpha\partial_\alpha\psi
    &=(\eta^{\rho\sigma}{\Gamma^\alpha}_{\rho\sigma}{\mathcal I}
       -2\hat\Gamma^\alpha)\partial_\alpha\psi
     =(\eta^{\theta\theta}{\Gamma^r}_{\theta\theta}
       +\eta^{\phi\phi}{\Gamma^r}_{\phi\phi})\partial_r\psi\\
    &=[-\tfrac{1}{r^2}(-r)-\tfrac{1}{r^2\sin^2\theta}(-r\sin^2\theta)]\partial_r\psi
     =\tfrac{2}{r}\partial_r\psi\\
  {\mathcal B}\psi
    &=(\eta^{\rho\sigma}{\Gamma^\beta}_{\rho\sigma}\hat\Gamma_\beta
       -\partial^\beta\hat\Gamma_\beta-\hat\Gamma^\beta\hat\Gamma_\beta
       +\tfrac{1}{2}\gamma^\rho\gamma^\sigma\hat\Omega_{\rho\sigma})\psi\\
    &=(\eta^{\phi\phi}{\Gamma^\theta}_{\phi\phi}\hat\Gamma_\theta
       -\eta^{\theta\theta}\hat\Gamma_\theta^2
       -\eta^{\phi\phi}\hat\Gamma_\phi^2)\psi\\
    &=-\tfrac{1}{r^2\sin^2\theta}(-\sin\theta\cos\theta)
         (\tfrac{1}{r}\psi^r{\bf e}_\theta+\tfrac{1}{r}\psi^{tr}{\bf e}_{t\theta})
      +\tfrac{1}{r^2}(-\psi^r{\bf e}_r-\psi^{tr}{\bf e}_{tr})\\
    &\quad
      +\tfrac{1}{r^2\sin^2\theta}\tfrac{1}{r}(-r\sin^2\theta\,\psi^r{\bf e}_r
                     -\sin\theta\cos\theta\,\psi^\theta{\bf e}_\theta)\\
    &\quad
      +\tfrac{1}{r^2\sin^2\theta}\tfrac{1}{r}(-r\sin^2\theta\,\psi^{tr}{\bf e}_{tr}
                     -\sin\theta\cos\theta\,\psi^{t\theta}{\bf e}_{t\theta})\\
    &=-\tfrac{2}{r^2}(\psi^r{\bf e}_r+\psi^{tr}{\bf e}_{tr}).\qedhere
\end{align*}
\end{proof}

\begin{fact}\label{fact:snonsing2}
The nonsingular solutions of $D_r^2\varphi=m^2\varphi$ are given by
$$\varphi^\emptyset=A^\emptyset j_0(pr),\quad
  \varphi^{tr}=A^{tr}j_1(pr),\quad
  \varphi^t=A^tj_0(pr),\quad
  \varphi^r=A^rj_1(pr)
$$
where $p\doteq\sqrt{E^2-m^2}$, and $j_k$ is the $k$--th order spherical Bessel function of the first kind.  That is, $j_0(u)=\tfrac{\sin{u}}{u}$ and $j_1(u)=\tfrac{\sin{u}-u\cos{u}}{u^2}$.
\end{fact}

\begin{proof}
From Fact \ref{fact:ssol2}, we see that there are only two different equations among the four components of the equation $D_r^2\varphi=m^2\varphi$.  Namely,
$$\partial_r^2f+\tfrac{2}{r}\partial_rf+p^2f=0
  \quad\text{and}\quad
  \partial_r^2f+\tfrac{2}{r}\partial_rf-\tfrac{2}{r^2}g+p^2g=0
$$
where $p^2=E^2-m^2$.  Or equivalently if we set $u\doteq pr$,
$$u^2\partial_u^2f+2u\partial_uf+u^2f=0
  \quad\text{and}\quad
  u^2\partial_u^2f+2u\partial_uf+(u^2-2)g=0.
$$
Solutions to the first equation are the zeroth order spherical Bessel functions $j_0(u)$ and $y_0(u)$ of the first and second kind.  And solutions to the second equation are the first order functions $j_1(u)$ and $y_1(u)$.  Note that $j_k(u)$ is nonsingular, whereas $y_k(u)$ is singular at $u=0$.
\end{proof}

\begin{fact}\label{fact:sprojsol}
The operator $\pi_s\doteq\tfrac{1}{2m}(m{\mathcal I}+D_r)$ is projection from the space of solutions of $D_r^2\varphi=m^2\varphi$ onto the space of solutions of $D_r\varphi=m\varphi$.
\end{fact}

\begin{proof}
Let ${\mathcal S}$ be the space of solutions to $D_r^2\varphi=m^2\varphi$.  Suppose that $\varphi'\in{\mathcal S}$, so that $D_r^2\varphi'=m^2\varphi'$.  Then
\begin{align*}
  \pi_s^2\varphi'
  &=\tfrac{1}{4m^2}(m{\mathcal I}+D_r)(m{\mathcal I}+D_r)\varphi'
   =\tfrac{1}{4m^2}(m^2\varphi'+2mD_r\varphi'+D_r^2\varphi')\\
  &=\tfrac{1}{4m^2}(m^2\varphi'+2mD_r\varphi'+m^2\varphi')
   =\tfrac{1}{2m}(m\varphi'+D_r\varphi')
   =\pi_s\varphi'.
\end{align*}
So that $\pi_s$ is a projection from ${\mathcal S}$.  It remains to show that $\varphi=\pi_s\varphi'$ solves the equation $D_r\varphi=m\varphi$.
\begin{align*}
  D_r\varphi
  &=D_r\tfrac{1}{2m}(m{\mathcal I}+D_r)\varphi'
   =\tfrac{1}{2m}(mD_r\varphi'+D_r^2\varphi')
   =\tfrac{1}{2m}(mD_r\varphi'+m^2\varphi')\\
  &=m\cdot\tfrac{1}{2m}(m\varphi'+D_r\varphi')
   =m\,\pi_s\varphi'
   =m\varphi.\qedhere
\end{align*}
\end{proof}

\begin{fact}\label{fact:sthetacomm}
The matrix $\Theta_t$ commutes with the Dirac operator.
\end{fact}

\begin{proof}
Note that $\Theta_t{\bf e}_I=\gamma_I{\bf e}_t$.  So if $t$ is not in the multi--index $I$, then $\Theta_t{\bf e}_I={\bf e}_I\wedge{\bf e}_t=(-1)^{|I|}{\bf e}_t\wedge{\bf e}_I$.  And if ${\bf e}_I={\bf e}_t\wedge{\bf e}_J$, then $\Theta_t{\bf e}_I=-\eta_{tt}(-1)^{|J|}{\bf e}_J=(-1)^{|I|}{\bf e}_J$.  In either case, $\Theta_t$ only has constant coefficients.  So $\Theta_t$ commutes with $D=\gamma^\alpha(\partial_\alpha+\hat\Gamma_\alpha)$ provided that it commutes with the $\hat\Gamma_\alpha$.  Indeed if $t\not\in I$, then $\hat\Gamma_\alpha{\bf e}_I=A^K{\bf e}_K$ with $|K|=|I|$ and $t\not\in K$, so that
we have
$$\Theta_t\hat\Gamma_\alpha{\bf e}_I
  =(-1)^{|I|}{\bf e}_t\wedge\hat\Gamma_\alpha{\bf e}_I
   =\hat\Gamma_\alpha[(-1)^{|I|}{\bf e}_t\wedge{\bf e}_I]
   =\hat\Gamma_\alpha\Theta_t{\bf e}_I
$$
as $\hat\Gamma_\alpha{\bf e}_t=0$.  And if ${\bf e}_I={\bf e}_t\wedge{\bf e}_J$, then
$$\Theta_t\hat\Gamma_\alpha{\bf e}_I
  =\Theta_t({\bf e}_t\wedge\hat\Gamma_\alpha{\bf e}_J)
  =(-1)^{|I|}\hat\Gamma_\alpha{\bf e}_J
  =\hat\Gamma_\alpha\Theta_t{\bf e}_I.
$$
In both cases, $\Theta_t\hat\Gamma_\alpha=\hat\Gamma_\alpha\Theta_t$.
\end{proof}

\begin{fact}\label{fact:snonsing}
There are two independent nonsingular solutions to the equation $D_r\varphi=m\varphi$ given by
\begin{align*}
  \varphi_0(E)
    &\doteq mj_0(pr){\bf e}_\emptyset-iEj_0(pr){\bf e}_t+pj_1(pr){\bf e}_r\\
  \varphi_1(E)
    &\doteq mj_1(pr){\bf e}_{tr}+pj_0(pr){\bf e}_t+iEj_1(pr){\bf e}_r
\end{align*}
(where $p=\sqrt{E^2-m^2}$).  Moreover, we have $\varphi_1(E)=(\tfrac{iE}{p}{\mathcal I}-\tfrac{m}{p}\Theta_t)\varphi_0(E)$.
\end{fact}

\begin{proof}
By Fact \ref{fact:snonsing2}, the space of nonsingular solutions to $D_r^2\varphi=m^2\varphi$ is a four--dimensional vector space ${\mathcal N}$.  The trace of $\pi_s$ when restricted to ${\mathcal N}$ is
$${\it tr}(\pi_s|{\mathcal N})
  =\tfrac{1}{2m}{\it tr}\bigl((m{\mathcal I}+D_r)|{\mathcal N}\bigr)
  =\tfrac{1}{2m}(4m)=2
$$
as $D_r|{\mathcal N}$ has no diagonal terms via equation \eqref{eq:sdirac}.  So that the image of $\pi_s|{\mathcal N}$ has dimension two.  And Fact \ref{fact:sprojsol} then implies that there are only two independent nonsingular solutions to $D_r\varphi=m\varphi$, and these are in the image of $\pi_s|{\mathcal N}$.  Using equation \eqref{eq:sdirac}, we compute
\begin{align*}
  2m\,\pi_s(j_0(pr){\bf e}_\emptyset)
  &=(m{\mathcal I}+D_r)j_0(pr){\bf e}_\emptyset\\
  &=mj_0(pr){\bf e}_\emptyset - iEj_0(pr){\bf e}_t - \partial_rj_0(pr){\bf e}_r\\
  &=mj_0(pr){\bf e}_\emptyset - iEj_0(pr){\bf e}_t + pj_1(pr){\bf e}_r
\end{align*}
as $\partial_uj_0(u)=-j_1(u)$.  And because $\partial_uj_1(u)=j_0(u)-\tfrac{2}{u}\,j_1(u)$, we also have
\begin{align*}
  2m\,\pi_s(j_1(pr){\bf e}_{tr})
  &=(m{\mathcal I}+D_r)j_1(pr){\bf e}_{tr}\\
  &=mj_1(pr){\bf e}_{tr}+[\partial_rj_1(pr)+\tfrac{2}{r}j_1(pr)]{\bf e}_t+iEj_1(pr){\bf e}_r\\
  &=mj_1(pr){\bf e}_{tr}+pj_0(pr){\bf e}_t+iEj_1(pr){\bf e}_r.
\end{align*}
These yield two solutions to $D_r\varphi=m\varphi$, which are evidently independent.  For the last statement in the fact, we compute
\begin{align*}
  (iE{\mathcal I}-m\Theta_t)\varphi_0(E)
  &=iE[mj_0(pr){\bf e}_\emptyset-iEj_0(pr){\bf e}_t+pj_1(pr){\bf e}_r]\\
  &\quad\quad
    -m[mj_0(pr){\bf e}_t+iEj_0(pr){\bf e}_\emptyset-pj_1(pr){\bf e}_{tr}]\\
  &=mpj_1(pr){\bf e}_{tr}+(E^2-m^2)j_0(pr){\bf e}_t+ipEj_1(pr){\bf e}_r\\
  &=p\varphi_1(E).\qedhere
\end{align*}
\end{proof}

We remark that by Fact \ref{fact:sthetacomm}, $\varphi_0(E)$ and $\Theta_t\varphi_0(E)$ are also two independent nonsingular solutions to $D_r\varphi=m\varphi$.  However, we choose $\varphi_1(E)$ instead of $\Theta_t\varphi_0(E)$ because the former has a simple plane wave decomposition into nice strictly negative spinors.

\begin{fact}\label{fact:sssol}
The singular spherically symmetric solutions to the Dirac equation are linear combinations of the two independent spinors
\begin{align*}
  \psi_0(E)
    &\doteq\tfrac{1}{\sqrt{2}m}e^{-iEt}[mj_0(pr){\bf e}_\emptyset
          -iEj_0(pr){\bf e}_t+pj_1(pr){\bf e}_r]\\
  \psi_1(E)
    &\doteq\tfrac{1}{\sqrt{2}m}e^{-iEt}[mj_1(pr){\bf e}_{tr}
          +pj_0(pr){\bf e}_t+iEj_1(pr){\bf e}_r]
\end{align*}
where $p=\sqrt{E^2-m^2}$.
\end{fact}

%::::::::::::::::::::::::::::::::::::::::::::::::::::::::::::::::
\subsection{Projections}

\noindent
[straightforward computation]

%::::::::::::::::::::::::::::::::::::::::::::::::::::::::::::::::
\subsection{Plane wave decomposition}

For the four--vector ${\bf q}=(q^\alpha)=(q^t,\vec{q})$, define the distribution
$$\Psi(E,{\bf q})
  \doteq\tfrac{1}{4\pi p^2}\,\delta(E-q^t)\,\delta(p-|\vec{q}|)
$$
where $p=\sqrt{E^2-m^2}$ ($m$ is considered a fixed quantity).

\begin{fact}\label{fact:splanedecomp0}
$\displaystyle\psi_0(E)=\int_{\mathbb M}\Psi(E,{\bf q})\,e^{-iq_\alpha x^\alpha}\chi_0^+({\bf q})\,d^4{\bf q}$.
\end{fact}

\begin{proof}
Fix $(x^\alpha)=(t,\vec{r})$.  If $r=|\vec{r}|\neq 0$, then we can write
$$\vec{q}=|\vec{q}|(\cos\mu\,{\bf e}_r+\sin\mu\cos\nu\,\vec{u}+\sin\mu\sin\nu\,\vec{v})$$
where ${\bf e}_r=\vec{r}/r$, $\mu$ is the angle between $\vec{q}$ and $\vec{r}$, the triple $\vec{u}$, $\vec{v}$, ${\bf e}_r$ forms a right--handed orthonormal frame, and $\nu$ is an angle with $0\leq\nu\leq 2\pi$.  Then
\begin{align*}
  &\int_{\mathbb M}\Psi(E,{\bf q})\,e^{-iq_\alpha x^\alpha}\chi_0^+({\bf q})\,d^4{\bf q}\\
  &=\tfrac{1}{4\pi p^2}\int_{q^t=-\infty}^\infty\int_{|\vec{q}|=0}^\infty
                       \int_{\mu=0}^\pi\int_{\nu=0}^{2\pi}
                          \delta(E-q^t)\,\delta(p-|\vec{q}|)
                          e^{-iq_\alpha x^\alpha}\chi_0^+({\bf q})
                          |\vec{q}|^2\sin\mu\,dq^t\,d|\vec{q}|\,d\mu\,d\nu\\
  &=\tfrac{1}{4\pi p^2}\int_{\mu=0}^\pi\int_{\nu=0}^{2\pi}
                          e^{-iEt}e^{ipr\cos\mu}
                          \tfrac{1}{\sqrt{2}m}(m{\bf e}_\emptyset-iE{\bf e}_t
                          -i\tfrac{p}{|\vec{q}|}\vec{q})
                          p^2\sin\mu\,d\mu\,d\nu\\
  &=\tfrac{1}{4\pi\sqrt{2}m}e^{-iEt}\int_{\mu=0}^\pi\int_{\nu=0}^{2\pi}
       e^{ipr\cos\mu}(m{\bf e}_\emptyset-iE{\bf e}_t-ip\cos\mu\,{\bf e}_r
                     -ip\sin\mu\cos\nu\,\vec{u}\\
  &\hspace{200pt}
                     -ip\sin\mu\sin\nu\,\vec{v})\sin\mu\,d\mu\,d\nu\\
  &=\tfrac{1}{4\pi\sqrt{2}m}e^{-iEt}\int_{\mu=0}^\pi\int_{\nu=0}^{2\pi}
       e^{ipr\cos\mu}(m{\bf e}_\emptyset-iE{\bf e}_t-ip\cos\mu\,{\bf e}_r)\sin\mu\,d\mu\,d\nu,
\end{align*}
as $\int_{\nu=0}^{2\pi}\sin\nu\,d\nu=0=\int_{\nu=0}^{2\pi}\cos\nu\,d\nu$.  Now,
\begin{align*}
  \int_{\mu=0}^\pi\int_{\nu=0}^{2\pi}e^{ipr\cos\mu}\sin\mu\,d\mu\,d\nu
    &=(2\pi)(-\tfrac{e^{ipr\cos\mu}}{ipr})|_{\nu=0}^\pi
     =\tfrac{i2\pi}{pr}(e^{-ipr}-e^{ipr})\\
    &=\tfrac{i2\pi}{pr}(-2i\sin{pr})
     =4\pi j_0(pr).
\end{align*}
And making the substitution $u\doteq\cos\mu$, we compute
\begin{align*}
  &\int_{\mu=0}^\pi\int_{\nu=0}^{2\pi}e^{ipr\cos\mu}\cos\mu\sin\mu\,d\mu\,d\nu
    =2\pi\int_{u=-1}^1e^{ipru}u\,du\\
  &=2\pi(\tfrac{u}{ipr}e^{ipru}-\tfrac{1}{(ipr)^2}e^{ipru})|_{u=-1}^1
   =2\pi[\tfrac{1}{ipr}(e^{ipr}+e^{-ipr})+\tfrac{1}{(pr)^2}(e^{ipr}-e^{-ipr})]\\
  &=4\pi\bigl(\tfrac{\cos{pr}}{ipr}+\tfrac{i\sin{pr}}{(pr)^2}\bigr)
   =4\pi i\,\tfrac{\sin{pr}-pr\cos{pr}}{(pr)^2}
   =4\pi i\,j_1(pr).
\end{align*}
It follows that
\begin{align*}
  \int_{\mathbb M}\Psi(E,{\bf q})\,e^{-iq_\alpha x^\alpha}\chi_0^+({\bf q})\,d^4{\bf q}
  &=\tfrac{1}{\sqrt{2}m}e^{-iEt}
        (mj_0(pr){\bf e}_\emptyset-iEj_0(pr){\bf e}_t+pj_1(pr){\bf e}_r)\\
  &=\psi_0(E) 
\end{align*}

On the other hand, if $r=0$ then $q_\alpha x^\alpha=q^tt$.  Using spherical polar coordinates:
$$\vec{q}=|\vec{q}|(\sin\mu\cos\nu\,{\bf e}_x+\sin\mu\sin\nu\,{\bf e}_y
                    +\cos\mu\,{\bf e}_z),$$
we compute
\begin{align*}
  &\int_{\mathbb M}\Psi(E,{\bf q})\,e^{-iq_\alpha x^\alpha}\chi_0^+({\bf q})\,d^4{\bf q}\\
  &=\tfrac{1}{4\pi p^2}\int_{q^t=-\infty}^\infty\int_{|\vec{q}|=0}^\infty
                       \int_{\mu=0}^\pi\int_{\nu=0}^{2\pi}
                          \delta(E-q^t)\,\delta(p-|\vec{q}|)
                          e^{-iq^tt}\chi_0^+({\bf q})
                          |\vec{q}|^2\sin\mu\,dq^t\,d|\vec{q}|\,d\mu\,d\nu\\
  &=\tfrac{1}{4\pi\sqrt{2}m}e^{-iEt}\int_{\mu=0}^\pi\int_{\nu=0}^{2\pi}
                          [m{\bf e}_\emptyset-iE{\bf e}_t
                           -ip(\sin\mu\cos\nu\,{\bf e}_x+\sin\mu\sin\nu\,{\bf e}_y
                           +\cos\mu\,{\bf e}_z)]\sin\mu\,d\mu\,d\nu\\
  &=\tfrac{1}{2\sqrt{2}m}e^{-iEt}\int_{\mu=0}^\pi
                         (m{\bf e}_\emptyset-iE{\bf e}_t-ip\cos\mu\,{\bf e}_z)
                         \sin\mu\,d\mu\\
  &=\tfrac{1}{\sqrt{2}m}e^{-iEt}(m{\bf e}_\emptyset-iE{\bf e}_t)
   =\psi_0(E)
\end{align*}
as $j_0(pr)=1$ and $j_1(pr)=0$ when $r=0$.
\end{proof}

\begin{lemma}
$(ip^t{\mathcal I}-m\Theta_t)\chi_0^+({\bf p})=ip^x\chi_1^-({\bf p})-ip^y\chi_2^-({\bf p})+ip^z\chi_3^-({\bf p})$.
\end{lemma}

\begin{proof}
From Fact \ref{fact:spindep}, we have $(p^t\Theta_t+p^k\Theta_k)\chi_0^+({\bf p})=im\chi_0^+({\bf p})$.  Therefore,
\begin{align*}
  (ip^t{\mathcal I}-m\Theta_t)\chi_0^+({\bf p})
  &=ip^t\chi_0^+({\bf p})+i\Theta_t(p^t\Theta_t+p^k\Theta_k)\chi_0^+({\bf p})\\
  &=ip^t\chi_0^+({\bf p})-ip^t\chi_0^+({\bf p})-i(p^x\Theta_{tx}
        +p^y\Theta_{ty}+p^z\Theta_{tz})\chi_0^+({\bf p})\\
  &=-i(-p^x{\mathcal J}\Theta_{yz}+p^y{\mathcal J}\Theta_{xz}
       -p^z{\mathcal J}\Theta_{xy})\chi_0^+({\bf p})\\
  &=ip^x\chi_1^-({\bf p})-ip^y\chi_2^-({\bf p})+ip^z\chi_3^-({\bf p}).\qedhere
\end{align*}
\end{proof}

\begin{fact}
$\displaystyle\psi_1(E)=\int_{\mathbb M}\Psi(E,{\bf q})\,e^{-iq_\alpha x^\alpha}\chi_S^-({\bf q})\,d^4{\bf q}$, where $\chi_S^-({\bf p})\doteq iu^x\chi_1^-({\bf p})-iu^y\chi_2^-({\bf p})+iu^z\chi_3^-({\bf p})$ and $\vec{u}=\vec{p}/|\vec{p}|$.
\end{fact}

\begin{proof}
Using the previous lemma, the fact that $\Psi(E,{\bf q})$ has factors $\delta(E-q^t)$ and $\delta(p-|\vec{q}|)$, Fact \ref{fact:splanedecomp0}, and Fact \ref{fact:snonsing}, we have
\begin{align*}
  \int_{\mathbb M}\Psi(E,{\bf q})\,e^{-iq_\alpha x^\alpha}\chi_S^-({\bf q})\,d^4{\bf q}
  &=\int_{\mathbb M}\Psi(E,{\bf q})\,e^{-iq_\alpha x^\alpha}
        (\tfrac{iq^t}{|\vec{q}|}{\mathcal I}-\tfrac{m}{|\vec{q}|}\Theta_t)
        \chi_0^+({\bf q})\,d^4{\bf q}\\
  &=(\tfrac{iE}{p}{\mathcal I}-\tfrac{m}{p}\Theta_t)
      \int_{\mathbb M}\Psi(E,{\bf q})\,e^{-iq_\alpha x^\alpha}
         \chi_0^+({\bf q})\,d^4{\bf q}\\
  &=(\tfrac{iE}{p}{\mathcal I}-\tfrac{m}{p}\Theta_t)\psi_0(E)
   =\psi_1(E).\qedhere
\end{align*}
\end{proof}

\begin{fact}
$\chi_S^-({\bf p})=\frac{1}{\sqrt{2}m}(-im{\bf e}_t\wedge\vec{u}+|\vec{p}|{\bf e}_t+p^t\vec{u})$, where $\vec{u}=\vec{p}/|\vec{p}|$.
\end{fact}

\begin{proof}
\begin{align*}
  \chi_S^-({\bf p})
  &=iu^x\chi_1^-({\bf p})-iu^y\chi_2^-({\bf p})+iu^z\chi_3^-({\bf p})\\
  &=\tfrac{i}{\sqrt{2}m}[
       u^x(-m{\bf e}_{tx}-ip^t{\bf e}_x-ip^x{\bf e}_t+ip^y{\bf e}_{txy}+ip^z{\bf e}_{txz})\\
  &\quad\quad\quad
      -u^y(m{\bf e}_{ty}+ip^t{\bf e}_y+ip^x{\bf e}_{txy}+ip^y{\bf e}_t-ip^z{\bf e}_{tyz})\\
  &\quad\quad\quad
      +u^z(-m{\bf e}_{tz}-ip^t{\bf e}_z-ip^x{\bf e}_{txz}-ip^y{\bf e}_{tyz}-ip^z{\bf e}_t)]\\
  &=\tfrac{i}{\sqrt{2}m}[-m{\bf e}_t\wedge(u^x{\bf e}_x+u^y{\bf e}_y+u^y{\bf e}_z)
                         -i(u^xp^x+u^yp^y+u^zp^z){\bf e}_t\\
  &\quad\quad\quad
                         -ip^t(u^x{\bf e}_x+u^y{\bf e}_y+u^z{\bf e}_x)
                         +i(u^xp^y-u^yp^x){\bf e}_{txy}\\
  &\quad\quad\quad
                         +i(u^xp^z-u^zp^x){\bf e}_{txz}
                         +i(u^yp^z-u^zp^y){\bf e}_{tyz}]\\
  &=\tfrac{i}{\sqrt{2}m}(-im{\bf e}_t\wedge\vec{u}+|\vec{p}|{\bf e}_t
                         +p^t\vec{u}).\qedhere
\end{align*}
\end{proof}

%::::::::::::::::::::::::::::::::::::::::::::::::::::::::::::::::
\subsection{Moving spherical solutions}

We apply a Lorentz boost to the solutions in Fact \ref{fact:sssol}.  Assume that a particle is moving at velocity $\vec{v}=(v^x,v^y,v^z)=v^k{\bf e}_k$ in the observer's frame.  Let $(x^\alpha)=(t,\vec{r})$ with $\vec{r}=(x,y,z)=x^k{\bf e}_x$ be coordinates in the observer's frame, and $(x'^\alpha)=(t',\vec{r}\,')$ coordinates in the rest frame.  Setting $\vec{u}\doteq\vec{v}/|\vec{v}|$, it is well--known thatu
\begin{equation}\label{eq:lorentzboost}
  t=\gamma(t'+\vec{r}\,'\cdot\vec{v})
  \quad\text{and}\quad
  \vec{r}=\vec{r}\,'+(\gamma-1)(\vec{r}\,'\cdot\vec{u})\vec{u}+\gamma t'\vec{v},
\end{equation}
and
\begin{equation}\label{eq:ilorentzboost}
  t'=\gamma(t-\vec{r}\cdot\vec{v})
  \quad\text{and}\quad
  \vec{r}\,'=\vec{r}+(\gamma-1)(\vec{r}\cdot\vec{u})\vec{u}-\gamma t\vec{v}.
\end{equation}
Here $\gamma\doteq 1/\sqrt{1-|\vec{v}|^2}$.

\begin{lemma}\label{lem:obsbasis}
If ${\bf e}_t$ and ${\bf e}_k$ are the basis vectors of Minkowski space in the observer's frame, the the basis vector in the rest frame are
$${\bf e}_{t'}=\gamma({\bf e}_t+\vec{v})
  \quad\text{and}\quad
  {\bf e}_{k'}={\bf e}_k+\gamma v^k{\bf e}_t+(\gamma-1)u^k\vec{u}.
$$
\end{lemma}

\begin{proof}
We compute the Lorentz transform of the basis vectors in the observer's frame.  From equation \eqref{eq:lorentzboost} we have
\begin{align*}
  {\bf e}_{t'}
  &=\tfrac{\partial t}{\partial t'}{\bf e}_t
    +\tfrac{\partial x}{\partial t'}{\bf e}_x
    +\tfrac{\partial y}{\partial t'}{\bf e}_y
    +\tfrac{\partial z}{\partial t'}{\bf e}_z
   =\gamma{\bf e}_t+\gamma v^x{\bf e}_x
    +\gamma v^y{\bf e}_y
    +\gamma v^z{\bf e}_z\\
  {\bf e}_{x'}
  &=\tfrac{\partial t}{\partial x'}{\bf e}_t
    +\tfrac{\partial x}{\partial x'}{\bf e}_x
    +\tfrac{\partial y}{\partial x'}{\bf e}_y
    +\tfrac{\partial z}{\partial x'}{\bf e}_z\\
  &=\gamma v^x{\bf e}_t
    +[1+(\gamma-1)u^xu^x]{\bf e}_x
    +(\gamma-1)u^xu^y{\bf e}_y
    +(\gamma-1)u^xu^z{\bf e}_z\\
  {\bf e}_{y'}
  &=\tfrac{\partial t}{\partial y'}{\bf e}_t
    +\tfrac{\partial x}{\partial y'}{\bf e}_x
    +\tfrac{\partial y}{\partial y'}{\bf e}_y
    +\tfrac{\partial z}{\partial y'}{\bf e}_z\\
  &=\gamma v^y{\bf e}_t
    +(\gamma-1)u^yu^x{\bf e}_x
    +[1+(\gamma-1)u^yu^y]{\bf e}_y
    +(\gamma-1)u^yu^z{\bf e}_z\\
  {\bf e}_{z'}
  &=\tfrac{\partial t}{\partial z'}{\bf e}_t
    +\tfrac{\partial x}{\partial z'}{\bf e}_x
    +\tfrac{\partial y}{\partial z'}{\bf e}_y
    +\tfrac{\partial z}{\partial z'}{\bf e}_z\\
  &=\gamma v^z{\bf e}_t
    +(\gamma-1)u^zu^x{\bf e}_x
    +(\gamma-1)u^zu^y{\bf e}_y
    +[1+(\gamma-1)u^zu^z]{\bf e}_z.\qedhere
\end{align*}
\end{proof}

\begin{fact}\label{fact:sboostbasis}
\begin{align*}
  {\bf e}_{r'}
    &=\tfrac{1}{\sigma}[\vec{r}
      +\gamma^2(\vec{r}\cdot\vec{v}-t)\vec{v}
      +\gamma^2(\vec{r}\cdot\vec{v}-|\vec{v}|^2t){\bf e}_t]\\
  {\bf e}_{t'r'}
    &=\tfrac{\gamma}{\sigma}
        [{\bf e}_t\wedge(\vec{r}-t\vec{v})+\vec{v}\wedge\vec{r}],
\end{align*}
where
$$\sigma\doteq\sqrt{r^2+\gamma^2(\vec{r}\cdot\vec{v})^2
                       -2\gamma^2(\vec{r}\cdot\vec{v})t
                       +\gamma^2|\vec{v}|^2t^2}.
$$
\end{fact}

\begin{proof}
Set $\sigma\doteq|\vec{r}\,'|=r'$.  Observe that from equation \eqref{eq:ilorentzboost},
\begin{align*}
  |\vec{r}\,'|^2
  &=|\vec{r}|^2+2(\gamma-1)(\vec{r}\cdot\vec{u})^2-2\gamma t(\vec{r}\cdot\vec{v})
    +(\gamma-1)^2(\vec{r}\cdot\vec{u})^2\\
  &\quad\quad
    -2(\gamma-1)\gamma t(\vec{r}\cdot\vec{u})(\vec{v}\cdot\vec{u})
    +\gamma^2t^2|\vec{v}|^2\\
  &=r^2+[2(\gamma-1)+(\gamma-1)^2](\vec{r}\cdot\vec{u})^2
    -2\gamma t[1+(\gamma-1)](\vec{r}\cdot\vec{v})
    +\gamma^2t^2|\vec{v}|^2\\
  &=r^2+\gamma^2(\vec{r}\cdot\vec{v})^2-2\gamma^2(\vec{r}\cdot\vec{v})t
    +\gamma^2|\vec{v}|^2t^2,
\end{align*}
as $(\vec{r}\cdot\vec{u})(\vec{v}\cdot\vec{u})=(\vec{r}\cdot\vec{u})|\vec{v}|(\vec{u}\cdot\vec{u})=(\vec{r}\cdot\vec{v})$ and $2(\gamma-1)+(\gamma-1)^2=\gamma^2-1=\gamma^2|\vec{v}|^2$.  Which gives the stated expression for $\sigma$.  Now ${\bf e}_{r'}=\tfrac{x'}{r'}{\bf e}_{x'}+\tfrac{y'}{r'}{\bf e}_{y'}+\tfrac{z'}{r'}{\bf e}_{z'}$, so from equation \eqref{eq:ilorentzboost} and lemma \ref{lem:obsbasis},
\begin{align*}
  \sigma{\bf e}_{r'}
  &=x'{\bf e}_{x'}+y'{\bf e}_{y'}+z'{\bf e}_{z'}\\
  &=[x+(\gamma-1)(\vec{r}\cdot\vec{u})u^x-\gamma tv^x]
      [{\bf e}_x+\gamma v^x{\bf e}_t+(\gamma-1)u^x\vec{u}]\\
  &\quad\quad
    +[y+(\gamma-1)(\vec{r}\cdot\vec{u})u^y-\gamma tv^y]
       [{\bf e}_y+\gamma v^y{\bf e}_t+(\gamma-1)u^y\vec{u}]\\
  &\quad\quad
    +[z+(\gamma-1)(\vec{r}\cdot\vec{u})u^z-\gamma tv^z]
       [{\bf e}_z+\gamma v^z{\bf e}_t+(\gamma-1)u^z\vec{u}]\\
  &=(x{\bf e}_x+y{\bf e}_y+z{\bf e}_z)
    +\gamma(xv^x+yv^y+zv^z){\bf e}_t\\
  &\quad\quad
    +(\gamma-1)(xu^x+yu^y+zu^z)\vec{u}
    +(\gamma-1)(\vec{r}\cdot\vec{u})(u^x{\bf e}_x+u^y{\bf e}_y+u^z{\bf e}_z)\\
  &\quad\quad
    +(\gamma-1)(\vec{r}\cdot\vec{u})\gamma(u^xv^x+u^yv^y+u^zv^z){\bf e}_t\\
  &\quad\quad
    +(\gamma-1)^2(\vec{r}\cdot\vec{u})(u^xu^x+u^yu^y+u^zu^z)\vec{u}
    -\gamma t(v^x{\bf e}_x+v^y{\bf e}_y+v^z{\bf e}_z)\\
  &\quad\quad
    -\gamma^2 t(v^xv^x+v^yv^y+v^zv^z){\bf e}_t
    -\gamma(\gamma-1)t(v^xu^x+v^yu^y+v^zu^z)\vec{u}\\
  &=\vec{r}+\gamma(\vec{r}\cdot\vec{v}){\bf e}_t
    +(\gamma-1)(\vec{r}\cdot\vec{u})\vec{u}
    +(\gamma-1)(\vec{r}\cdot\vec{u})\vec{u}\\
  &\quad\quad
    +(\gamma-1)(\vec{r}\cdot\vec{u})\gamma(\vec{u}\cdot\vec{v}){\bf e}_t
    +(\gamma-1)^2(\vec{r}\cdot\vec{u})\vec{u}
    -\gamma t\vec{v}
    -\gamma^2 t|\vec{v}|^2{\bf e}_t\\
  &\quad\quad
    -\gamma(\gamma-1)t(\vec{v}\cdot\vec{u})\vec{u}\\
  &=\vec{r}+[2(\gamma-1)(\vec{r}\cdot\vec{u})+(\gamma-1)^2(\vec{r}\cdot\vec{u})
             -\gamma t|\vec{v}|-\gamma(\gamma-1)t(\vec{v}\cdot\vec{u})]\vec{u}\\
  &\quad\quad
    +[\gamma(\vec{r}\cdot\vec{v})+\gamma(\gamma-1)(\vec{r}\cdot\vec{u})(\vec{u}\cdot\vec{v})
      -\gamma^2 t|\vec{v}|^2]{\bf e}_t\\
  &=\vec{r}+\gamma^2(\vec{r}\cdot\vec{v}-t)\vec{v}
    +\gamma^2(\vec{r}\cdot\vec{v}-t|\vec{v}|^2){\bf e}_t
\end{align*}
Finally, using this equation and lemma \ref{lem:obsbasis} again, we get
\begin{align*}
  \sigma{\bf e}_{t'r'}
  &={\bf e}_{t'}\wedge(\sigma{\bf e}_{r'})
   =\gamma({\bf e}_t+\vec{v})
    \wedge[\vec{r}+\gamma^2(\vec{r}\cdot\vec{v}-t)\vec{v}
           +\gamma^2(\vec{r}\cdot\vec{v}-t|\vec{v}|^2){\bf e}_t]\\
  &=\gamma[{\bf e}_t\wedge\vec{r}+\gamma^2(\vec{r}\cdot\vec{v}-t){\bf e}_t\wedge\vec{v}
           +\vec{v}\wedge\vec{r}+\gamma^2(\vec{r}\cdot\vec{v}-t|\vec{v}|^2)\vec{v}\wedge{\bf e}_t]\\
  &=\gamma[{\bf e}_t\wedge\vec{r}+\vec{v}\wedge\vec{r}
           +\gamma^2(\vec{r}\cdot\vec{v}-t-\vec{r}\cdot\vec{v}+t|\vec{v}|^2){\bf e}_t\wedge\vec{v}]\\
  &=\gamma({\bf e}_t\wedge\vec{r}+\vec{v}\wedge\vec{r}-t{\bf e}_t\wedge\vec{v})
   =\gamma[{\bf e}_t\wedge(\vec{r}-t\vec{v})+\vec{v}\wedge\vec{r}].\qedhere
\end{align*}
\end{proof}

\begin{fact}
In the observer's frame, the solutions in Fact \ref{fact:sssol} are
\begin{align*}
  \psi_0(E,\vec{v})
  &=e^{-iE\gamma(t-\vec{r}\cdot\vec{v})}
    \{mj_0(p\sigma){\bf e}_\emptyset
      -iE\gamma j_0(p\sigma)({\bf e}_t+\vec{v})\\
  &\hspace{84pt}
      +\tfrac{p}{\sigma}j_1(p\sigma)
         [\vec{r}+\gamma^2(\vec{r}\cdot\vec{v}-t)\vec{v}
          +\gamma^2(\vec{r}\cdot\vec{v}-|\vec{v}|^2t){\bf e}_t]\}\\
  \psi_1(E,\vec{v})
  &=e^{-iE\gamma(t-\vec{r}\cdot\vec{v})}
    \{\tfrac{m\gamma}{\sigma}j_1(p\sigma)
        [{\bf e}_t\wedge(\vec{r}-t\vec{v})+\vec{v}\wedge\vec{r}]
      +p\gamma j_0(p\sigma)({\bf e}_t+\vec{v})\\
  &\hspace{84pt}
      +\tfrac{iE}{\sigma}j_1(p\sigma)
       [\vec{r}+\gamma^2(\vec{r}\cdot\vec{v}-t)\vec{v}
          +\gamma^2(\vec{r}\cdot\vec{v}-|\vec{v}|^2t){\bf e}_t]\}
\end{align*}
\end{fact}

\begin{proof}
These follow immediately from equation \eqref{eq:ilorentzboost} and Fact \ref{fact:sboostbasis}.
\end{proof}

%%%%%%%%%%%%%%%%%%%%%%%%%%%%%%%%%%%%%%%%%%%%%%%%%%%%%%%%%%%%%%%%
\appendix

%::::::::::::::::::::::::::::::::::::::::::::::::::::::::::::::::
\section{Theta matrix multiplication table}\label{sec:appendixA}

In rectangular coordinates, the product of the theta matrices $\Theta_I\Theta_J$, with $I$ and $J$ multi--indices, will also be a theta matrix up to sign.  That is,
\begin{equation}\label{eq:IJK}
  \Theta_I\Theta_J=\pm\Theta_K
\end{equation}
for some multi--index $K$.  For example, we compute the product $\Theta_{xy}\Theta_{tyz}$.  From Fact \ref{fact:thetaunique}, we only need to compute its effect on ${\bf e}_\emptyset$:
\begin{align*}
  \Theta_{xy}\Theta_{tyz}{\bf e}_\emptyset
  &=\Theta_{xy}{\bf e}_{tyz}
   =\gamma_{tyz}{\bf e}_{xy}
   =\gamma_{ty}\gamma_z{\bf e}_{xy}
   =\gamma_{ty}({\bf e}_z\wedge{\bf e}_{xy}-\iota_z{\bf e}_{xy})\\
  &=\gamma_{ty}{\bf e}_{zxy}
   =\gamma_{ty}{\bf e}_{xyz}
   =\gamma_t\gamma_y{\bf e}_{xyz}
   =\gamma_t({\bf e}_y\wedge{\bf e}_{xyz}-\iota_y{\bf e}_{xyz})\\
  &=-\gamma_t[-{\bf e}_x\wedge(\iota_y{\bf e}_y)\wedge{\bf e}_z]
   =\eta_{yy}\gamma_t{\bf e}_{xz}
   =\eta_{yy}({\bf e}_t\wedge{\bf e}_{xz}-\iota_t{\bf e}_{xz})\\
  &=\eta_{yy}{\bf e}_{txz}
   =-{\bf e}_{txz}
   =-\Theta_{txz}{\bf e}_\emptyset.
\end{align*}
So that $\Theta_{xy}\Theta_{tyz}=-\Theta_{txz}$.

The entry in the $I$--row and $J$--column of tables \ref{table:thetatable1} and \ref{table:thetatable2} gives the (signed) value of $K$ in equation \eqref{eq:IJK}.

\begin{table}
\begin{center}
{\small
\begin{tabular}{c||c|c|c|c|c|c|c|c|}
$I\backslash J$ & $\emptyset$ & $t$ & $x$ & $y$ & $z$
   & ${tx}$ & ${ty}$ & ${tz}$\\\hline\hline
$\emptyset$ & $\emptyset$ & $t$ & $x$ & $y$ & $z$ & $tx$ & $ty$ & $tz$\\\hline
$t$ & $t$ & $-\emptyset$ & $-tx$ & $-ty$ & $-tz$ & $x$ & $y$ & $z$\\\hline
$x$ & $x$ & $tx$ & $\emptyset$ & $-xy$ & $-xz$ & $t$ & $-txy$ & $-txz$\\\hline
$y$ & $y$ & $ty$ & $xy$ & $\emptyset$ & $-yz$ & $txy$ & $t$ & $-tyz$\\\hline
$z$ & $z$ & $tz$ & $xz$ & $yz$ & $\emptyset$ & $txz$ & $tyz$ & $t$\\\hline
$tx$ & $tx$ & $-x$ & $-t$ & $txy$ & $txz$ & $\emptyset$ & $-xy$ & $-xz$\\\hline
$ty$ & $ty$ & $-y$ & $-txy$ & $-t$ & $tyz$ & $xy$ & $\emptyset$ & $-yz$\\\hline
$tz$ & $tz$ & $-z$ & $-txz$ & $-tyz$ & $-t$ & $xz$ & $yz$ & $\emptyset$\\\hline
$xy$ & $xy$ & $txy$ & $y$ & $-x$ & $xyz$ & $ty$ & $-tx$ & $txyz$\\\hline
$xz$ & $xz$ & $txz$ & $z$ & $-xyz$ & $-x$ & $tz$ & $-txyz$ & $-tx$\\\hline
$yz$ & $yz$ & $tyz$ & $xyz$ & $z$ & $-y$ & $txyz$ & $tz$ & $-ty$\\\hline
$txy$ & $txy$ & $-xy$ & $-ty$ & $tx$ & $-txyz$ & $y$ & $-x$ & $xyz$\\\hline
$txz$ & $txz$ & $-xz$ & $-tz$ & $txyz$ & $tx$ & $z$ & $-xyz$ & $-x$\\\hline
$tyz$ & $tyz$ & $-yz$ & $-txyz$ & $-tz$ & $ty$ & $xyz$ & $z$ & $-y$\\\hline
$xyz$ & $xyz$ & $txyz$ & $yz$ & $-xz$ & $xy$ & $tyz$ & $-txz$ & $txy$\\\hline
$txyz$ & $txyz$ & $-xyz$ & $-tyz$ & $txz$ & $-txy$ & $yz$ & $-xz$ & $xy$\\\hline
\end{tabular}
}
\end{center}
\caption{Theta matrix multiplication table, part 1.}
\label{table:thetatable1}
\end{table}

\begin{table}
\begin{center}
{\small
\begin{tabular}{c||c|c|c|c|c|c|c|c|}
$I\backslash J$ & $xy$ & $xz$ & $yz$ & $txy$ & $txz$
   & ${tyz}$ & ${xyz}$ & ${txyz}$\\\hline\hline
$\emptyset$ & $xy$ & $xz$ & $yz$ & $txy$ & $txz$ & $tyz$ & $xyz$ & $txyz$\\\hline
$t$ & $txy$ & $txz$ & $tyz$ & $-xy$ & $-xz$ & $-yz$ & $-txyz$ & $xyz$\\\hline
$x$ & $-y$ & $-z$ & $xyz$ & $-ty$ & $-tz$ & $txyz$ & $yz$ & $tyz$\\\hline
$y$ & $x$ & $-xyz$ & $-z$ & $tx$ & $-txyz$ & $-tz$ & $-xz$ & $-txz$\\\hline
$z$ & $xyz$ & $x$ & $y$ & $txyz$ & $tx$ & $ty$ & $xy$ & $txy$\\\hline
$tx$ & $-ty$ & $-tz$ & $txyz$ & $y$ & $z$ & $-xyz$ & $-tyz$ & $yz$\\\hline
$ty$ & $tx$ & $-txyz$ & $-tz$ & $-x$ & $xyz$ & $z$ & $txz$ & $-xz$\\\hline
$tz$ & $txyz$ & $tx$ & $ty$ & $-xyz$ & $-x$ & $-y$ & $-txy$ & $xy$\\\hline
$xy$ & $-\emptyset$ & $yz$ & $-xz$ & $-t$ & $tyz$ & $-txz$ & $-z$ & $-tz$\\\hline
$xz$ & $-yz$ & $-\emptyset$ & $xy$ & $-tyz$ & $-t$ & $txy$ & $y$ & $ty$\\\hline
$yz$ & $xz$ & $-xy$ & $-\emptyset$ & $txz$ & $-txy$ & $-t$ & $-x$ & $-tx$\\\hline
$txy$ & $-t$ & $tyz$ & $-txz$ & $\emptyset$ & $-yz$ & $xz$ & $tz$ & $-z$\\\hline
$txz$ & $-tyz$ & $-t$ & $txy$ & $yz$ & $\emptyset$ & $-xy$ & $-ty$ & $y$\\\hline
$tyz$ & $txz$ & $-txy$ & $-t$ & $-xz$ & $xy$ & $\emptyset$ & $tx$ & $-x$\\\hline
$xyz$ & $-z$ & $y$ & $-x$ & $-tz$ & $ty$ & $-tx$ & $-\emptyset$ & $-t$\\\hline
$txyz$ & $-tz$ & $ty$ & $-tx$ & $z$ & $-y$ & $x$ & $t$ & $-\emptyset$\\\hline
\end{tabular}
}
\end{center}
\caption{Theta matrix multiplication table, part 2.}
\label{table:thetatable2}
\end{table}

%::::::::::::::::::::::::::::::::::::::::::::::::::::::::::::::::
\section{Square of the extended Dirac operator}

Here we assume a general Lorentz metric $g_{\alpha\beta}$ on a general spacetime manifold.  The extended Dirac operator is $D=\gamma^\alpha\hat\nabla_\alpha$, where $\hat\nabla_\alpha=\partial_\alpha+\hat\Gamma_\alpha$.  The extended curvature operator is $\hat\Omega_{\alpha\beta}=\hat\nabla_\alpha\hat\nabla_\beta-\hat\nabla_\beta\hat\nabla_\alpha=\partial_\alpha\hat\Gamma_\beta-\partial_\beta\hat\Gamma_\alpha+[\hat\Gamma_\alpha,\hat\Gamma_\beta]$.

\begin{fact}\label{fact:DD}
$D^2\psi=-g^{\alpha\beta}\partial_{\alpha\beta}\psi+{\mathcal A}^\alpha\partial_\alpha\psi+{\mathcal B}\psi$, where ${\mathcal A}^\alpha\doteq g^{\rho\sigma}{\Gamma^\alpha}_{\rho\sigma}{\mathcal I}-2\hat\Gamma^\alpha$, and ${\mathcal B}\doteq g^{\rho\sigma}{\Gamma^\beta}_{\rho\sigma}\hat\Gamma_\beta-\partial^\beta\hat\Gamma_\beta-\hat\Gamma^\beta\hat\Gamma_\beta+\tfrac{1}{2}\gamma^\rho\gamma^\sigma\hat\Omega_{\rho\sigma}$.
\end{fact}

\begin{proof}
In {\it Coupling the Dirac and Einstein equation through geometry} [Found. Phys. 52, 1 (2022)], it is shown that $[\gamma^\beta,\hat\Gamma_\alpha]=\partial_\alpha\gamma^\beta+{\Gamma^\beta}_{\alpha\mu}\gamma^\mu$.  Thus,
\begin{align*}
  D^2\psi
  &=(\gamma^\alpha\partial_\alpha+\gamma^\alpha\hat\Gamma_\alpha)
      \gamma^\beta\hat\nabla_\beta\psi\\
  &=\gamma^\alpha(\partial_\alpha\gamma^\beta)\hat\nabla_\beta\psi
    +\gamma^\alpha\gamma^\beta\partial_\alpha\hat\nabla_\beta\psi
    +\gamma^\alpha\hat\Gamma_\alpha\gamma^\beta\hat\nabla_\beta\psi\\
  &=\gamma^\alpha(\partial_\alpha\gamma^\beta
                  +\hat\Gamma_\alpha\gamma^\beta)\hat\nabla_\beta\psi
    +\gamma^\alpha\gamma^\beta\hat\nabla_\alpha\hat\nabla_\beta\psi
    -\gamma^\alpha\gamma^\beta\hat\Gamma_\alpha\hat\nabla_\beta\psi\\
  &=\gamma^\alpha(\gamma^\beta\hat\Gamma_\alpha
                  -{\Gamma^\beta}_{\alpha\mu}\gamma^\mu)\hat\nabla_\beta\psi
    +\gamma^\alpha\gamma^\beta\hat\nabla_\alpha\hat\nabla_\beta\psi
    -\gamma^\alpha\gamma^\beta\hat\Gamma_\alpha\hat\nabla_\beta\psi\\
  &=-{\Gamma^\beta}_{\alpha\mu}\gamma^\mu\gamma^\alpha\hat\nabla_\beta\psi
    +\gamma^\alpha\gamma^\beta\hat\nabla_\alpha\hat\nabla_\beta\psi\\
  &=g^{\alpha\mu}{\Gamma^\beta}_{\alpha\mu}\hat\nabla_\beta\psi
    +\gamma^\alpha\gamma^\beta\hat\nabla_\alpha\hat\nabla_\beta\psi\\
  &=g^{\alpha\mu}{\Gamma^\beta}_{\alpha\mu}\partial_\beta\psi
    +g^{\alpha\mu}{\Gamma^\beta}_{\alpha\mu}\hat\Gamma_\beta\psi
    +\gamma^\alpha\gamma^\beta\hat\nabla_\alpha\hat\nabla_\beta\psi
\end{align*}
(making use of the Clifford algebra relation, Fact \ref{fact:gammaclifford}).  Now, the last term can be written as
\begin{align*}
  \gamma^\alpha\gamma^\beta\hat\nabla_\alpha\hat\nabla_\beta\psi
  &=\tfrac{1}{2}(\gamma^\alpha\gamma^\beta-\gamma^\beta\gamma^\alpha
                 -2g^{\alpha\beta}{\mathcal I})\hat\nabla_\alpha\hat\nabla_\beta\psi\\
  &=\tfrac{1}{2}\gamma^\alpha\gamma^\beta(\hat\nabla_\alpha\hat\nabla_\beta
                                          -\hat\nabla_\beta\hat\nabla_\alpha)\psi
    -g^{\alpha\beta}\hat\nabla_\alpha\hat\nabla_\beta\psi\\
  &=\tfrac{1}{2}\gamma^\alpha\gamma^\beta\hat\Omega_{\alpha\beta}\psi
    -g^{\alpha\beta}\hat\nabla_\alpha\hat\nabla_\beta\psi.
\end{align*}
Moreover, we have
\begin{align*}
  g^{\alpha\beta}\hat\nabla_\alpha\hat\nabla_\beta\psi
  &=g^{\alpha\beta}(\partial_\alpha+\hat\Gamma_\alpha)
     (\partial_\beta\psi+\hat\Gamma_\beta\psi)\\
  &=g^{\alpha\beta}[\partial_{\alpha\beta}\psi
                    +(\partial_\alpha\hat\Gamma_\beta)\psi
                    +\hat\Gamma_\beta\partial_\alpha\psi
                    +\hat\Gamma_\alpha\partial_\beta\psi
                    +\hat\Gamma_\alpha\hat\Gamma_\beta\psi]\\
  &=g^{\alpha\beta}\partial_{\alpha\beta}\psi
    +(\partial^\beta\hat\Gamma_\beta)\psi
    +2\hat\Gamma^\alpha\partial_\alpha\psi
    +\hat\Gamma^\beta\hat\Gamma_\beta\psi.\qedhere
\end{align*}
\end{proof}

\begin{lemma}\label{lem:curveleibniz}
The extendend curvature satisfies the Leibniz rule: $\hat\Omega_{\alpha\beta}(\psi\wedge\phi)=(\hat\Omega_{\alpha\beta}\psi)\wedge\phi+\psi\wedge(\hat\Omega_{\alpha\beta}\phi)$.
\end{lemma}

\begin{proof}
Because the extended connection satisfies the Leibniz rule, we have
\begin{align*}
  \hat\nabla_\alpha\hat\nabla_\beta(\psi\wedge\phi)
  &=\hat\nabla_\alpha[(\hat\nabla_\beta\psi)\wedge\phi
                      +\psi\wedge(\hat\nabla_\beta\phi)]\\
  &=(\hat\nabla_\alpha\hat\nabla_\beta\psi)\wedge\phi
    +(\hat\nabla_\beta\psi)\wedge(\hat\nabla_\alpha\phi)\\
  &\quad\quad
    +(\hat\nabla_\alpha\psi)\wedge(\hat\nabla_\beta\phi)
    +\psi\wedge(\hat\nabla_\alpha\hat\nabla_\beta\phi).
\end{align*}
So that $(\hat\nabla_\alpha\hat\nabla_\beta-\hat\nabla_\beta\hat\nabla_\alpha)(\psi\wedge\phi)$ is equal to
$$[(\hat\nabla_\alpha\hat\nabla_\beta-\hat\nabla_\beta\hat\nabla_\alpha)\psi]\wedge\phi
  +\psi\wedge[(\hat\nabla_\alpha\hat\nabla_\beta-\hat\nabla_\beta\hat\nabla_\alpha)\phi].
  \qedhere
$$
\end{proof}

\begin{fact}
The curvature is trivial if and only if the extended curvature is.
\end{fact}

\begin{proof}
The extended curvature restricts to the curvature: $\hat\Omega_{\alpha\beta}{\bf e}_\alpha=\Omega_{\alpha\beta}{\bf e}_\alpha$.  This established the sufficiency.  The necessity follows from lemma \ref{lem:curveleibniz}.
\end{proof}